\documentclass[twocolumn,showpacs,prd,aps,nofootinbib,showkeys,unsortedaddress,10pt]{revtex4-2}

\usepackage{amsmath}
\usepackage{amssymb}
\usepackage{slashed}
\usepackage{hyperref}
\usepackage{color}
\hypersetup{
    bookmarksnumbered=true,
    colorlinks=true,       % false: boxed links; true: colored links
    linkcolor=red,          % color of internal links
    citecolor=darkgreen,        % color of links to bibliography
    filecolor=magenta,      % color of file links
    urlcolor=cyan,           % color of external links
    menucolor=blue,
}
\usepackage{comment}
\usepackage{simpler-wick}
\usepackage[compat=1.1.0]{tikz-feynhand}
\usepackage[capitalise]{cleveref}
\crefname{section}{Sec.}{Sec.}
\crefname{appendix}{App.}{App.}
\creflabelformat{appendix}{#2#1#3}

\definecolor{darkgreen}{rgb}{0,.7,0}

\newcommand{\be}{\begin{equation}}
\newcommand{\ee}{\end{equation}}
\newcommand{\infinity}{\infty}
\newcommand{\infinty}{\infty}

\newcommand{\ess}{\hspace{0.1em}}

\newcommand{\msbar}{$\overline{\mathrm{MS}}$}

\renewcommand{\Im}{\mathrm{Im}\,}
\newcommand{\im}{\mathrm{Im}\,}
\renewcommand{\Re}{\mathrm{Re}\,}
\newcommand{\re}{\mathrm{Re}\,}

\newcommand{\sech}{\mathrm{sech}}

\DeclareMathOperator{\sinc}{sinc}

\DeclareMathOperator{\arctanh}{arctanh}
\DeclareMathOperator{\arccoth}{arccoth}

\makeatletter
\newcommand{\pushright}[1]{\ifmeasuring@#1\else\omit\hfill$\displaystyle#1$\fi\ignorespaces}
\newcommand{\pushleft}[1]{\ifmeasuring@#1\else\omit$\displaystyle#1$\hfill\fi\ignorespaces}
\makeatother

\begin{document}

\title{Finite System Size Correction to the Effective Coupling in \texorpdfstring{$\phi^4$}{phi4} Scattering}

\author{W.\ A.\ Horowitz}
\email{wa.horowitz@uct.ac.za}
\affiliation{%
 Department of Physics, University of Cape Town, Rondebosch 7701, South Africa
}

\author{J.\ F.\ Du Plessis}
\email{dpljea028@myuct.ac.za}
\affiliation{%
 Department of Mathematics and Applied Mathematics, University of Cape Town, Rondebosch 7701, South Africa
}

\date{\today}

\begin{abstract}
    We compute and explore numerically the finite system size correction to NLO $2\to2$ scattering in massive scalar $\phi^4$ theory.  The derivation uses ``denominator regularization'' (instead of the usual dimensional regularization) on a spacetime with spatial directions compactified to a torus, with characteristic lengths not necessarily of equal size.  We determine a useful analytic continuation of the generalized Epstein zeta function to isolate the usual UV divergence.  Self-consistently, the renormalized finite system size correction reduces to zero as the system size goes to infinity and, further, satisfies the optical theorem.  One of our checks of unitarity leads to a generalization of a number theoretic result from Hardy and Ramanujan.  Precise numerical exploration of the finite system size correction to the amplitude and coupling when two spatial dimensions are finite requires the exploitation of the analytic structure of the finite system size result via a dispersion relation.  We find that the finite system size scattering amplitude exhibits ``geometric'' bound states.  Even away from these bound states, the finite system size correction to the effective coupling can be large.
\end{abstract}
\maketitle

%%%%%%%%%%%%%%%%%%%%%%%%%%%%%%%%%%%%%%%%%%%%%%%%%%%%%%%%%%%%%%%%%%%%%%%%%%%%%%%%%%%%%%%%%%%%%%%%%%%%%%
\section{Introduction}
%%%%%%%%%%%%%%%%%%%%%%%%%%%%%%%%%%%%%%%%%%%%%%%%%%%%%%%%%%%%%%%%%%%%%%%%%%%%%%%%%%%%%%%%%%%%%%%%%%%%%%
Observations from the Large Hadron Collider (LHC) reveal that the indications of quark-gluon plasma (QGP) formation, observed in large nucleus-nucleus collisions, are also present in high multiplicity proton-proton (p+p) and proton-nucleus (p+A) collisions \cite{Abelev:2012ola,Aad:2012gla,Adare:2013piz,Aad:2015gqa}. Remarkably, the distribution and correlations of low momentum particles in these smaller collision systems can be effectively described using relativistic, nearly inviscid hydrodynamics \cite{Bzdak:2013zma,Weller:2017tsr}. This hydrodynamic framework employs an equation of state computed through lattice Quantum Chromodynamics (QCD) and extrapolated to infinite spatial volume \cite{Borsanyi:2013bia}. The inference drawn from the comparison between these hydrodynamics predictions and the measured data is that the medium generated in these high multiplicity small systems is a nearly inviscid QGP, similar to that formed in large nucleus-nucleus collisions.

A recent study investigating finite system size effects in a massless free scalar thermal field theory with Dirichlet boundary conditions showed that the finite system size corrections to thermodynamic properties can effectively mimic the temperature dependence of the energy density and pressure of full QCD \cite{Mogliacci:2018oea}. The findings revealed substantial finite system size corrections of $\sim40$\% to the usual thermodynamic quantities such as pressure and entropy for systems of the size of p+p collisions. Even for systems the size of mid-central nucleus-nucleus collisions, the corrections were on the order of 10\%. Quenched lattice QCD calculations using periodic boundary conditions confirmed the qualitative impact of these finite system size effects, especially in systems with asymmetric finite lengths \cite{Kitazawa:2019otp}.

The equation of state, which encompasses the speed of sound and the trace anomaly, plays a crucial role in hydrodynamic simulations of relativistic hadronic collisions. Surprisingly, despite the breaking of conformal symmetry due to Dirichlet boundary conditions, it was observed that the free massless scalar field theory yields an identically traceless energy-momentum tensor \cite{Horowitz:2021dmr}.  In general, for any infinite volume system, a trace anomaly can \emph{only} come from the quantum scale breaking anomaly in the coupling.  To estimate the effect of finite system size corrections on the trace anomaly in QCD, the finite system size correction to the coupling in a massive scalar theory on the lattice \cite{Montvay:1994cy} was inserted \cite{Horowitz:2021dmr} into the QCD coupling in the hard thermal loop pressure and energy density computed to $\mathcal O(g^5)$ in QCD \cite{Andersen:2004fp}. The analysis showed a significant reduction in the size of the trace anomaly due to finite system size effects \cite{Horowitz:2021dmr}. Such a substantial reduction in the trace anomaly would have a considerable impact on the extraction of shear and bulk viscosities when comparing hydrodynamics simulations to experimental data, presumably imperiling the interpretation of the medium created in high-multiplicity p+p collisions as a nearly inviscid perfect fluid.

We are therefore interested in computing analytically the finite system size corrections to the trace anomaly of QCD induced through the finite system size corrections to the QCD running coupling.  This program presents significant challenges that will require understanding several important techniques.  The two most important conceptual difficulties to overcome are to determine how to regularize and renormalize the thermal field theory in a finite size system and to include the effect of torons, non-trivial vacuum gauge configurations on a torus \cite{tHooft:1979rtg,Coste:1985mn}.  

The work presented here provides a step in the direction of the first challenge by computing the finite system size correction to the running coupling in massive $\phi^4$ theory for $2\rightarrow2$ scattering at next-to-leading order (NLO).  In order to perform this computation, we use the technique of ``denominator regularization,'' which is more appropriate for regularizing field theories in a finite system than dimensional regularization \cite{Horowitz:2022rpp}.  Following \cite{Elizalde:1997jv} we derive an analytic continuation of the generalized Epstein zeta function that is needed to extract the UV divergence in the denominator regularized finite system, and may be readily applied in future thermal field theory derivations.  We perform a non-trivial self-consistency check of our result by confirming that our NLO contribution satisfies unitarity.  One of our methods of checking unitarity led us to generalize a number theoretic result from Hardy and Ramanujan \cite{Hardy1978RamanujanTL}.  We then provide some sample plots of the real and imaginary parts of the finite system size NLO correction to the amplitude for $2\rightarrow2$ scattering.  The numerics for the $s$ channel contribution in a system with two finite length sides is highly non-trivial and required us to utilize dispersion relation techniques.  We further confirmed our numerical results by an asymptotic analysis on the $t$ channel contribution in a system fully confined in three spatial dimensions.  We then show the impact of the finite system size corrections on the effective coupling in our theory.  For moderate and larger momenta and lengths, and except for ``geometric bound states'' in which the final state of the system is confined to the finite direction, the finite system size corrections are modest for the coupling for systems with only one finite spatial direction.  Corrections to the coupling for systems with two compact spatial directions can be large.  Avoiding questions about the physical setup and necessary corrections to the usual scattering program in quantum field theories, we see that corrections to scattering in systems with three finite spatial directions are maximal: the total cross section is either zero for unphysical modes or infinite for physical modes.

%%%%%%%%%%%%%%%%%%%%%%%%%%%%%%%%%%%%%%%%%%%%%%%%%%%%%%%%%%%%%%%%%%%%%%%%%%%%%%%%%%%%%%%%%%%%%%%%%%%%%%
\section{Review of the Usual Infinite Volume NLO \texorpdfstring{$2\rightarrow2$}{2 to 2} Scattering}\label{sec:infVol}
%%%%%%%%%%%%%%%%%%%%%%%%%%%%%%%%%%%%%%%%%%%%%%%%%%%%%%%%%%%%%%%%%%%%%%%%%%%%%%%%%%%%%%%%%%%%%%%%%%%%%%
In order to warm up to the finite system size calculation, to fix some notation, and to provide a valuable reference to compare to, we compute $2\rightarrow2$ scattering at NLO in massive $\phi^4$ theory using dimensional regularization and  the \msbar{} renormalization scheme.  

We begin with the bare Lagrangian for $\phi^4$ theory:
\begin{align}
    \mathcal L = \frac12\partial_\mu\phi_0\partial^\mu\phi_0 - \frac12m_0^2\phi_0^2-\frac{\lambda_0}{4!}\phi_0^4.
\end{align}
We choose as usual to multiplicatively renormalize.  We'll work in $d = 4-\epsilon$ spacetime dimensions.  In order to fix our coupling constant to be dimensionless in any number of spacetime dimensions, we introduce a scale $\mu$, with dimensions of energy,
\begin{align}
    \phi_r & \equiv Z_\phi^{-1/2}\phi_0 \nonumber\\
    m_r^2 & \equiv Z_m^{-1}Z_\phi m_0^2 \\
    \lambda_r & \equiv Z_\lambda^{-1}Z_\phi^2\mu^\epsilon\lambda_0. \nonumber
\end{align}
Further defining
\begin{align}
    Z_\phi & \equiv 1+\delta_\phi \nonumber\\
    Z_m & \equiv 1+\frac{1}{m^2}\delta_m^2 \\
    Z_\lambda & \equiv 1 + \frac{1}{\lambda_r}\delta_\lambda \nonumber
\end{align}
we arrive at the renormalized Lagrangian
\begin{multline}
    \mathcal L = \frac12\partial_\mu\phi\partial^\mu\phi - \frac12m^2\phi^2-\frac{\lambda}{4!}\mu^\epsilon\phi^4 \\
    +\delta_\phi\frac12\partial_\mu\phi\partial^\mu\phi - \delta_m^2\frac12\phi^2 - \delta_\lambda\mu^\epsilon\frac{1}{4!}\phi^4,
\end{multline}
where we've dropped all $r$ subscripts on renormalized quantities for notational convenience.

\newcommand{\hi}{0.5}
\begin{figure}[!t]
  \centering
  \begin{tikzpicture}[baseline=(c.base)]
    \begin{feynhand}
        \setlength{\feynhandarrowsize}{4pt}
        \setlength{\feynhanddotsize}{0mm}
        \vertex [particle] (a) at (-1.5,1) {$p_A$};
        \vertex [particle] (b) at (-1.5,-1) {$p_B$};
        \vertex [dot] (c) at (-.5,0) {};
        \vertex [dot] (d) at (.5,0) {};
        \vertex [particle] (e) at (1.5,1) {$p_1$};
        \vertex [particle] (f) at (1.5,-1) {$p_2$};
        \propag [fer] (a) to (c);
        \propag [fer] (b) to (c);
        \propag [fer] (c) to [out = 45, in = 135, looseness = 1.75] (d);
        \propag [antfer] (c) to [out = 315, in = 225, looseness = 1.75, edge label' = $k$] (d);
        \propag [fer] (d) to (e);
        \propag [fer] (d) to (f);
    \end{feynhand}
  \end{tikzpicture}
  \begin{tikzpicture}[baseline=(a.base)]
    \begin{feynhand}
        \vertex (a) at (-.3,0) {};
        \vertex (b) at (.3,0) {};
        \vertex (c) at (0,-.3) {};
        \vertex (d) at (0,.3) {};
        \draw (a) to (b);
        \draw (c) to (d);
    \end{feynhand}
  \end{tikzpicture}
  \begin{tikzpicture}[baseline=(g.base)]
    \begin{feynhand}
        \setlength{\feynhanddotsize}{0mm}
        \vertex (a) at (-1,\hi) {};
        \vertex (b) at (-1,-\hi) {};
        \vertex (c) at (1,\hi) {};
        \vertex (d) at (1,-\hi) {};
        \vertex [dot] (e) at (0,\hi) {};
        \vertex [dot] (f) at (0,-\hi) {};
        \vertex (g) at (0,0) {};
        \draw (a) to (c);
        \draw (b) to (d);
        \draw (e) to [out = 355, in = 5, looseness = 1.75] (f);
        \draw (e) to [out = 185, in = 175, looseness = 1.75] (f);
    \end{feynhand}
  \end{tikzpicture}\\
  \begin{tikzpicture}[baseline=(a.base)]
    \begin{feynhand}
        \vertex (a) at (-.3,0) {};
        \vertex (b) at (.3,0) {};
        \vertex (c) at (0,-.3) {};
        \vertex (d) at (0,.3) {};
        % \vertex (e) at (0,0) {};
        \draw (a) to (b);
        \draw (c) to (d);
    \end{feynhand}
  \end{tikzpicture}
  \begin{tikzpicture}[baseline=(g.base)]
    \begin{feynhand}
        \setlength{\feynhanddotsize}{0mm}
        \vertex (a) at (-1,\hi) {};
        \vertex (b) at (-1,-\hi) {};
        \vertex [dot] (c) at (1,\hi) {};
        \vertex [dot] (d) at (1,-\hi) {};
        \vertex [dot] (e) at (0,\hi) {};
        \vertex [dot] (f) at (0,-\hi) {};
        \vertex (g) at (0,0) {};
        \draw (a) to (e);
        \draw (b) to (f);
        \draw (f) to [out = 0, in = 265, looseness = 1.] (c);
        \draw [top] (e) to [out = 0, in = 95, looseness = 1.] (d);
        \draw (e) to [out = 355, in = 5, looseness = 1.75] (f);
        \draw (e) to [out = 185, in = 175, looseness = 1.75] (f);
    \end{feynhand}
  \end{tikzpicture}
  \begin{tikzpicture}[baseline=(a.base)]
    \begin{feynhand}
        \vertex (a) at (-.3,0) {};
        \vertex (b) at (.3,0) {};
        \vertex (c) at (0,-.3) {};
        \vertex (d) at (0,.3) {};
        \draw (a) to (b);
        \draw (c) to (d);
    \end{feynhand}
  \end{tikzpicture}
  \begin{tikzpicture}[baseline=(g.base)]
    \begin{feynhand}
        \vertex (a) at (-\hi,\hi) {};
        \vertex (b) at (-\hi,-\hi) {};
        \vertex (c) at (\hi,\hi) {};
        \vertex (d) at (\hi,-\hi) {};
        \vertex [crossdot] (g) at (0,0) {};
        \draw (a) to (g);
        \draw (b) to (g);
        \draw (g) to (c);
        \draw (g) to (d);
    \end{feynhand}
  \end{tikzpicture}
  \caption{
  \label{f:NLO}
  The four diagrams contributing to the NLO correction to $2\rightarrow2$ scattering in our renormalized $\phi^4$ theory.  Momenta are labelled on the $s$ channel diagram but are suppressed for simplicity on the $t$ and $u$ channel diagrams.
  }
\end{figure}
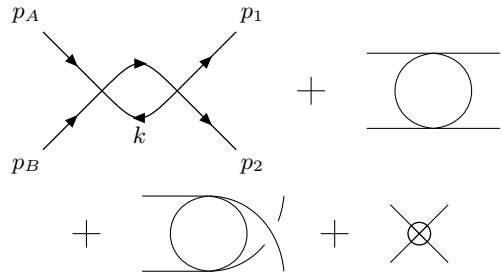

When we perform LSZ reduction, we must take care with $R_\phi$, the residue of the propagator at the physical mass.  However, for $\phi^4$ theory in $4-\epsilon$ spacetime dimensions, one has that the first order correction to the scalar self energy is
\begin{align}
    -i\Sigma^{[1]}(p^2) 
    & = \begin{tikzpicture}[baseline=(c.base)]
        \begin{feynhand}
            \setlength{\feynhanddotsize}{0mm}
            \vertex (a) at (-0.85,0) {};
            \vertex (b) at (0.85,0) {};
            \vertex [dot] (c) at (0,0) {};
            \vertex [dot] (d) at (0,0.5) {};
            \draw (a) to (b) {};
            \draw (c) to [out=180, in=180, looseness=1.75] (d);
            \draw (d) to [out=0, in=0, looseness=1.75] (c);
        \end{feynhand}
    \end{tikzpicture}
    +
    \begin{tikzpicture}[baseline=(c.base)]
        \begin{feynhand}
            \vertex (a) at (-0.85,0) {};
            \vertex (b) at (0.85,0) {};
            \vertex [crossdot] (c) at (0,0) {};
            \draw (a) to (c);
            \draw (c) to (b);
        \end{feynhand}
    \end{tikzpicture} \nonumber\\[5pt]
    & = \frac{-i\lambda\mu^\epsilon}{2}\frac{1}{(4\pi)^{d/2}}\frac{\Gamma\big(1-\frac d2\big)}{(m^2)^{1-\frac d2}} + i(p^2\delta_\phi-\delta_m^2).
\end{align}
Since the divergence is momentum independent, we may take $\delta_\phi=0$, and we have that $R_\phi=1 + \mathcal O(\lambda^2)$.

Thus when we compute the leading order amplitude for $2\rightarrow2$ scattering, we find
\begin{align}
    i\mathcal M
     =     \begin{tikzpicture}[baseline=(e.base)]
        \begin{feynhand}
            \vertex (a) at (-.5,-.5) {};
            \vertex (b) at (-.5,.5) {};
            \vertex (c) at (.5,.5) {};
            \vertex (d) at (.5,-.5) {};
            \vertex (e) at (0,0) {};
            \draw (a) to (c);
            \draw (b) to (d);
        \end{feynhand}
    \end{tikzpicture}
    = & -i\lambda\mu^\epsilon + \mathcal O(\lambda^2) \nonumber\\
    \xrightarrow[\epsilon\rightarrow0]{} & -i\lambda + \mathcal O(\lambda^2);
\end{align}
i.e.\ there's no symmetry factor in this particular case.

The diagrams associated with the NLO correction to $2\rightarrow2$ scattering are shown in \cref{f:NLO}.  Let's focus on the $s$ channel diagram.  A detailed examination of the contracted fields shows that the symmetry factor is $S=1/2$.  Then defining $p\equiv p_A+p_B$ and $V(p^2;\mu,\epsilon)$ by
\renewcommand{\hi}{0.5}
\newcommand{\wi}{0.25}
\newcommand{\wid}{0.55}
\begin{align}
    (-i\lambda)^2\mu^\epsilon iV(p^2;\mu,\epsilon)
    \equiv
    \begin{tikzpicture}[baseline=(c.base)]
        \begin{feynhand}
            \setlength{\feynhandarrowsize}{4pt}
            \setlength{\feynhanddotsize}{0mm}
            \vertex (a) at (-\wid,\hi) {};
            \vertex (b) at (-\wid,-\hi) {};
            \vertex [dot] (c) at (-\wi,0) {};
            \vertex [dot] (d) at (\wi,0) {};
            \vertex (e) at (\wid,\hi) {};
            \vertex (f) at (\wid,-\hi) {};
            \draw (a) to (c);
            \draw (b) to (c);
            \draw (c) to [out = 65, in = 115, looseness = 1.75] (d);
            \draw (c) to [out = 295, in = 245, looseness = 1.75] (d);
            \draw (d) to (e);
            \draw (d) to (f);
        \end{feynhand}
    \end{tikzpicture}\nonumber
\end{align}
we have that
\begin{multline}
    (-i\lambda)^2\mu^\epsilon iV(p^2;\mu,\epsilon) = \frac{(-i\lambda\mu^\epsilon)^2}{2} \\ \times\int\frac{d^dk}{(2\pi)^d}\frac{i}{k^2-m^2+i\varepsilon}\frac{i}{(p+k)^2-m^2+i\varepsilon}.
\end{multline}
Note the $\epsilon>0$ that guarantees convergence of the integral on the RHS of the expression, which is different from the $\varepsilon>0$ that enforces the Feynman propagator pole prescription.  Following the usual procedure to combine denominators using a Feynman $x$ parameter, Wick rotating, and evaluating the resulting Euclidean integral, and expanding to $\mathcal O(\epsilon)$ yields
\begin{multline}
    \label{e:Vp2}
    V(p^2;\mu,\epsilon) = -\frac12\frac{1}{(4\pi)^2}\left\{ \int_0^1dx\Big[ \frac2\epsilon-\gamma_E + \ln4\pi \right. \\\left. + \ln\frac{\mu^2}{-x(1-x)p^2+m^2-i\varepsilon} +\mathcal O(\epsilon) \Big] \right\}.
\end{multline}
Note that close examination of the location of the poles of the denominator (combined in the usual way using the Feynman $x$ parameter) shows that the poles are restricted to the second and fourth quadrants in the complex plane for both $p^2>0$ and for $p^2<0$; hence the Wick rotation, and thus \cref{e:Vp2}, are valid for $p^2\in\mathbb R$.

Applying the \msbar{} renormalization scheme fixes the coupling counterterm to
\begin{align}
    \delta_\lambda = 3\lambda^2\frac12\frac{1}{(4\pi)^2}\Big[ \frac2\epsilon - \gamma_E + \ln4\pi \Big]
\end{align}
and the renormalized NLO contribution to
\begin{align}
    \label{e:infNLO}
    \overline V(p^2;\mu) = -\frac12\frac{1}{(4\pi)^2} \int_0^1dx \ln\frac{\mu^2}{-x(1-x)p^2+m^2-i\varepsilon}.
\end{align}
Thus up to NLO the $2\rightarrow2$ amplitude is
\begin{align}
    i\mathcal M
    & = -i\lambda\mu^\epsilon \big[ 1 + \lambda\big( V(s;\mu) + V(t;\mu) + V(u;\mu) \big) + \frac1\lambda\delta_\lambda \big] \nonumber\\
    \label{e:NLOth}
    & \xrightarrow[\epsilon\rightarrow0]{}-i\lambda\big[ 1 + \lambda\big( \overline V(s;\mu) +\overline V(t;\mu) +\overline V(u;\mu) \big)]. 
\end{align}

One may perform a non-trivial cross check of \cref{e:infNLO} by confirming that the Optical Theorem is satisfied.  For $p^2>4m^2$ and $x_-<x<x_+$, where
\begin{align}
    -x_\pm(1-x_\pm)p^2+m^2 = 0 \\ \Rightarrow \quad \label{e:xpm} x_\pm = \frac12\left[ 1\pm\sqrt{1-\frac{4m^2}{p^2}} \right],
\end{align}
we have that
\begin{align}
    \Im\ln\frac{\mu^2}{-x(1-x)p^2+m^2-i\varepsilon} = \pi.
\end{align}
Thus to order $\mathcal O(\lambda^2)$
\begin{align}
    \Im\mathcal M = -\lambda^2\pi\Big[-\frac12\frac{1}{(4\pi)^2}\Big]\int_{x_-}^{x_+}dx.
\end{align}
Therefore
\begin{align}\label{eq:infVolImpart}
    2\ess\Im\mathcal M = \frac{\lambda^2}{16\pi}\sqrt{1-\frac{4m^2}{s}}\theta(s-4m^2) + \mathcal O(\lambda^3).
\end{align}
One may straightforwardly show that one arrives at the exact same result from evaluating
\begin{multline}
    \sigma_{\mathrm{tot}} = \frac12\int\frac{d^3p_2}{(2\pi)^32E_{p_2}}\int\frac{d^3p_1}{(2\pi)^32E_{p_1}}|\mathcal M|^2 \\ \times(2\pi)^4\delta^{(4)}(p_A+p_B-p_1-p_2),
\end{multline}
where $\mathcal M = \lambda + \mathcal O(\lambda^2)$ is the leading order cross section, and one must be careful of the overall factor of $1/2$ due to the presence of two identical particles in the final state.

The four point Green function is given by $i\mathcal M$.  Since to $\mathcal O(\lambda^2)$ there's no contribution from an anomalous dimension $\gamma$, we have the Callan-Symanzik equation
\begin{align}
    (\mu\partial_\mu + \beta\partial_\lambda)\mathcal M = 0 \nonumber\\
    (\mu\partial_\mu + \beta\partial_\lambda)\big[{-}\lambda - \lambda^2\big(\overline V(s;\mu) + \overline V(t;\mu) + \overline V(u;\mu) \big)\big] = 0
\end{align}
Since $\mu\partial_\mu\overline V(p^2;\mu) = -(4\pi)^{-2}$, we find the standard result \cite{Peskin:1995ev} that
\begin{align}
    \beta = \frac{3\lambda^2}{(4\pi)^2}.
\end{align}
And since $\beta\equiv\partial\lambda/\partial\ln\mu$, we must have that
\begin{align}
    \int_{\lambda(\mu_0)}^{\lambda(\mu)}\frac{d\lambda}{\lambda^2} = \int_{\ln\mu_0}^{\ln\mu}d\ln\mu'\frac{3}{4\pi^2} \\
    \label{e:runningcoupling}
    \Rightarrow\qquad \lambda(\mu) = \frac{\lambda(\mu_0)}{1+\lambda(\mu_0)\frac{3}{(4\pi)^2}\ln\mu_0/\mu}.
\end{align}
We'll refer to \cref{e:runningcoupling} as the \emph{running} coupling. 

In, e.g., $e^++e^-\rightarrow\mu^++\mu^-$ scattering in QED, one may exactly resum the geometric series of 1PI fermion loops to generate an \emph{effective} coupling.  For $p^2\gg m_e^2$ the leading logarithmic behavior of the effective coupling is identical to the running coupling from the Callan-Symanzik equation evaluated at the center of mass energy $s$, $\alpha_{EM}(s)$.  This resummation procedure can be more difficult for different processes and different theories.  Nonetheless, the idea of an effective coupling is a valuable one.  For example, in QCD one has the BLM scale setting procedure, in which one resums the fermion bubbles and then completes the beta function \cite{Brodsky:1982gc}.

For our $\phi^4$ theory, the correct geometric resummation that captures the leading logarthmic physics of the Callan-Symanzik equation for the $2\rightarrow2$ process involves the $s,\,t$, and $u$ channels.  Recalling that one may write $u(s,t)$,
\begin{align}
    -i\lambda_{\mathrm{eff}}(s,t)
    & \equiv -i\lambda(\mu) \nonumber\\
    & \quad\times\big[ 1 + \lambda(\mu) \big( \overline V(s;\mu) + \overline V(t;\mu) + \overline V(u;\mu) \big) \nonumber\\
    & \qquad+ \lambda^2(\mu) \big( \overline V(s;\mu) + \overline V(t;\mu) + \overline V(u;\mu) \big)^2 \nonumber\\
    & \qquad + \ldots \big] \nonumber\\
    \label{e:effectivecoupling}
    & = \frac{-i\lambda(\mu)}{1 - \lambda(\mu)\big( \overline V(s;\mu) + \overline V(t;\mu) + \overline V(u;\mu) \big)},
\end{align}
where $\lambda(\mu)$ is the running coupling given by \cref{e:runningcoupling}.  At NLO, all $\mu$ dependence in \cref{e:effectivecoupling} exactly cancels between the running coupling and in the $\overline V$'s, leaving $\lambda_{\mathrm{eff}}$ with no $\mu$ dependence at all.  For $Q^2 \equiv -p^2>0$, one may easily evaluate the $x$ integral in \cref{e:infNLO} to find
\begin{align}
    \overline V(Q^2;\mu)
    & = -\frac12\frac{1}{(4\pi)^2}\bigg[ 2-2\sqrt{1+\frac{4m^2}{Q^2}} + \ln\frac{\mu^2}{m^2} \bigg] \\
    & = -\frac12\frac{1}{(4\pi)^2}\ln\frac{\mu^2}{Q^2} + \mathcal O \big( (m^2/Q^2) \big).
\end{align}
For $s\sim -t\sim -u\sim E^2 \gg m^2$ and we take $\mu^2=E^2$ in the running coupling \cref{e:runningcoupling}, then we once again have to leading log that $\lambda_{\mathrm{eff}}(E^2) = \lambda(E^2)$.

%%%%%%%%%%%%%%%%%%%%%%%%%%%%%%%%%%%%%%%%%%%%%%%%%%%%%%%%%%%%%%%%%%%%%%%%%%%%%%%%%%%%%%%%%%%%%
\section{Denominator Regularization}
%%%%%%%%%%%%%%%%%%%%%%%%%%%%%%%%%%%%%%%%%%%%%%%%%%%%%%%%%%%%%%%%%%%%%%%%%%%%%%%%%%%%%%%%%%%%%
\label{s:denreg}
When we move to the setup in which the spacetime is $\mathbb R\times T^n$, the integrals over spatial momenta will become sums.  The potential asymmetry of the various spatial directions and the discrete sums mean that the usual techniques of dimensional regularization are no longer applicable.  Instead of making the calculation a function of the number of spacetime dimensions and analytically continuing to $d=4$, we rather make the \emph{power} of the denominator in the loop integral for $V$ a function and analytically continue to the log divergent value of 2.  In particular, in denominator regularization the number of spacetime dimensions is fixed to 4.  As a result, one doesn't need to introduce $\mu$ to keep the coupling dimensionless as a function of $\epsilon$ (a $\mu$ will still be introduced to keep the dimensions of $V$ independent of the power of the denominator); the denominator regularization renormalized Lagrangian is
\begin{multline}
    \mathcal L = \frac12\partial_\mu\phi\partial^\mu\phi - \frac12m^2\phi^2-\frac{\lambda}{4!}\phi^4 \\
    +\delta_\phi\frac12\partial_\mu\phi\partial^\mu\phi - \delta_m^2\frac12\phi^2 - \delta_\lambda\frac{1}{4!}\phi^4.
\end{multline}

One must of course still have that $R_\phi=1+\mathcal O(\lambda^2)$ in denominator regularization since the regularization procedure cannot introduce any new divergences.  Nevertheless, one may easily check explicitly that in denominator regularization the one loop self energy is still momentum independent, and thus one has in denominator regularization that $R_\phi=1+\mathcal O(\lambda^2)$.  One trivially, then, has that $i\mathcal M = -i\lambda+\mathcal O(\lambda^2)$.  At NLO, one has after combining denominators with a Feynman parameter and Wick rotating
\begin{align}
    \label{e:V}
    (-i\lambda)^2iV(p^2) & = i\frac{\lambda^2}{2}\int_0^1dx\int\frac{d^4\ell_E}{(2\pi)^4}\frac{1}{(\ell_E^2+\Delta^2)^{2}}, \\
    \Delta^2 & \equiv -x(1-x)p^2+m^2-i\varepsilon.
\end{align}
The integral in \cref{e:V} diverges logarithmically in the UV as the numerator scales like $\ell_E^3$ and the denominator like $\ell_E^4$.  We now perform denominator regularization by allowing $V$ to depend on the power of the denominator.  We introduce the denominator regulator $\epsilon$ (again distinct from $\varepsilon$) such that the integral converges for all $\epsilon>0$.  In order to maintain the dimensionlessness of $V$ we introduce a scale $\mu$ with dimensions of energy.  We then have
\begin{multline}
    V(p^2;\mu,\epsilon) = -\frac12\int_0^1dx\int\frac{d^4\ell_E}{(2\pi)^4}\frac{\mu^{2\epsilon}}{(\ell_E^2+\Delta^2)^{2+\epsilon}}.
\end{multline}
Similar to the dimensional regularization procedure, one may perform the spherical integration over the full 4 dimensional spacetime.  In order to better make contact with our future finite-sized spatial spacetime calculation, we rather perform the ``temporal'' integration first (temporal in quotes as in this case no direction is different from another).  Then one has that
\begin{align}
    V(p^2;\mu,\epsilon)
    & = -\frac12\frac{1}{(2\pi)^4} \int_0^1dx\frac{\sqrt\pi\Gamma\big( \frac32+\epsilon \big)}{\Gamma(2+\epsilon)} \nonumber\\
    & \qquad\qquad\qquad\times\int d^3\ell_E \frac{\mu^{2\epsilon}}{(\ell_E^2+\Delta^2)^{\frac32+\epsilon}}.
\end{align}
The remaining momentum integrals evaluate to
\begin{align}
    4\pi\int_0^\infty \ell_E^2d\ell_E \frac{\mu^{2\epsilon}}{(\ell_E^2+\Delta^2)^{\frac32+\epsilon}} = \frac{\sqrt\pi\Gamma(\epsilon)}{4\Gamma\big(\frac32+\epsilon\big)}\bigg(\frac{\mu^2}{\Delta^2}\bigg)^\epsilon.
\end{align}
Thus
\begin{align}
    V(p^2;\mu,\epsilon)
    & = -\frac12\frac{1}{(4\pi)^2}\int_0^1dx\frac{\Gamma(\epsilon)}{\Gamma(2+\epsilon)}\bigg(\frac{\mu^2}{\Delta^2}\bigg)^\epsilon \nonumber\\
    & = -\frac12\frac{1}{(4\pi)^2}\int_0^1dx \left[ \frac1\epsilon-1+\ln\big( \frac{\mu^2}{\Delta^2} \big) \right] + \mathcal O(\epsilon).
\end{align}
It's interesting that the $-1$ of the finite part from denominator regularization is identical to the $-1$ that one finds when regularizing through an explicit UV cutoff, c.f.\ the $-\gamma_E + \ln4\pi$ one finds from dimensional regularization.  
A similar calculation in which one performs the usual spherical integration over the full 4D space yields an identical result.

One can see that the denominator regularization procedure reproduces the same $1/\epsilon$ divergence and $\ln\mu^2/\Delta^2$ dependence (with the same overall coefficient) for $V(p^2;\mu,\epsilon)$ as in dimensional regularization (as it must); the only difference comes in the coefficient of the divergence (which is halved, but can simply be absorbed by a rescaling of $\epsilon$) and the finite correction of $-1$ compared to $-\gamma_E+\ln4\pi$.  If we modify our \msbar{} prescription for denominator regularization to absorb the $1/\epsilon$ divergence and the $-1$ instead of the $2/\epsilon$ divergence and the $-\gamma_E+\ln4\pi$, then the renormalized NLO $2\rightarrow2$ cross section from denominator regularization agrees exactly with that from the usual dimensional regularization procedure.  In particular, for our denominator regularization procedure we take
\begin{align}
    \delta_\lambda = 3\lambda^2\frac12\frac{1}{(4\pi)^2}\Big[ \frac1\epsilon - 1 \Big].
\end{align}

%%%%%%%%%%%%%%%%%%%%%%%%%%%%%%%%%%%%%%%%%%%%%%%%%%%%%%%%%%%%%%%%%%%%%%%%%%%%%%%%%%%%%%%%%%%%%
\section{Finite Size Corrections}
%%%%%%%%%%%%%%%%%%%%%%%%%%%%%%%%%%%%%%%%%%%%%%%%%%%%%%%%%%%%%%%%%%%%%%%%%%%%%%%%%%%%%%%%%%%%%
\subsection{Field Theory Defined}
%%%%%%%%%%%%%%%%%%%%%%%%%%%%%%%%%%%%%%%%%%%%%%%%%%%%%%%%%%%%%%%%%%%%%%%%%%%%%%%%%%%%%%%%%%%%%
It's just as easy to consider a real scalar field theory in only 4 spacetime dimensions with all three spatial dimensions periodic as to consider a very general scalar field theory with $n$ directions periodically identified and $m$ directions of infinite extent.  Let the $i^\mathrm{th}$ compact spatial dimensions have size $[-\pi L_i,\pi L_i]$ and take
\begin{align}
    \phi(x)
    & = \sum_{\vec k\in\mathbb Z^n}\frac{1}{(2\pi)^nL_1\cdots L_n}\int\frac{d^mp}{(2\pi)^m}\frac{1}{2E_{\vec p}} \nonumber\\
    & \qquad\qquad\times\bigg[ e^{-i\ess p\cdot x}a(\vec p) + e^{i\ess p\cdot x}a^\dagger(\vec p) \bigg], \nonumber\\
    p^i & = \frac{k^i}{L_i}, \quad i=1,\ldots,n \nonumber\\
    p^j & = p^j, \quad j = 1,\ldots, m \nonumber\\
    p^ 0 & = E_{\vec p} = \sqrt{\vec p^2+m^2}. 
\end{align}
Then
\begin{align}
    [a(\vec p),a^\dagger(\vec q)] = (2\pi)^{n+m}2E_{\vec p} L_1\cdots L_n \delta_{\vec k_{\vec p},\vec k_{\vec q}}\delta^{(m)}(\vec p-\vec q)
\end{align}
gives a field that obeys the usual canonical commutation relation
\begin{align}
    [\phi(x),\Pi(y)]_{x^0=y^0} = i\delta^{(n+m)}(\vec x- \vec y).
\end{align}
Further, the total momentum in the field is
\begin{align}
    P^\mu & = \sum_{\vec k\in\mathbb Z^n}\frac{1}{(2\pi)^nL_1\cdots L_n}\int\frac{d^mp}{(2\pi)^m}\frac{1}{2E_{\vec p}}p^\mu a^\dagger(\vec p)a(\vec p),
\end{align}
and hence we have the usual interpretation of an excitation of a mode of the field as a particle of mass $m$ and momentum $\vec p$.

Of crucial importance, one may compute the contraction
\begin{multline}
    \wick[offset=1.2em]{\c\phi (x) \c\phi (y)} = \sum_{\vec k\in\mathbb Z^n}\frac{1}{(2\pi)^nL_1\cdots L_n}\int\frac{d^mp}{(2\pi)^m}\int\frac{dp^0}{2\pi} \\ e^{-i\ess p\cdot(x-y)}\frac{i}{p^2-m^2+i\varepsilon}.
\end{multline}

%%%%%%%%%%%%%%%%%%%%%%%%%%%%%%%%%%%%%%%%%%%%%%%%%%%%%%%%%%%%%%%%%%%%%%%%%%%%%%%%%%%%%%%%%%%%%
\subsection{LO Scattering}
%%%%%%%%%%%%%%%%%%%%%%%%%%%%%%%%%%%%%%%%%%%%%%%%%%%%%%%%%%%%%%%%%%%%%%%%%%%%%%%%%%%%%%%%%%%%%
\label{s:LOfinite}
We may straightforwardly compute the LO scattering amplitude in the usual way.  Since we're working at leading order, we don't need to consider any subtleties due to finite system size corrections to LSZ reduction; the asymptotic states are trivially the single particle states of the free theory.  Thus the $\mathcal T$ matrix can be readily found to be
\begin{align}
    \label{e:LOfinite}
    \langle p_1 p_2 & |T| p_A p_B\rangle \nonumber\\
    & = -i\ess\lambda\int d^{m+n+1}z e^{i\ess z\cdot(p_A+p_B-p_1-p_2)} + \mathcal O(\lambda^2) \nonumber\\
    & = -i\ess\lambda\ess2\pi \delta(p_A^0+p_B^0-p_1-p_2) \nonumber\\
    & \qquad\times(2\pi)^m\delta^{(m)}(\vec p_A+\vec p_B-\vec p_1-\vec p_2) \nonumber\\
    & \qquad\times (2\pi)^n L_1\cdots L_n\delta_{\vec \ell_{p_A}+\vec \ell_{p_B},\vec \ell_{p_1}+\vec \ell_{p_2}} + \mathcal O(\lambda^2),
\end{align}
where the $m$ Dirac deltas provide the spatial momentum conservation in the directions of infinite extent of the momenta to be the same and the $n$ Kronecker deltas provide the spatial momentum conservation in the directions of finite extent.  There is also an overall Dirac delta providing energy conservation.  Note that even if we work in a spacetime that is fully compact in the spatial directions, energy conservation is given by a Dirac delta function as we've assumed that the extent of the time direction is infinite.  Note further that we will at times consider (unphysical) momenta in the finite spatial directions that are not integer modes but that the Kronecker deltas associated with momentum conservation in those directions are only non-zero when the modes are properly integer.

The choice of allowing time to flow from $-\infinity$ to $+\infinity$ is perhaps somewhat inconsistent with the spirit of a scattering experiment one might imagine in a finite space.  In infinite space, a scattering experiment consists of wavepackets that start off infinitely separated in the infinite past that are allowed to propagate to definite momentum states in the infinite future \cite{Peskin:1995ev}.  Implicitly, in this infinite volume picture, the particles interact at a time $\sim t=0$ only.  There are two pathologies in the finite volume case when all spatial dimensions are finite: first, the wavepackets can never be infinitely separated; second, the particles interact an infinite number of times.  The issue of initial separation isn't an issue at leading order because there's no renormalization; beyond leading order, so long as the space is large compared to the interaction length, the wavepackets can be considered well separated.  Further, the NLO contribution to the one particle irreducible self energy diagram will remain independent of the particle's momentum; hence at NLO, wavefunction renormalization is still trivial.  That the particles interact an infinite number of times over an infinite time interval will be reflected in an infinite total cross section.  

One could alternatively consider a finite time for propagation, for example of the order of the size of the system.  However, the integral over $z^0$ in \cref{e:LOfinite} then yields a complicated expression that only converges to a Dirac delta function in the limit of the time for propagation going to infinity.  Then the subsequent calculation of the NLO contribution is sufficiently far away from usual scattering problems or a thermal field theoretic calculation that its utility appears limited.

%%%%%%%%%%%%%%%%%%%%%%%%%%%%%%%%%%%%%%%%%%%%%%%%%%%%%%%%%%%%%%%%%%%%%%%%%%%%%%%%%%%%%%%%%%%%%
\subsection{NLO Scattering}
%%%%%%%%%%%%%%%%%%%%%%%%%%%%%%%%%%%%%%%%%%%%%%%%%%%%%%%%%%%%%%%%%%%%%%%%%%%%%%%%%%%%%%%%%%%%%
If we now restrict ourselves to three periodic spatial directions and no spatial directions of infinite extent, $n=3$ and $m=0$, one may immediately write down the quantity we need to evaluate for the NLO correction to $2\rightarrow2$ scattering:
\begin{multline}
    V(p^2,\{L_i\};\mu,\epsilon)
    = -\frac12\int_0^1dx\int\frac{d\ell_E^0}{2\pi}\sum_{\vec k\in\mathbb Z^3}\\
    \frac{1}{(2\pi)^3L_1L_2L_3}\frac{\mu^{2\epsilon}}{[\ell_E^2+\Delta^2]^{2+\epsilon}},
\end{multline}
where
        $\Delta^2 \equiv -x(1-x)p^2 + m^2 - i\varepsilon$ and $
        \ell^\mu_E = (\ell_E^0,\frac{k^i}{L_i}+x\ess p^i)^\mu$.
Notice that in contrast to the infinite volume case we cannot simply shift the spatial integration to remove the $+x\ess p^i$ shift in $\ell_E^\mu$.  As in \cref{s:denreg} we may evaluate the $\ell_E^0$ integral to find
\begin{multline}
    \label{e:infsum}
    V(p^2,\{L_i\};\mu,\epsilon)
    = -\frac12\frac{1}{2\pi}\frac{1}{(2\pi)^3L_1L_2L_3}\int_0^1dx \\ \times\frac{\sqrt\pi\Gamma\big(\frac32+\epsilon\big)}{\Gamma(2+\epsilon)} %\nonumber\\
    \sum_{\vec k\in\mathbb Z^3}\frac{\mu^{2\epsilon}}{\left( \sum_{i=1}^3\big(\frac{k^i}{L_i}+x\ess p^i\big)^2 + \Delta^2 \right)^{\frac32+\epsilon}}.
\end{multline}

Our result includes a generalized Epstein zeta function \cite{Kirsten:1994yp},
\begin{align}
    \label{e:epsteinzeta}
    \zeta(\{a_i\},\{b_i\},c;s) & \equiv \sum_{\vec n \in \mathbb Z^p} \big[ a_i^2 n_i^2 + b_i n_i + c \big]^{-s},
\end{align}
where repeated indices are implicitly summed over.  The generalized Epstein zeta function converges for $s>d$.  As per usual we wish to isolate the pole occurring at $s=d$ and determine the finite remainder.  To do so, we utilize the Poisson summation formula to provide an analytic continuation of the generalized Epstein zeta function; we detail the derivation in \cref{s:appsum}.  We may immediately apply \cref{e:epsteinanalytic} with $s=\frac 32+\epsilon$ to find
\begin{align}
    V&(p^2,\{L_i\};\mu,\epsilon) = -\frac12\frac{1}{(4\pi)^2}\int_0^1dx\left\{ \frac1\epsilon - 1 + \ln\frac{\mu^2}{\Delta^2} \right. \nonumber\\
    &\left.+2\sideset{}{'}\sum_{\vec m\in\mathbb Z^3}e^{-2\pi\ess i \ess x \sum m_i p^i L_i} K_0\Big( 2\pi\sqrt{\Delta^2\sum m_i^2L_i^2} \Big)  \right\} \nonumber\\
    & + \mathcal O(\epsilon),
\end{align}
where the suppressed limits of the sums run from $i=1\ldots3$.  One may find similar expressions using \cref{e:epsteinanalytic} for different numbers of spatial dimensions; for $n<3$ there's no divergence.

Using our modified \msbar{} convention, the renormalized NLO contribution to $2\rightarrow2$ scattering in 3 periodic spatial dimensions is
\begin{multline}
    \label{e:renormfiniteV}
    \overline V(p^2,\{L_i\};\mu) = -\frac12\frac{1}{(4\pi)^2}\int_0^1dx\left\{ \ln\frac{\mu^2}{\Delta^2} \right. \\
        \left. +2\sideset{}{'}\sum_{\vec m\in\mathbb Z^3}e^{-2\pi\ess i \ess x \sum m_i p^i L_i} K_0\Big( 2\pi\sqrt{\Delta^2\sum m_i^2L_i^2} \Big)  \right\},
\end{multline}
and the counterterm is unchanged from the infinite volume case.

Since asymptotically $K_0(z)\sim \exp(-z)/\surd z$ we see that the finite system size corrections naturally go to zero as the system size grows.  Notice further that the UV divergence is unaffected by the finite system size corrections.  We should have expected this lack of sensitivity of the UV divergence to the finite system size, since a finite system size acts as an IR cutoff; the infinitely small distances probed at the infinite UV are insensitive to the global existence of any boundary conditions for the manifold (effectively) infinitely far away.  As a result, a leading logarithmic analysis such as from an application of the Callan-Symanzik equation won't be able to capture the finite system size effects on the running coupling; rather, we must explicitly perform the resummation of the 1PI diagrams to see the subleading $1/L$ corrections to the running coupling.  Even though this analysis is subleading log in the limit of large $p$, we're interested in the momentum region in which the finite system size effects aren't vanishingly small; i.e., we're interested in the case of $p\lesssim1/L$.

\section{Unitarity Check}
We should check the consistency of our finite system size corrections against the optical theorem: one should find self-consistently that the time evolution captured up to NLO is unitary.  For self-consistency, then, we should find that
\begin{align}
    \label{e:unitarity}
     2\ess\im\mathcal M = \sigma_{\mathrm{tot}}.
 \end{align}

We have already shown that unitarity holds in the infinite volume limit with $m=3$ spatial dimensions of infinite extent and $n=0$ spatial dimensions of finite extent in \cref{sec:infVol}.  We will show that unitarity holds as we increase $n$ incrementally to 3.  However, instead of rederiving $\overline V$ for each case, we may simply take the appropriate number of $L_i\rightarrow\infinity$.  Since the modified Bessel function of the second kind decreases exponentially with argument, for each $i$ for which $L_i\rightarrow\infinity$ we may take the iterator $k_i\equiv0$\footnote{There is a slight subtlety here.  We will see that for configurations such that all outgoing momenta are in the finite spatial directions the amplitude diverges (since the particles interact an infinite number of times).  These ``geometric bound states'' only occur for countably discrete number of momentum configurations, so the limit of the finite system size correction going to the infinite volume result will hold everywhere but on a set of measure zero.}.

In general, the left hand side of \cref{e:unitarity} yields
\begin{multline}
    2\ess\im\mathcal M
     = -2\lambda^2\ess\im\big( \overline V(s,\{L_i\};\mu) \\ + \overline V(t,\{L_i\};\mu) + \overline V(u,\{L_i\};\mu) \big), 
\end{multline}
where $\overline V(p^2,\{L_i\};\mu)$ is given by \cref{e:renormfiniteV}.  As noted in \cref{s:appsum}, one may organize the sum for the finite system size correction such that the phases are only cosines.  Therefore the only contribution to the imaginary part of the amplitude may come from values of $x$ such that $\Delta^2<0$, in which case there are contributions from evaluating the logarithm of negative numbers and from evaluating the modified Bessel function for arguments with an imaginary part.  Since $t$ and $u$ are non-positive, we therefore have that $\im\mathcal M$ only comes from $\overline V(s,\{L_i\};\mu)$.  We will in general work in the center of mass frame, in which case $p^i = 0$ for the $s$ channel $\overline V(s,\{L_i\};\mu)$.  Recall from the infinite volume case reviewed in \cref{sec:infVol} that $\re\Delta^2<0$ for $s>4m^2$ and $x_-<x<x_+$, where $0<x_\pm<1$ are given by \cref{e:xpm}.  As was done in the infinite volume review, we self-consistently align the branch cuts of $\arg$, $\log$, $K_0$, and the square root along the negative real axis.  Then the small imaginary part from the propagators in the loop means that for $s>4m^2$ and $x_-<x<x_+$ we have that $\Delta^2$ is in the third quadrant of the complex plane; thus $\sqrt{\Delta^2}$ is in the fourth quadrant of the complex plane.  Therefore for $x_-<x<x_+$ we have, for $\Delta^2\equiv-x(1-x)s+m^2-i\varepsilon$
\begin{align}
    \im K_0\big( 2\pi&\sqrt{\Delta^2\sum m_i^2L_i^2} \big) \nonumber\\
    & \qquad \qquad = \im K_0\big(\varepsilon - 2\pi i |\Delta|\sqrt{\sum m_i^2 L_i^2}\big) \nonumber\\
    & \qquad \qquad = \frac\pi2J_0\big(2\pi|\Delta|\sqrt{\sum m_i^2 L_i^2}\big),
\end{align}
where $J_0$ is the usual Bessel function of the first kind, and we've dropped the irrelevant terms linear and higher order in $\varepsilon$ on the right hand side (and can take $|\Delta| = x(1-x)s - m^2$).  

Defining
    $\tilde Q \equiv \sqrt{1-(4m^2/s)}$
we have
\begin{align}
    2\, \im\mathcal M
    & = \frac{\lambda^2}{16\pi}\tilde Q\theta(\tilde Q^2) \bigg[ 1 \nonumber\\ 
    & \qquad + \frac{1}{\tilde Q}\sideset{}{'}\sum_{\vec m\in\mathbb Z^n} \int_{x_-}^{x^+}dx\,J_0(2\pi |\Delta|\sqrt{\sum m_i^2L_i^2}) \bigg] \nonumber\\
    & = \frac{\lambda^2}{16\pi}\tilde Q\theta(\tilde Q^2) \bigg[ 1 \nonumber\\ 
    & \qquad + \sideset{}{'}\sum_{\vec m\in\mathbb Z^n} \int_{0}^{1}dv\,J_0(\pi \sqrt{s\sum m_i^2L_i^2}\tilde Q\sqrt{1-v^2}) \bigg] \nonumber\\
    & = \frac{\lambda^2}{16\pi}\tilde Q\theta(\tilde Q^2)\sum_{\vec m\in\mathbb Z^n}\sinc(\pi \sqrt{s\sum m_i^2L_i^2}\tilde Q) \nonumber\\
    \label{e:opticalLHS1}
    & = \frac{\lambda^2}{16\pi}\tilde Q\theta(\tilde Q^2)\sum_{\vec{\tilde m}\in\Lambda^n}\sinc(\pi \sqrt{s}\tilde Q|\vec{\tilde m}|).
\end{align}
In the first line, the 1 in the square brackets is the contribution from $\int dx\,\im\ln\mu^2/\Delta^2$; we have also already taken the imaginary part of the $K_0$ Bessel function.  In the second line we defined the new variable $v\equiv (2x-1)/\tilde Q$ and exploited the symmetry of the integrand about $v=0$.  In the third line we evaluated $\int_0^1 dv J_0(a\sqrt{1-v^2})=\sinc(a)$ \cite{Gradshteyn}.  In the fourth line, $\sinc$ is the usual unnormalized $\sinc$ function,
\begin{align}
    \sinc(x) \equiv \frac{\sin(x)}{x}.
\end{align}
In the last line, we exchanged the sum over integers $\vec m$ (which are weighted by $L_i^2$ in the summand) with a sum over the lattice $\Lambda^n$ defined by the $n$ lengths $L_i$.  We make this last change to a sum over a lattice in anticipation of exploiting the Poisson summation formula over lattices \cite{cohn2013formal}.  The Poisson summation over lattices is given by
\begin{align}
    \sum_{\vec{\tilde m}\in\Lambda^n}f(\vec{\tilde m}) = \frac{1}{\det \Lambda}\sum_{\vec{\tilde k}\in\Lambda^{*n}}\tilde F(\vec{\tilde k}),
\end{align}
where $\Lambda^*$ is the lattice dual to $\Lambda$ and $\tilde F$ is the usual Fourier transform of $f$,
\begin{align}
    \tilde F(\vec {\tilde k} )\equiv \int d^n m \, e^{2\pi\ess i \ess \vec k\cdot \vec m} f(\vec {\tilde m}).
\end{align}
Following exactly the method of performing the $n$ dimensional Fourier transform as shown in \cref{s:appsum} with now $f(\vec{\tilde m}) = \sinc(\pi \sqrt{s}\tilde Q|\vec{ \tilde m}|)$, we have that
\begin{align}
    \tilde F(\vec {\tilde k})
    & = \Omega_{n-2}\sqrt\pi\frac{\Gamma\big(\frac{n-1}{2}\big)}{\Gamma\big(\frac{n}{2}\big)} \int_0^\infinity \tilde m^{n-1}d\tilde m \nonumber\\
    & \qquad\qquad \times \sinc(\pi\sqrt{s}\tilde Q\tilde m){}_0F_1(;\,\frac n2;\,-(\pi\tilde k\tilde m)^2) \nonumber\\
    & = \Omega_{2-n}\big( \frac s4\tilde Q^2-\tilde k^2 \big)^{\frac{1-n}{2}}\theta\big( \frac s4\tilde Q^2-\tilde k^2 \big).
\end{align}
Thus
\begin{multline}
    \label{e:opticalLHS}
    2\, \im \mathcal M = \frac{\lambda^2}{2(4\pi)^2\sqrt s}\theta(\tilde Q^2) \Omega_{2-n} \\ \times\frac{1}{\prod L_i}\sum_{\vec{\tilde k}\in\Lambda^{*n}}\big( \frac s4\tilde Q^2-\tilde k^2 \big)^{\frac{1-n}{2}}\theta\big( \frac s4\tilde Q^2-\tilde k^2 \big).
\end{multline}

Consider now the total cross section,
    \begin{align}
        \sigma_{\mathrm{tot}}
        & = \frac12\sum_{\vec k_1\in\mathbb Z^n}\frac{1}{(2\pi)^n\prod L_i}\int\frac{d^mp_1}{(2\pi)^m2E_1} \nonumber\\
        & \qquad\times \sum_{\vec k_2\in\mathbb Z^n}\frac{1}{(2\pi)^n\prod L_i}\int\frac{d^mp_2}{(2\pi)^m2E_2} \nonumber\\
        & \qquad \times\lambda^2(2\pi)^4\prod L_i\delta(p_A^0+p_B^0-p_1^0-p_2^0) \nonumber\\
        & \qquad \times \delta^{(m)}(\vec p_A + \vec p_B - \vec p_1 - \vec p_2)\delta^{(n)}_{\vec k_A+\vec k_B,\vec k_1+\vec k_2}.
    \end{align}
We may immediately collapse the $p_2$ integrals with the Dirac delta functions and the $k_2$ sums with the Kronecker deltas.  Then 
\begin{align}
    \sigma_{\mathrm{tot}}
    & = \frac{\lambda^2}{2(2\pi)^2}\sum_{\vec k_1\in\mathbb Z^n}\frac{1}{\prod L_i}\int \frac{d^mp_1}{(2E_1)^2} \nonumber\\ 
    & \qquad\qquad\qquad\qquad\times \delta(\sqrt{s} - 2\sqrt{p_1^2 + \sum \frac{k_i^2}{L_i^2}+m^2}) \nonumber\\
    & = \frac{\lambda^2}{2(4\pi)^2\sqrt s}\theta(\tilde Q^2)\Omega_{m-1}\nonumber\\
    & \qquad \times \frac{1}{\prod L_i} \sum_{\vec k\in\mathbb Z^n}\big( \frac s4\tilde Q^2 - \sum{k_i^2}{L_i^2} \big)^{\frac{m-2}{2}} \nonumber\\
    & \qquad \times \theta\big( \frac s4 - m^2 - \sum{k_i^2}{L_i^2} \big) \nonumber\\
    & = \frac{\lambda^2}{2(4\pi)^2\sqrt s}\theta(\tilde Q^2)\Omega_{2-n} \nonumber\\
    & \qquad \times \frac{1}{\prod L_i}\sum_{\vec {\tilde k}\in\Lambda^{*n}}\big( \frac s4\tilde Q^2 - \vec{\tilde k}^2 \big)^{\frac{1-n}{2}}
    \label{e:opticalRHS}
    \theta\big( \frac s4\tilde Q^2 - \vec{\tilde k}^2 \big).
\end{align}
In the second line we integrate out the Dirac delta function.  In the third line, we use $n+m=3$ to set $m = 3-n$ and also change the sum over the integers to a sum over the lattice dual to the lattice from \cref{e:opticalLHS1}.

One can readily see that \cref{e:opticalLHS,e:opticalRHS} are equal, and therefore the optical theorem (i.e.\ unitarity) is satisfied for our newly derived result for $n = 0,\,1,\,2,\,$ or $3$ compact spatial dimensions of lengths $\{L_i\}$.  

We provide an alternative check on unitarity for \cref{e:renormfiniteV} for $n=2$ or 3 compact spatial dimensions that utilizes a novel generalization of a conjectured identity from Ramanujan and Hardy involving the square counting function \cite{Hardy1978RamanujanTL} in \cref{s:altunitarity}.

%%%%%%%%%%%%%%%%%%%%%%%%%%%%%%%%%%%%%%%%%%%%%%%%%%%%%%%%%%%%%%%%%%%%%%%%%%%%%%%%%%%%%%%%%%%%%
\section{Sample Plots of \texorpdfstring{$\overline V$}{Vbar}}\label{sec:vbarplots}
%%%%%%%%%%%%%%%%%%%%%%%%%%%%%%%%%%%%%%%%%%%%%%%%%%%%%%%%%%%%%%%%%%%%%%%%%%%%%%%%%%%%%%%%%%%%%
We would like to have some sense of the scale and type of correction due to the presence of the finite size of the system; in particular, we would like to compare the magnitude and sign of the contribution that remains when all lengths are taken to infinity to the magnitude and the sign of the contribution that goes to zero smoothly as all lengths are taken to infinity.

Since the full result we derived earlier for $2\rightarrow2$ scattering \cref{e:renormfiniteV} captures both the infinite volume and finite volume contributions, and we may take any one or more finite lengths to infinity, we will do so now.  In particular, since the modified Bessel function of the second kind decays exponentially for large argument, for any length $L_i$ taken to infinity, we may simply take only the $m_i=0$ contribution associated with that length to find the result for only $n = 1$ or $2$ compact directions.

If we restrict ourselves to considering only $n= 1$ or $2$, then we may avoid any questions about the applicability of or corrections to the usual LSZ and Gell-Mann--Low formalisms associated with scattering in quantum field theories so long as we have our incoming (outgoing) particles come from one of the directions of infinite length. It is important to consider that in compactified directions the physical momentum modes are discretized, so instead of considering scattering angles in the finite directions, we consider the mode number instead.

In general, one may straightforwardly evaluate \cref{e:renormfiniteV} by brute force for $p^2<0$, i.e.\ for the $t$ or, equivalently, the $u$ channel contribution to the amplitude.  In addition to the obvious numerical check of doubling the number of terms included in the sum and verifying that the results are unchanged, we further confirm that the brute force method is accurate by comparing to an asymptotic analytic analysis in \cref{sec:asymptotic}.

The numerics are significantly more difficult in the $s$ channel, since the integral over $x$ yields both positive and negative values of the argument of the modified Bessel function.  As a result, the $s$ channel has both real and imaginary parts.  One may show that the imaginary part is straightforward and involves only a sum of a finite number of terms.  The real part involves an infinite sum over oscillating (not modified) Bessel functions.  For one compact dimension, one may make significant analytic progress on the $s$ channel.  Using
\begin{align}
    \re K_0(-x-i\ess\varepsilon) = -\frac\pi2Y_0(x), \qquad x > 0,
\end{align}
and
\begin{align}
    \int_0^z Y_0(\sqrt{z^2-x^2})dx = \frac2\pi\big[ \operatorname{Ci}(z)\sin(z) - \operatorname{Si}(z)\cos(z) \big],
\end{align}
where $Y_0$ is the usual Bessel function of the second kind, $\operatorname{Ci}(x)$ is the usual cosine integral, and $\operatorname{Si}(x)$ is the usual sine integral\footnote{The integral over the $Y_0$ modified Bessel function may be derived by integrating its Froebenius expansion term by term and then performing the infinite sum.}, and noting that $p_i\equiv 0$ in the $s$ channel.  We then have that the real part of the finite volume contribution is
\begin{align}
    \re & \overline V_{\mathrm{fin}}(s,\,L;\,\mu) \nonumber\\
    & = 4\sum_{\ell=1}^\infinity\Big[2 \int_0^{x_-}
    K_0\big(2\pi\ess \ell\ess L\sqrt{-4x(1-x)E^2+m^2}\big)dx \nonumber\\
    & \qquad + \frac{1}{2\pi\ess L\ess E\ess \ell}\Big( \operatorname{Ci}\!\big(2\pi\ess \ell\ess L\ess q\big)\sin(2\pi\ess \ell \ess L \ess q) \nonumber\\
    \label{eq:vbarnum1}
    & \qquad \qquad\qquad- \operatorname{Si}\!\big(2\pi\ess \ell\ess L\ess q\big)\cos(2\pi\ess \ell \ess L \ess q) \Big)\Big],
\end{align}
where $s=4E^2$ and $q\equiv \surd(s-4m^2)$.  The integral over the modified Bessel function is numerically safe as the argument is pure real over $x\in(0,x_-)$.  Then, from \cref{e:opticalLHS1} and \cref{eq:1Dsinc}, we have that
\begin{align}
    \im & \overline V_{\mathrm{fin}}(s,\,L;\,\mu) = \pi\frac{ q}{E}\Big( \frac{1}{2L q}+\frac{\lfloor L q\rfloor}{L q}-1\Big).
\end{align}

Although the cosine and sine integral special functions oscillate in \cref{eq:vbarnum1}, the overall $1/\ell$ factor leads to a numerically stable solution.  In two (or more) compact dimensions, a similar analysis as done to arrive at \cref{eq:vbarnum1} yields a sum that does not behave well numerically for small $q$ or $L$.  A direct evaluation of \cref{e:renormfiniteV} similarly does not yield a numerically convergent result.  

In order to obtain plots of the real part of the $s$ channel contribution in two compact dimensions, we tried to analytically perform the complete infinite sum first, then numerically evaluate the integral over $x$.  In one compact dimension, one can factor out the iterator in the sum from the square root in the argument of the modified Bessel function.  One may then perform the complete infinite sum by using a common integral representation of the modified Bessel function,
\begin{align}
    \sum_{\ell = 1}^\infinity K_0(\ell x) 
    & = \sum_{\ell = 1}^\infinity \int_0^\infinity e^{-\ell x\cosh t}dt \nonumber\\
    & = \int_0^\infinity \frac{dt}{e^{x\cosh t}-1}.
\end{align}
However, in two compact dimensions, in order to perform the infinite sum one needs an integral representation of the modified Bessel function that, e.g., looks like the exponential of the square of the argument of the modified Bessel function.  We were unable to find any such useful representations in the literature.

Rather, the method we found that could provide numerically stable results relied on a Kramers-Kronig type dispersion relation we derived relating the (derivative of the) real part of $\overline V(s)$ to the imaginary part of $\overline V(s)$.  The detailed derivation and consistency checks of this result is outside the main scope of this work; we leave the details to \cref{s:dispersion}.  We provide several sample plots for the cases of 1 and 2 compact dimensions in the following subsections.

\subsection{One Compact Dimension}
We will consider scattering in the center of mass frame, where 
\begin{align}
    \label{e:onedvariables}
   p_{\text{in}}^i & =\left(\frac{n_i}{L},0,p_\text{inf}\right)^i \nonumber\\
   p_{\text{out}}^i & =\left(\frac{n_f}{L},\sin(\theta) p_{\text{inf},\,f},\cos(\theta)p_{\text{inf},\,f}\right)^i \\
   p_{\text{inf},\,f} & \equiv \sqrt{p_\text{inf}^2+\frac{n_i^2-n_f^2}{L}}, \nonumber
\end{align}
where $n_i,\,n_f\in\mathbb Z$ are the number of modes in the compact dimension such that the Mandelstam variables are given by
\begin{align}
   s & = 4(m^2+\vec{\,p}_\text{in}^2\!) \nonumber\\
   t & =-(\vec{\,p}_\text{in}\!-\vec{\,p}_\text{out}\!)^2 \\
   u & =-(\vec{\,p}_\text{in}\!+\vec{\,p}_\text{out}\!)^2. \nonumber
\end{align}
\subsubsection{\texorpdfstring{$s$}{s} Channel}\label{sec:1dschannel}

\begin{figure}[!tbp]
    \centering
    \includegraphics[width=\columnwidth]{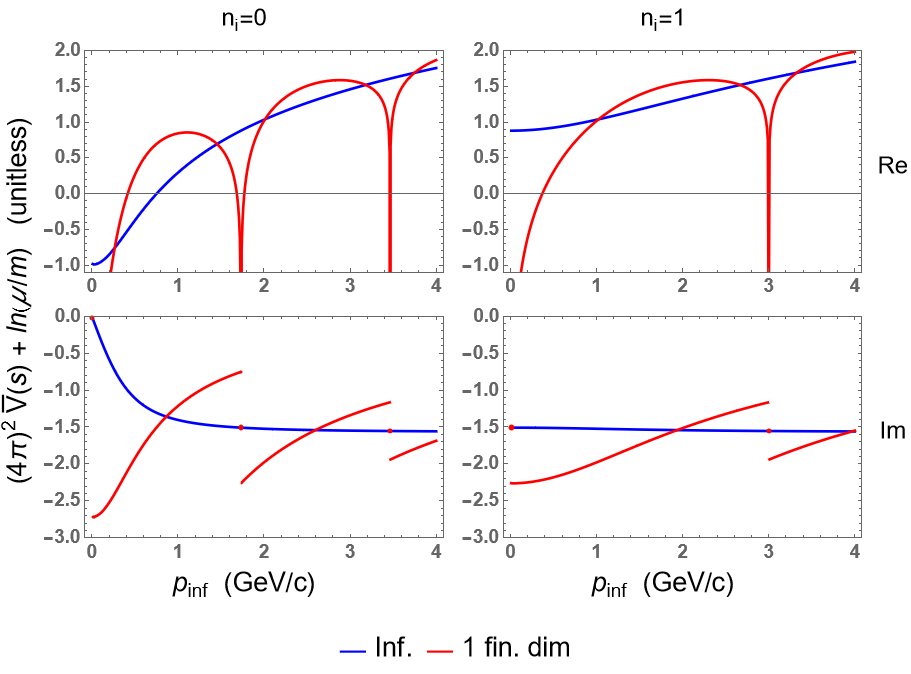}
    \caption{
    A plot of the infinite volume (blue) and finite length (red) real part (top) and imaginary part (bottom) of $\overline{V}_1(s,p_\text{inf},\mu)$ as a function of $p_\text{inf}$ for 1 compact dimension; left has no initial momentum in the compact direction, and right has one mode in the compact direction.  $\mu = 1$ GeV, $m=0.5$ GeV, and $L=1/\surd 3$ GeV$^{-1}$.  
    }
    \label{f:1DVscomparisonp}
\end{figure}

\begin{figure}[!tbp]
    \centering
    \includegraphics[width=\columnwidth]{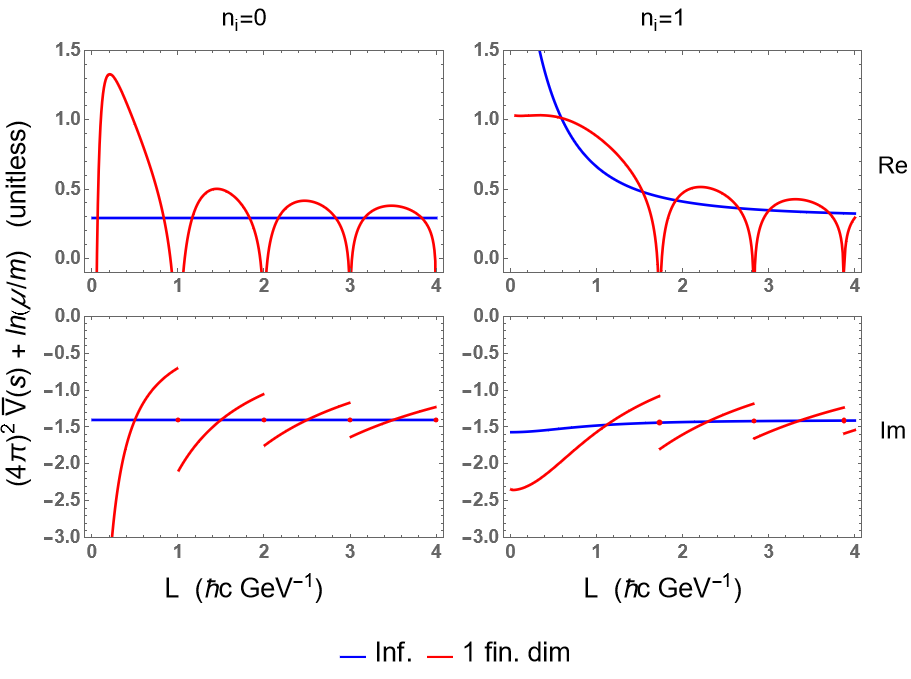}
    \caption{
    A plot of the infinite volume (blue) and finite length (red) real part (top) and imaginary part (bottom) of $\overline{V}_1(s,L,\mu)$ as a function of $L$ for 1 compact dimension; left has no initial momentum in the compact direction, and right has one mode in the compact direction.  $\mu = 1$ GeV, $m=0.5$ GeV, and $p_\text{inf} = 1$ GeV.  
    }
    \label{f:1DVscomparisonL}
\end{figure}

We use \cref{eq:g1new,eq:vbarg} to obtain
\begin{align}\label{eq:vs1}
    &\overline{V}_1(s,L;\mu)=-\frac{1}{32\pi^2}\Bigg[\ln\left(\frac{\mu^2}{m^2}\right)+a_1(L m)\\
    &+\sum_{l=-\infty}^\infty\left(\frac{2}{L }\frac{\arctanh\left(\frac{L }{2}\sqrt{\frac{s +i \varepsilon }{(L m)^2+l^2}}\right)}{\sqrt{s +i \varepsilon }}-\frac{1}{\sqrt{l^2+(L m)^2}}\right)\Bigg],\nonumber
\end{align}
where $a_1$ is given by \cref{eq:a1}.

We show a comparison between the infinite volume contribution to $\overline{V}_1(s,L,\mu)$ and the finite volume contribution when one of the three directions is finite as a function of $p_\text{inf}$  in \cref{f:1DVscomparisonp} and as a function of $L$ in \cref{f:1DVscomparisonL} for the $s$ channel. Note that the center of mass energy depends only on the magnitude of the incoming (equivalently outgoing) momentum, and not the scattering angle; thus we only plot against the value of the incoming finite system size momentum mode $n_i$. We take $\mu = 1$ GeV and $m=0.5$ GeV.  For the scan in $p_\text{inf}$ we take $L=1/\surd 3$ GeV$^{-1}$ and for the scan in $L$ we take $p_\text{inf} = 1$ GeV.  For the scan in $p_\text{inf}$, the length $L\sim0.1$ fm is rather small phenomenologically; however, the value is useful for illustrative purposes.  For $s>4m^2\Leftrightarrow |\vec p|>0$, both the infinite volume and finite volume corrections have non-vanishing imaginary parts.  

One can see that the finite system size correction does indeed converge to zero as either $|\vec p|\rightarrow\infinity$ or $L\rightarrow\infinity$.  However, the convergence is non-trivial.  For $|\vec p|\times L$ integer, the finite system size correction to the real part of $\overline{V}_1(s,L,\mu)$ receives a contribution from a term proportional to $-\arctanh(1)=-\infty$, while one can show that the finite system size correction to the imaginary part is identically zero at these integer values of $|\vec p|\times L$.  Further, the imaginary part is discontinuous across values of $|\vec p|\times L$ integer due to the floor function in \cref{eq:1Dsinc}.  Thus for large argument, the convergence of the correction to zero is only almost everywhere.  

We can interpret the divergences in the real part of $\overline{V}_1(s,L,\mu)$ in the $s$-channel as due to ``geometric bound states,'' in which \emph{all} the momentum of each of the outgoing particles is in the finite direction.  Since all the momentum is in the finite direction, the particles propagate back and forth a divergent number of times in the finite direction, with a divergent number of interactions over an arbitrarily long time.   With such an $out$ state, the particles take an arbitrarily long time to infinitely separate.  Therefore the amplitude picks up a resonance associated with the particles being in a ``geometric" bound state, with a corresponding pole in the $S$ matrix as one is used to for more generic bound states.  

It's important to note that as either $p_\text{inf}\rightarrow0$ or $L\rightarrow 0$ the divergences in the amplitude are not due to a geometric bound state: in neither case is there enough momentum to fill even one mode in the finite direction.  Rather, these IR-like divergences are precisely the effect of the small size of the system compared to the momentum ($p_\text{inf}\times L\ll 1$) we sought to capture in this work. We see, in fact, that the effect of placing the theory in a finite sized system \emph{diverges} as the effective size of the system (as measured in the length units set by the momentum in the system) goes to zero, $p_\text{inf}\times L\ll 1$.

In particular, by noting that $s\,L^2/4\to (L m)^2+n_i^2$ as $p_\text{inf}\to0$, and that $\arctanh(1)$ diverges, one can easily show that the real part of \cref{eq:vs1} diverges (to negative infinity) for all $n_i\in\mathbb{Z}$. This behavior can be seen then in \cref{f:1DVscomparisonp}. 

\begin{figure}[!tbp]
    \centering
    \includegraphics[width=\columnwidth]{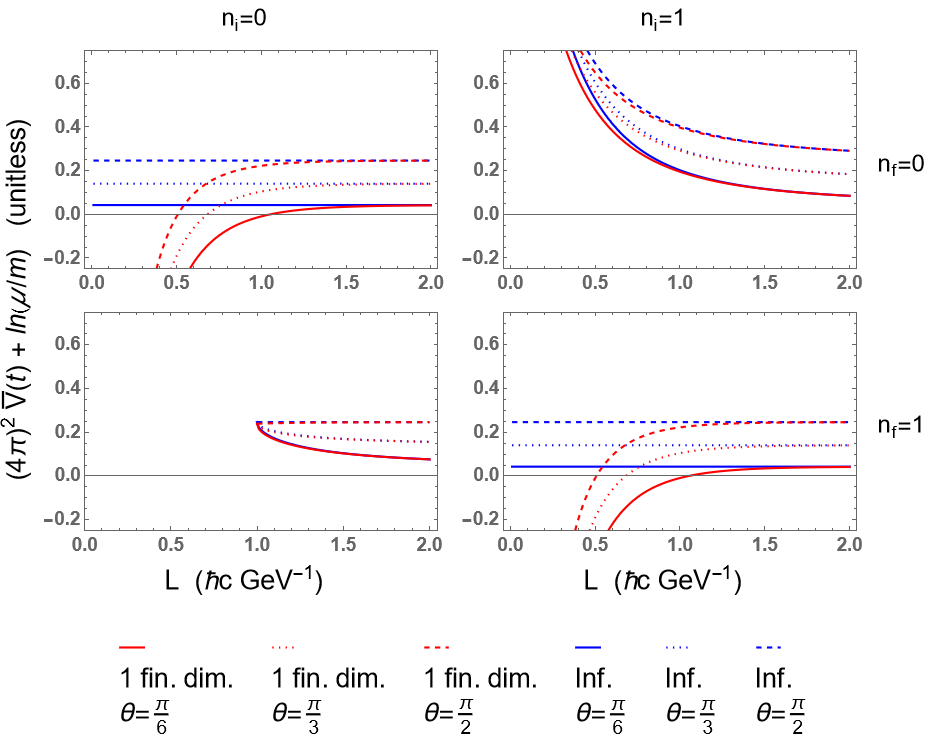}
    \caption{
    A plot of the infinite volume (blue) and finite length (red) $\overline V_1(t,L,\mu)$ as a function of $p_\text{inf}$ for $n=1$ compact dimension for three values of the scattering angle $\theta$ among the two remaining infinite directions, $\theta = \pi/6,\,\pi/3,$ and $\pi/2$, for solid, dotted, and dashed curves, respectively.  The initial momentum has zero (one) modes in the finite direction in the left (right) column; the final momentum has zero (one) modes in the finite direction in the top (bottom) row.  $\mu = 1$ GeV, $m=0.5$ GeV, and $p_\text{inf} = 1$ GeV.
    }
    \label{f:1DVtcomparisonL}
\end{figure}

In order to fully understand the $L\to0$ limit, we need a good handle on the asymptotics of $a_1$, which is non-trivial. We can however gain intuition from the observation that $s\,L^2/4\to n_i^2$.  If $n_i=0$ then one finds a $(L m)^{-1}$ divergence from the $l=0$ term in \cref{eq:vs1}. For $n_i>0$, the only divergence may come from $a_1'(L\,m)$, where $a_1'$ is given by \cref{eq:a1} excluding the $l=0$ term.  One can see in \cref{f:1DVtcomparisonL} the divergences in the real and imaginary parts of $\overline V_1$ as $L\to0$ for $n_i = 0$.

%_______________________________________________________________
\begin{figure}[!tbp]
    \centering
    \includegraphics[width=\columnwidth]{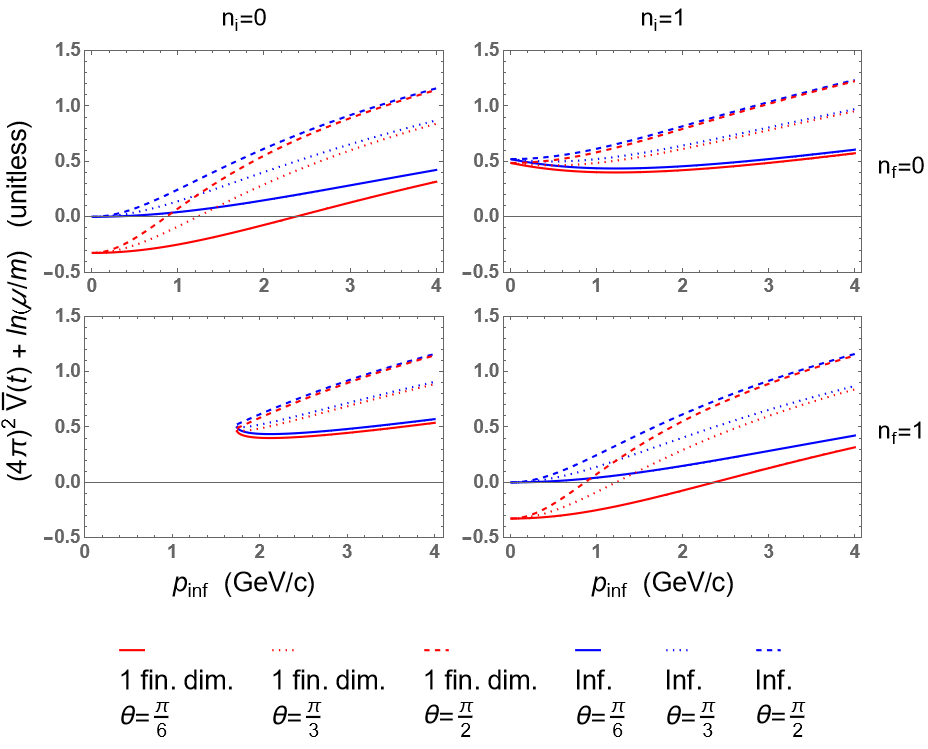}
    \caption{
    A plot of the infinite volume (blue) and finite length (red) $\overline V_1(t,p_\text{inf},\mu)$ as a function of $p_\text{inf}$ for $n=1$ compact dimension for three values of the scattering angle $\theta$ among the two remaining infinite directions, $\theta = \pi/6,\,\pi/3,$ and $\pi/2$, for solid, dotted, and dashed curves, respectively.  The initial momentum has zero (one) modes in the finite direction in the left (right) column; the final momentum has zero (one) modes in the finite direction in the top (bottom) row.  $\mu = 1$ GeV, $m=0.5$ GeV, and $L=1/\surd 3$ GeV$^{-1}$.
    }
    \label{f:1DVtcomparisonp}
\end{figure}
%_______________________________________________________________
%_______________________________________________________________
%_______________________________________________________________

\subsubsection{\texorpdfstring{$t$}{t} Channel}

\begin{figure*}[htbp]
    \centering
    \includegraphics[width=0.75\textwidth]{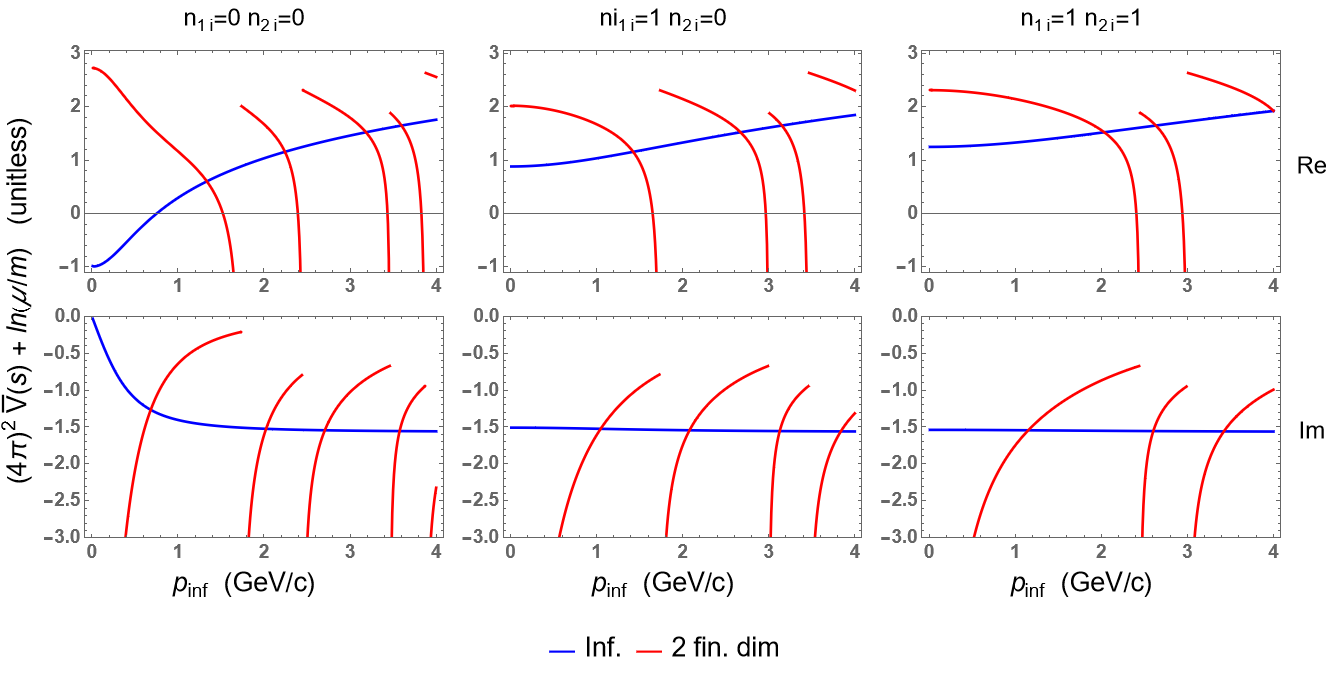}
    \caption{
    A plot of the infinite volume (blue) and finite length (red) real part (top) and imaginary part (bottom) of $\overline{V}_2(s,p_\text{inf},\mu)$ as a function of $p_\text{inf}$ for 2 compact dimensions.  Left has no initial momentum in either of the compact directions, middle has one mode in one finite direction, and right has one mode in both compact directions.  $\mu = 1$ GeV, $m=0.5$ GeV, and $L=1/\surd 3$ GeV$^{-1}$ for both compact directions.  
    }
    \label{f:2DVscomparisonp}
\end{figure*}

We show in \cref{f:1DVtcomparisonp,f:1DVtcomparisonL} a similar comparison as in \cref{f:1DVscomparisonp,f:1DVscomparisonL} but for the $t$ channel contribution for three values of scattering angle among the infinite directions $\theta=\pi/6,\,\pi/3,$ and $\pi/2$.  In this case, the convergence to zero of the finite system size correction is smooth as a function of either momentum or length going to infinity. If one, instead of fixing the modes in the finite direction, fix the scattering angle into the finite direction, and consider the discrete set of incoming momenta for which such a scattering angle is physical, one more clearly sees the slight oscillations from the phases in \cref{e:renormfiniteV}.  The finite system size correction is finite for fixed $L$ as $|\vec p|\rightarrow0$ for $m>0$; the non-zero mass sets an IR cutoff that limits the influence of the finite system size correction.  Not surprisingly, the finite system size correction becomes larger and larger, and subsequently more and more important, as $L$ decreases; the finite system size correction diverges for fixed $|\vec p|$ as $L\rightarrow0$.  The non-trivial dependence of $\overline V$ on scattering angle $\theta$ implies that the effective coupling will also in general depend on the scattering angle $\theta$.

In \cref{f:1DVtcomparisonp,f:1DVtcomparisonL} for $n_i<n_f$ $\overline V_1$ ceases to exist below a certain $p_\text{inf}$ or $L$ since there is no way to maintain conservation of energy in these regions.  

\cref{f:1DVtcomparisonL} shows a divergence to negative infinity for the finite volume correction as $L\to0$ for the $n_i=0=n_f$ and the $n_i=1=n_f$ cases and a divergence to positive infinity for both the infinite volume and the finite system size $\overline V_1$ in the $n_i = 1,\, n_f = 0$ case.  We may understand these divergences as follows.  For $L\to0$, $-t\,L/2\to n_i(n_i-n_f)$.  If $n_i\ne0$ and $n_i\ne n_f$, then we must have that $t\rightarrow-\infinity$ as $L\to0$.  Then one sees that the $-\ln\mu^2/\Delta^2$ term of \cref{e:renormfiniteV} will diverge to positive infinity.  Further the finite system size correction goes to zero as $K_0(z)\sim \exp(-z)/\surd z$ for $z\gg 1$, leading to the behavior seen for $n_i=1$ and $n_f=0$ case in \cref{f:1DVtcomparisonL}.  For $n_i=0=n_f$ and $n_i=1=n_f$, when $L\to0$ we have that $-t<\infinity$.  In this case the log remains finite, but the argument of the $K_0$ modified Bessel function in \cref{e:renormfiniteV} goes to 0, which leads to the divergence as $K_0(z)\sim \ln z$ for $z\ll1$.

%%%%%%%%%%%%%%%%%%%%%%%%%%%%%%%%%%%%%%%%%%%%%%%%%%%%%%%%%%%%%%%%%%%%%%%%%%%%%%%%%%%%%%%%%%%%%
\subsection{Two Compact Dimensions}

\begin{figure*}[htbp]
    \centering
    \includegraphics[width=0.75\textwidth]{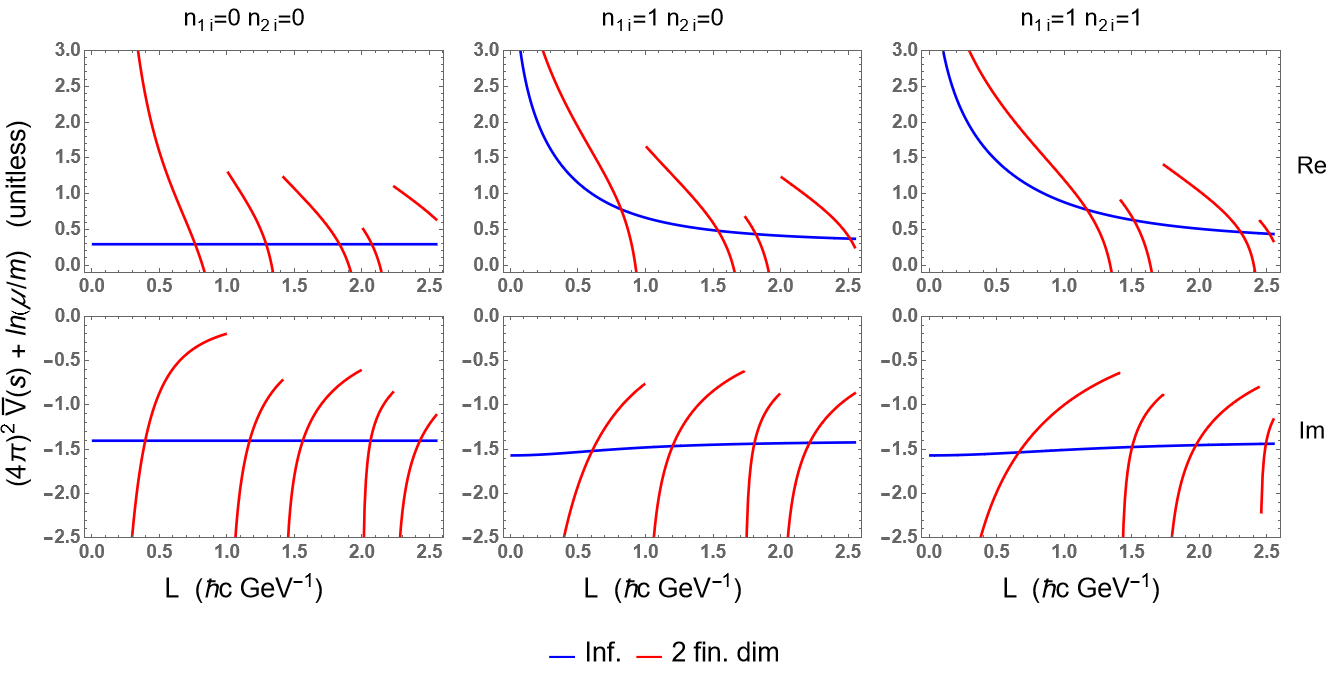}
    \caption{
    A plot of the infinite volume (blue) and finite length (red) real part (top) and imaginary part (bottom) of $\overline{V}_2(s,L,\mu)$ as a function of the length $L$ for 2 compact dimensions.  Left has no initial momentum in either of the compact directions, middle has one mode in one finite direction, and right has one mode in both compact directions.  $\mu = 1$ GeV, $m=0.5$ GeV, and $p_\text{inf} = 1$ GeV.  
    }
    \label{f:2DVscomparisonL}
\end{figure*}

We will consider scattering in the center of mass frame, where 
\begin{align}
   p^i_\text{in} & = \left(\frac{n_{1i}}{L},\frac{n_{2i}}{L},p_\text{inf}\right)^i \nonumber\\
   p^i_\text{out} & = \left(\frac{n_{1f}}{L},\frac{n_{1f}}{L},\sqrt{p_\text{inf}^2+\frac{n_{1i}^2-n_{1f}^2}{L}+\frac{n_{2i}^2-n_{2f}^2}{L}}\right)^i
\end{align}
such that the Mandelstam variables are again given by
\begin{align}
   s & =4(m^2+\vec{\,p}_\text{in}^2\!) \nonumber\\
   t & =-(\vec{\,p}_\text{in}\!-\vec{\,p}_\text{out}\!)^2 \\
   u & =-(\vec{\,p}_\text{in}\!+\vec{\,p}_\text{out}\!)^2. \nonumber
\end{align}
%%%%%%%%%%%%%%%%%%%%%%%%%%%%%%%%%%%%%%%%%%%%%%%%%%%%%%%%%%%%%%%%%%%%%%%%%%%%%%%%%%%%%%%%%%%%%
\subsubsection{\texorpdfstring{$s$}{s} Channel}
%%%%%%%%%%%%%%%%%%%%%%%%%%%%%%%%%%%%%%%%%%%%%%%%%%%%%%%%%%%%%%%%%%%%%%%%%%%%%%%%%%%%%%%%%%%%%
%_______________________________________________________________

%_______________________________________________________________
\begin{figure*}[tbp]
    \centering
    \includegraphics[width=0.75\textwidth]{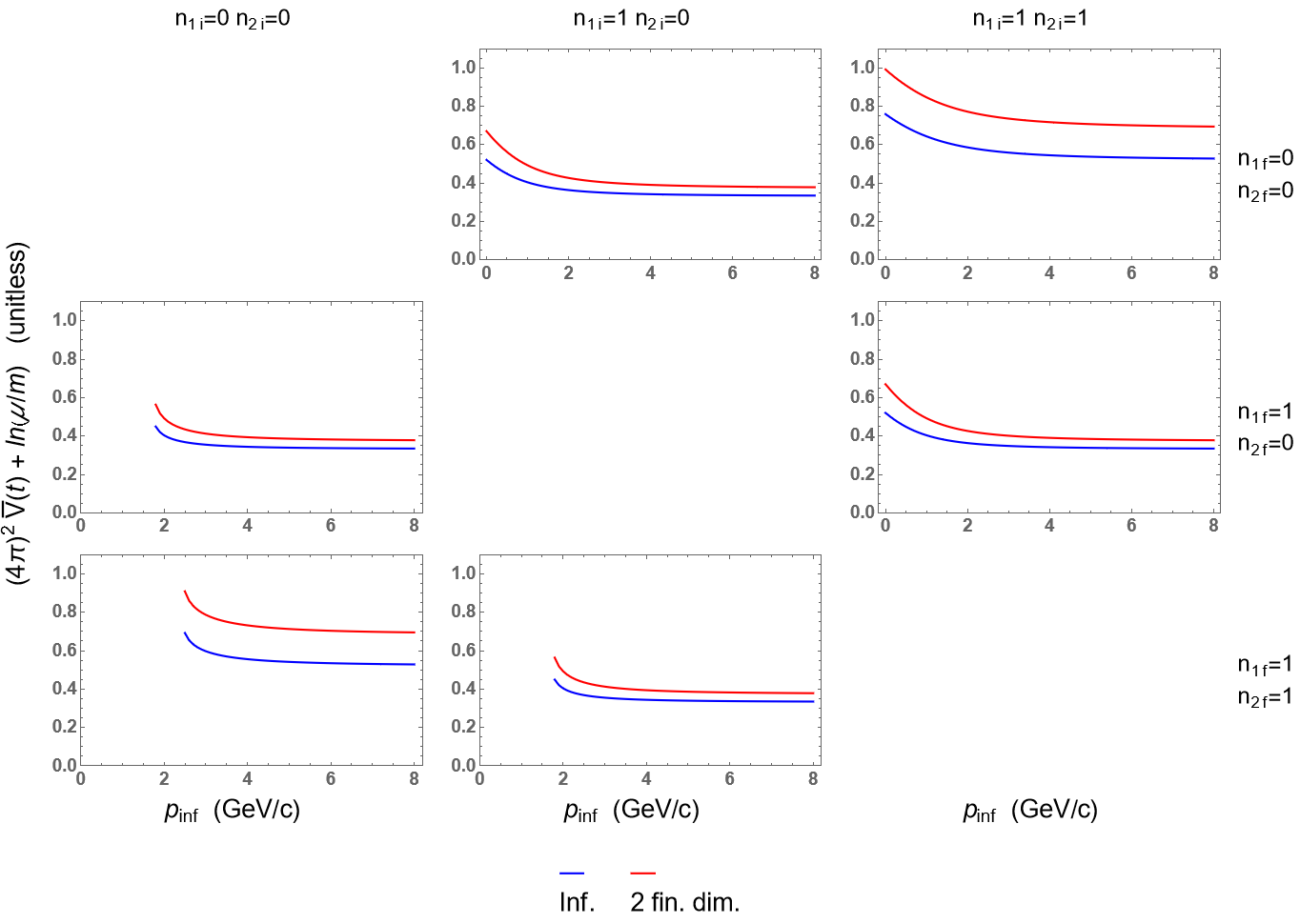}
    \caption{
    A plot of the infinite volume (blue) and finite length (red) $\overline V_2(t,p_\text{inf},\mu)$ as a function of $p_\text{inf}$ for 2 compact dimensions.  The plots are shown for various numbers of modes in the initial and final finite directions.  Plots are not shown for when the particles do not scatter, i.e.\ for $p_i=p_f$.  $\mu = 1$ GeV, $m=0.5$ GeV, and $L=1/\surd 3$ GeV$^{-1}$.
    }
    \label{f:2DVtcomparisonp}
\end{figure*}
%_______________________________________________________________

We use \cref{eq:g2new,eq:vbarg} to obtain
\begin{multline}\label{eq:vs2}
    \overline{V}_2(s,L;\mu)=-\frac{1}{32\pi^2}\Bigg[\ln\left(\frac{\mu^2}{m^2}\right)+a_2(L m)\\
    +\frac1\pi\sum_{l=0}^\infty r_2(l)\left(\frac{\frac{4}{L }\arcsin\left(\frac{L}{2}   \sqrt{\frac{s +i \varepsilon }{(L m) ^2+l}}\right)}{\sqrt{s+i \varepsilon } \sqrt{4 l-L ^2 (s-4m^2+i \varepsilon)}}\right.\\
    -\left.\frac{1}{l+(L m)^2}\right)\Bigg],
\end{multline}
where $r_2(k)$ is the square counting function, which counts the number of 2D vectors $\vec r$ of integer components whose length is $|\vec r| = k$ (e.g.\ $r_2(0)=1$, $r_2(1) = 4 = r_2(2)$), and $a_2$ is given by \cref{eq:a2}.

We show a comparison between the infinite volume contribution to $\overline{V}_2(s,L,\mu)$ and the finite volume contribution when two of the three directions are finite as a function of $p_\text{inf}$  in \cref{f:2DVscomparisonp} and as a function of $L$ in \cref{f:2DVscomparisonL} for the $s$ channel. Note again that the center of mass energy depends only on the magnitude of the incoming (equivalently outgoing) momentum, and not the scattering angle. We take $\mu = 1$ GeV and $m=0.5$ GeV.  For the scan in $p_\text{inf}$ we take $L=1/\surd 3$ GeV$^{-1}$ and for the scan in $L$ we take $p_\text{inf} = 1$ GeV.  

One sees that---like the one compact dimension case---the real part of $\overline V$ diverges at particular values of $|\!\vec{\,p}|\times L$.  For the two dimensional case, the imaginary part also diverges at these special values; also, the divergences are one-sided in the two dimensional case.  Although difficult to see in \cref{f:2DVscomparisonp,f:2DVscomparisonL}, one may show that as either $|\!\vec{\,p}|$ or $L$ goes to infinity, the finite system size correction converges almost everywhere to the infinite volume result.  It's easiest to see the special values of $|\!\vec{\,p}|$ and $L$ at which both the imaginary and real parts diverge from \cref{eq:vs2}. When $r_2\left(\frac14L ^2 (s-4m^2)\right)=r_2\left(\frac14L ^2 |\!\vec{\,p}|^2\right)\neq0$ the sum picks up a term that is divided by 0, giving a divergence; i.e.\ when $\frac14L ^2 \vec{\,p}^2\!$ is an integer that can be written as the sum of two squares (or equivalently is the length of a 2D integer vector), $\overline V$ will diverge.  When $r_2\left(\frac14L ^2 |\!\vec{\,p}|^2\right)=0$, the term falls outside of the sum over the lattice in the original expression, and the contribution is identically 0.  

We can also see in the figures that the divergences are one-sided for the real and imaginary parts, and that the divergences are from the opposite directions in the real and imaginary parts.  We may understand this one-sided divergence behavior by considering the square root in the denominator of \cref{eq:vs2}.  Suppose a particular $L_*,\,s_*,$ and $l_*$ exist such that $4 l_*-L_*^2 (s_*-4m^2+i \varepsilon) = 0$.  Now if $L\lesssim L_*$ (or, equivalently, $s\lesssim s_*$), then at $l_*$ the argument of the square root is nearly zero and positive.  Thus $\overline V_2$ will pick up a large real part.  But if we rather have that $L\gtrsim L_*$ (or, equivalently, $s\gtrsim s_*$), then at $l_*$ the argument of the square root is nearly zero and negative.  Thus $\overline V_2$ will pick up a large imaginary part.  We may use this reasoning further to understand the $p_\text{inf}\to0$ behavior seen in \cref{f:2DVscomparisonp}.  As $p_\text{inf}\to0$, $s\to4m^2$.  As $s\to4m^2$, the $\sqrt{4 l-L ^2 (s-4m^2+i \varepsilon)}$ denominator in \cref{eq:vs2} becomes very small and imaginary for $l=0$ (and only $l=0$), which leads to the real (imaginary) part of $\overline V_2$ going to a constant (diverging) as $p_\text{inf}\to0$.

The divergences away from $p_\text{inf}\rightarrow0$ and $L\rightarrow0$ shown in the figures can again be interpreted as ``geometric" bound states, where there is now the possibility for an $out$ state with all momentum distributed between the two finite dimensions. Unlike in the case of only one finite dimension, one may see from the imaginary part of $\overline V_2$ that the measure of the divergence is large enough that the total cross section is infinite at NLO.  

Again the divergences at $p_\text{inf}\rightarrow0$ and $L\rightarrow0$ are not due to geometric bound states but to the maximal influence of the small system size.  From \cref{eq:g2new}, $g_2\sim |\Lambda^*|$ and thus like $1/L^2$ for $L\to0$.  It is thus unsurprising that \cref{f:2DVscomparisonL} shows divergences in the real and imaginary parts of $\overline V_2$ as $L\to0$.

%%%%%%%%%%%%%%%%%%%%%%%%%%%%%%%%%%%%%%%%%%%%%%%%%%%%%%%%%%%%%%%%%%%%%%%%%%%%%%%%%%%%%%%%%%%%%
\subsubsection{\texorpdfstring{$t$}{t} Channel}
%%%%%%%%%%%%%%%%%%%%%%%%%%%%%%%%%%%%%%%%%%%%%%%%%%%%%%%%%%%%%%%%%%%%%%%%%%%%%%%%%%%%%%%%%%%%%

%_______________________________________________________________
\begin{figure*}[tbp]
    \centering
    \includegraphics[width=0.75\textwidth]{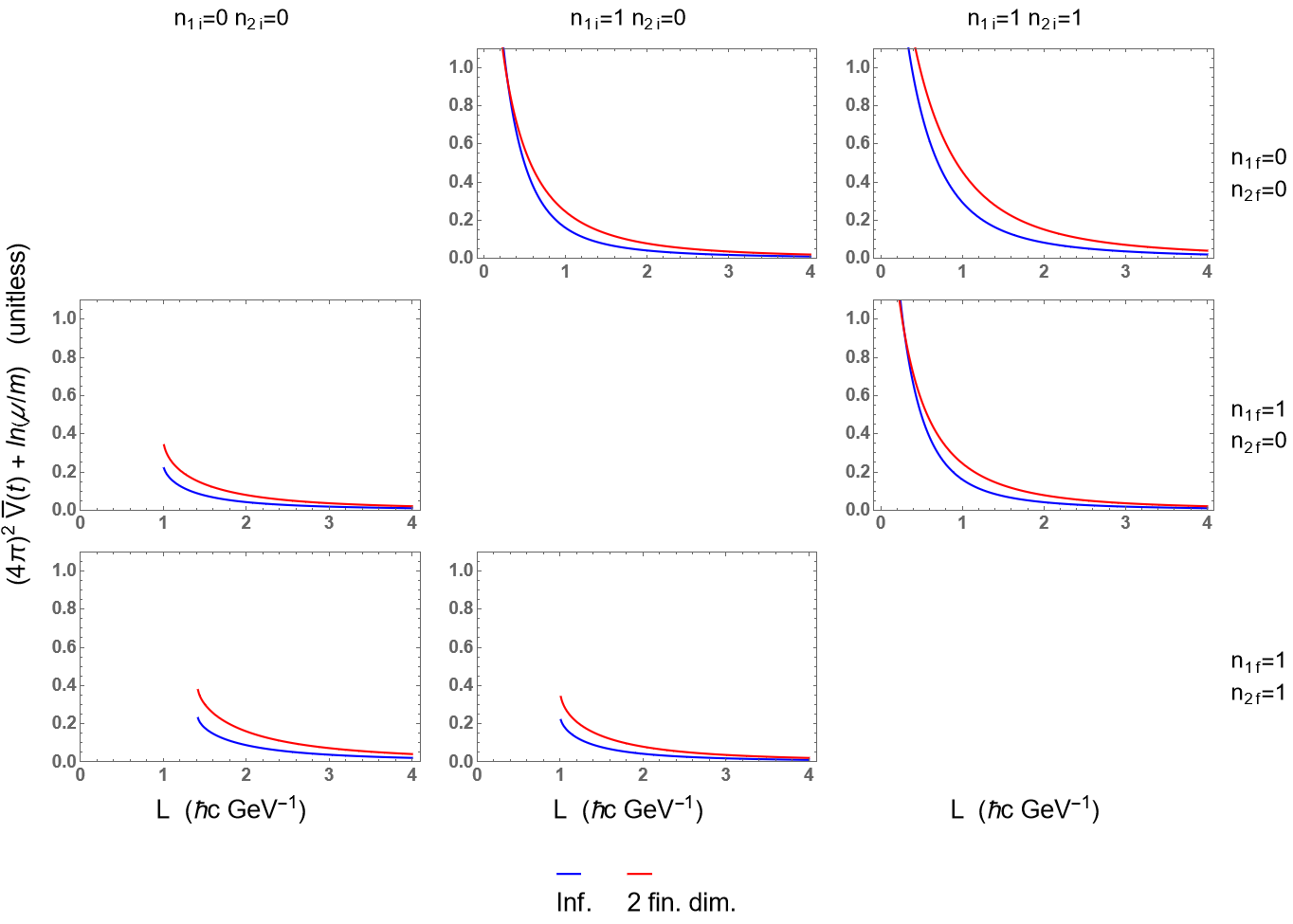}
    \caption{
    A plot of the infinite volume (blue) and finite length (red) $\overline V_2(t,L,\mu)$ as a function of $L$ for 2 compact dimensions.  The plots are shown for various numbers of modes in the initial and final finite directions.  Plots are not shown for when the particles do not scatter, i.e.\ for $p_i=p_f$.  $\mu = 1$ GeV, $m=0.5$ GeV, and $p_\text{inf} = 1$ GeV.
    }
    \label{f:2DVtcomparisonL}
\end{figure*}
%_______________________________________________________________

We compare the infinite volume and finite volume results for the $t$ channel $\overline V_2(t,L,\mu)$ as a function of $p_\text{inf}$ in \cref{f:2DVtcomparisonp} and $L$ in \cref{f:2DVtcomparisonL} for a variety of modes in the initial and final momenta.  We do not display plots for configurations for which no scattering occurs, i.e.\ for when the initial momenta are equal the final momenta $p_i = p_f$.  For a fixed momentum transfer in the finite directions, $t$ approaches a finite, non-zero value as $p_\text{inf}\to\infty$ limit.  We are thus unsurprised to see in \cref{f:2DVtcomparisonp} that the finite system size correction remains non-zero even in the limit of $p_\text{inf}\to\infty$.  However the real parts of $\overline V_2(s,L,\mu)$ and $\overline V_2(u,L,\mu)$ diverge in this limit, meaning the relative influence of the finite system size correction to the amplitude still vanishes as $p_\text{inf}\to\infty$.  Similar to the 1D case, for $n_{1i}^2+n_{2i}^2<n_{f1}^2+n_{f2}^2$ we have that the plots of $\overline V_2$ cut off for small $p_\text{inf}$ or $L$ where an imaginary outgoing momentum in the infinite direction would be necessary to maintain conservation of energy. 

We see that in contrast to the scan in momentum, we have convergence of the finite system size result in the $L\to\infty$ limit, as we would expect. As in 1D, when $n_{1i}(n_{1i}-n_{1f})+n_{2i}(n_{2i}-n_{2f})>0$, the $L\to0$ limit implies $t\to-\infty$.  Exactly as in the 1D case, $t\rightarrow-\infinity$ will lead to a divergence from the infinite volume contribution and a finite system size correction that goes to 0.

%%%%%%%%%%%%%%%%%%%%%%%%%%%%%%%%%%%%%%%%%%%%%%%%%%%%%%%%%%%%%%%%%%%%%%%%%%%%%%%%%%%%%%%%%%%%%
\subsection{Three Compact Dimensions}
%%%%%%%%%%%%%%%%%%%%%%%%%%%%%%%%%%%%%%%%%%%%%%%%%%%%%%%%%%%%%%%%%%%%%%%%%%%%%%%%%%%%%%%%%%%%%
The case of three compact dimensions presents a number of conceptual and numerical challenges.  Conceptually, unless the finite lengths are very large, the incoming and outgoing wavepackets are never widely separated.  Since the incoming and outgoing states can no longer be asymptotically separated, presumably one would have to modify the usual LSZ treatment in order to connect to a physically realizable setup.  Even though the connection to a real physical setup is unclear, one may still---in principle---formally evaluate \cref{e:renormfiniteV}.  

However, as is shown in detail in \cref{s:threecompact}, the $s$ channel contribution to $\overline V$ is actually trivial.  For momentum configurations that are physical, i.e.\ the momenta have an integer number of modes along any finite direction, one has that $\overline V(s_{physical}) = \infinity$, since the particles are ``geometrically bound''; the particles interact an infinite number of times inside the fully compact space.  Otherwise, for unphysical momenta, $\overline V(s_{unphysical}) = 0$.  

We would like to perform another non-trivial check of our analytics and numerics through a comparison with an analytic expression for an asymptotic expansion of \cref{e:renormfiniteV}.  We perform this check on the $t$ channel contribution to $\overline V$ in three compact dimensions in the next section.  

%%%%%%%%%%%%%%%%%%%%%%%%%%%%%%%%%%%%%%%%%%%%%%%%%%%%%%%%%%%%%%%%%%%%%%%%%%%%%%%%%%%%%%%%%%%%%
\section{Asymptotic Analysis}
\label{sec:asymptotic}
\begin{figure}
    \centering
    \includegraphics[width=\columnwidth]{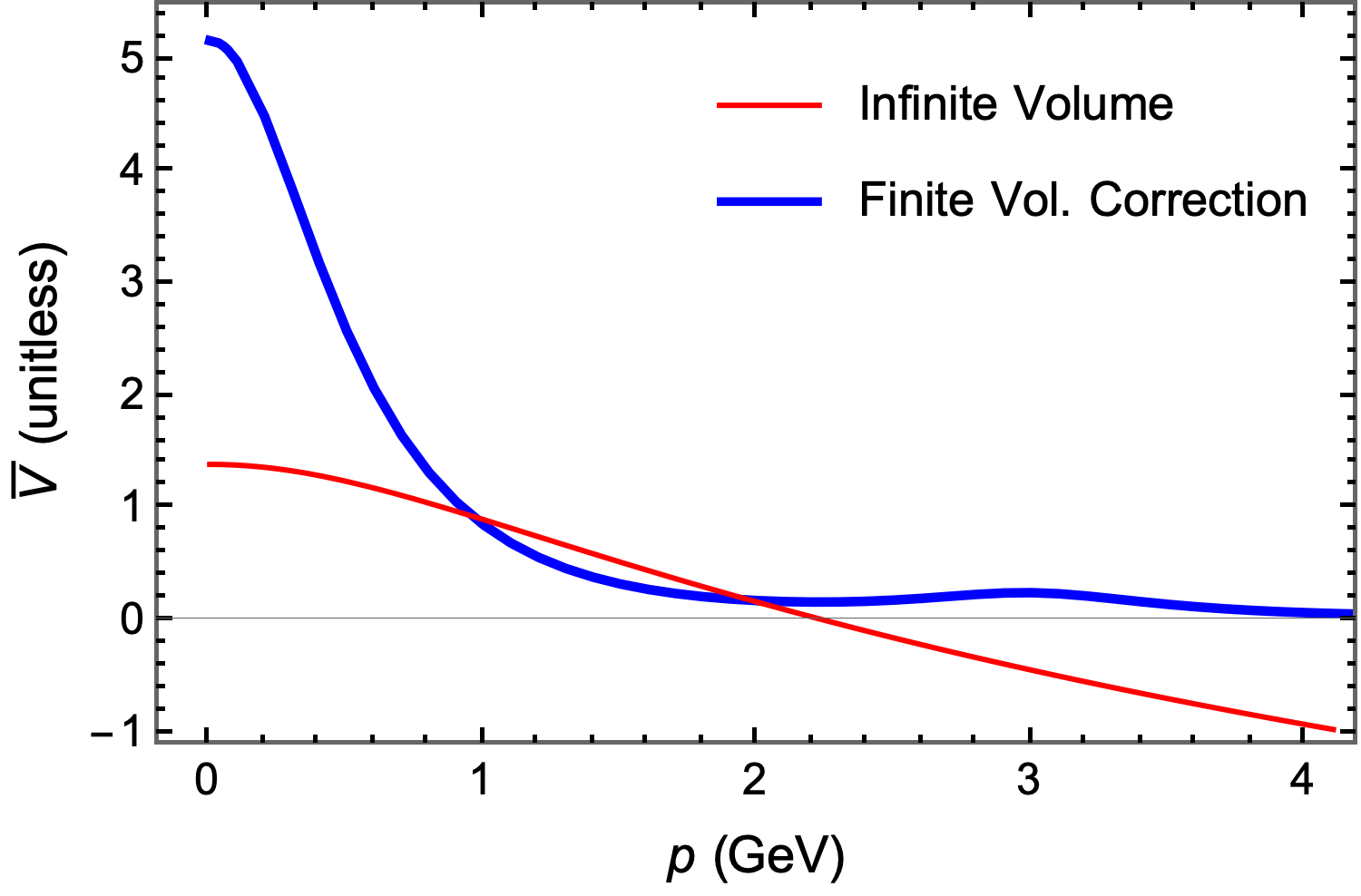} \\[0.1in]
    \includegraphics[width=\columnwidth]{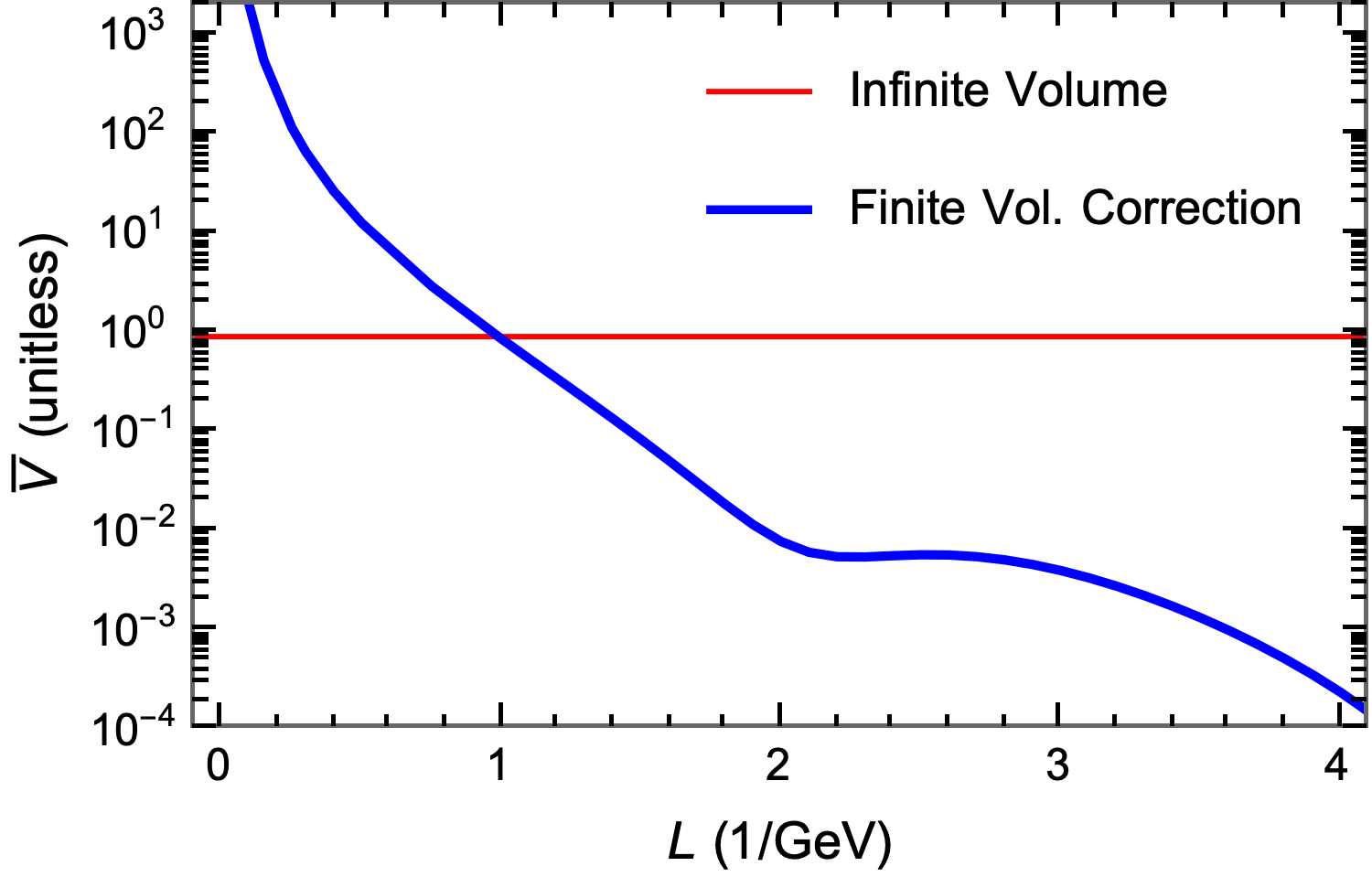}
    \caption{
    (Top) A comparison of the infinite volume contribution to $\overline V(p^2,\{L_i\};\mu,\epsilon)$ to the finite volume corrections (without the overall prefactor of $-1/32\pi^2$) as a function of $p\equiv|\vec p|$ for $L_i = 1/\surd 3$ (GeV$^\mathrm{-1}$), $m=0.5$ GeV, and $\mu = 1$ GeV for the $t$ channel contribution with the assumption of $p^0=0$ GeV.  (Bottom) Same comparison but as a function of $L_i = L/\surd 3$ for $|\vec p| = 1$ GeV, $p^0=0$ GeV, $m=0.5$ GeV, and $\mu = 1$ GeV.  The $y$-axes of both plots are dimensionless.  
    }
    \label{f:comparison}
\end{figure}

%%%%%%%%%%%%%%%%%%%%%%%%%%%%%%%%%%%%%%%%%%%%%%%%%%%%%%%%%%%%%%%%%%%%%%%%%%%%%%%%%%%%%%%%%%%%%
We show plots comparing the infinite volume contribution to the finite system size correction to $\overline V(t,\{L_i\};\mu,\epsilon)$ for $\vec{\,p}^2>0$ and $p^0=0$ in \cref{f:comparison} as a function of the momentum $|\!\vec{\,p}|$ and as a function of the finite system size $L$ for $m=0.5$ GeV and $\mu=1$ GeV.  We note that computing $\overline V(t)$ is something of a hypothetical exercise as the three compact dimension case is pathological.  Nevertheless, it's interesting to consider the $t$ channel contribution in three compact dimensions as a smooth function of $|\!\vec{\,p}|$ or $L$ because the phases in \cref{e:renormfiniteV} cause non-trivial oscillations as a function of these parameters and because the finite system size corrections are maximal in a fully compactified space.  One can see that for $p\sim 1/L$ the finite system size corrections are on the same order as the infinite volume contribution.  Note that the oscillations seen in the finite system size corrections as a function of both $|\!\vec{\,p}|$ and $L$ are real and due to the oscillating phases multiplying the modified Bessel function, as we shall show.

%_______________________________________________________________
%_______________________________________________________________

%%%%%%%%%%%%%%%%%%%%%%%%%%%%%%%%%%%%%%%%%%%%%%%%%%%%%%%%%%%%%%%%%%%%%%%%%%%%%%%%%%%%%%%%%%%%%
\subsection{Large Argument Analysis}
%%%%%%%%%%%%%%%%%%%%%%%%%%%%%%%%%%%%%%%%%%%%%%%%%%%%%%%%%%%%%%%%%%%%%%%%%%%%%%%%%%%%%%%%%%%%%
%_______________________________________________________________
\begin{figure}
    \centering
    \includegraphics[width=\columnwidth]{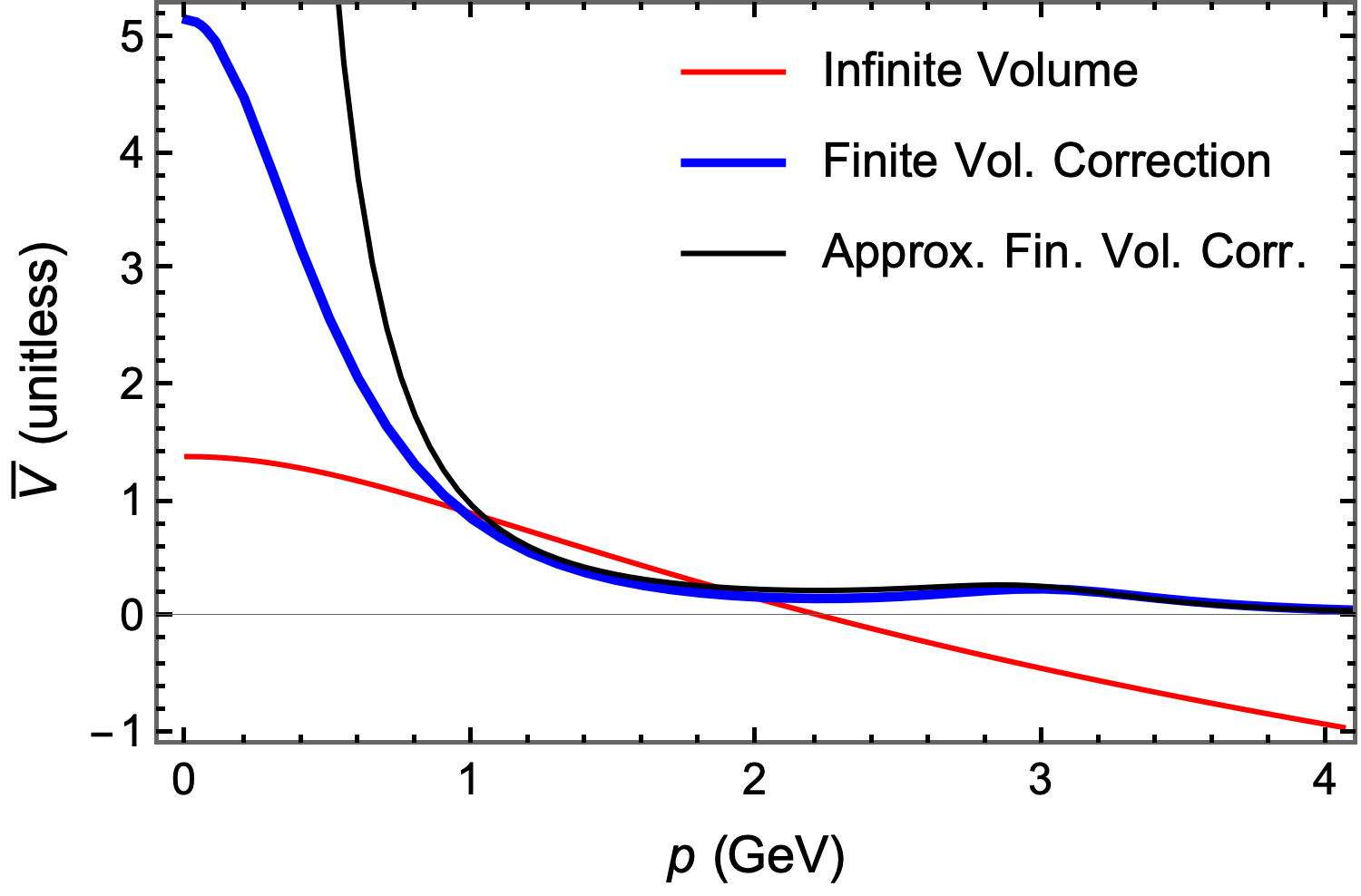} \\[0.1in]
    \includegraphics[width=\columnwidth]{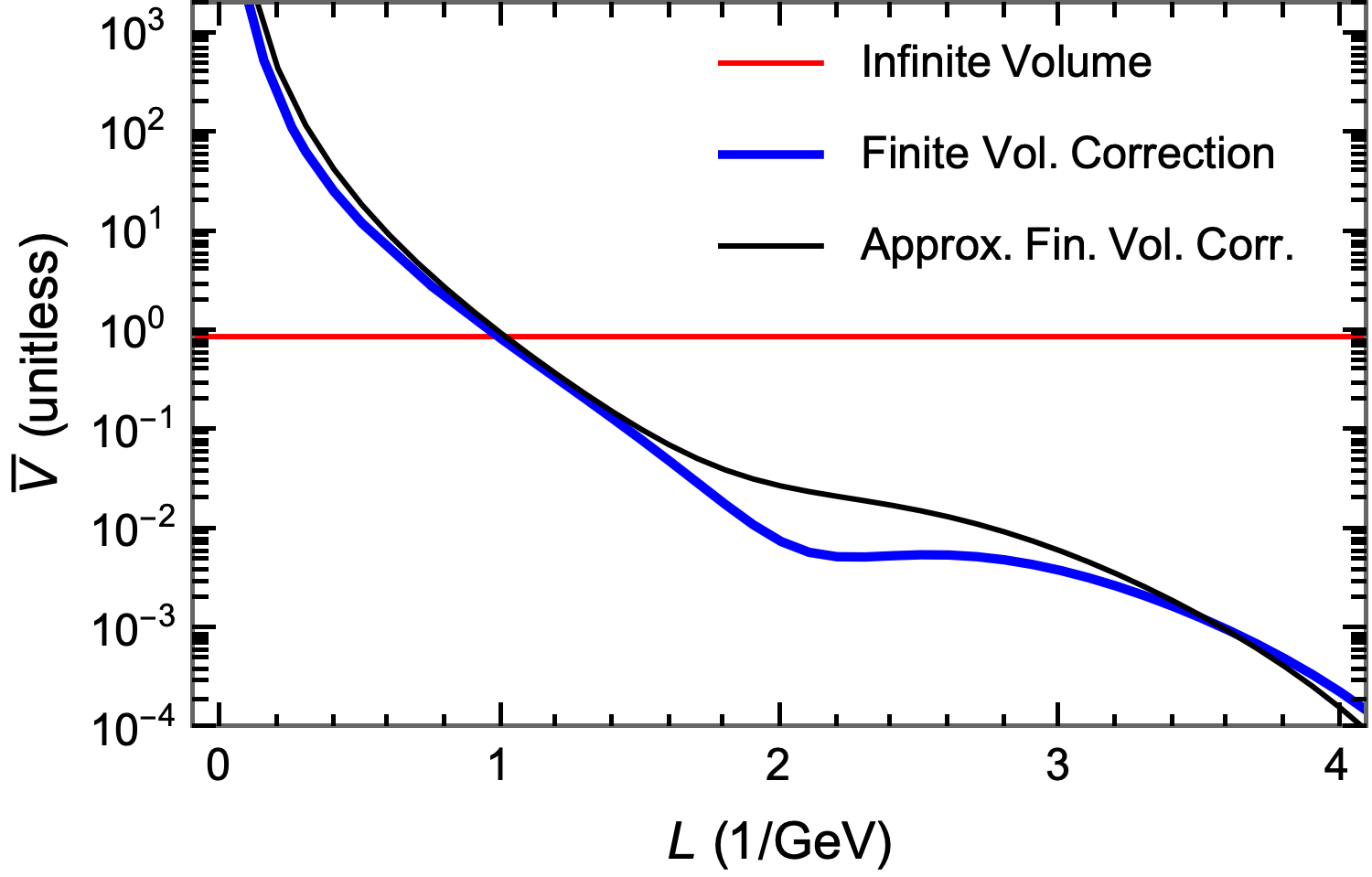}
    \caption{
    (Top) A comparison of the infinite volume contribution to $\overline V(p^2,\{L_i\};\mu,\epsilon)$ to the finite volume corrections (without the overall prefactor of $-1/32\pi^2$) and its asymptotic approximation for large argument as a function of $p\equiv|\vec p|$ for $L_i = 1/\surd 3$ (GeV$^\mathrm{-1}$), $m=0.5$ GeV, and $\mu = 1$ GeV for the $t$ channel contribution with the assumption of $p^0=0$ GeV.  (Bottom) Same comparison but as a function of $L_i = L/\surd 3$ for $|\vec p| = 1$ GeV, $p^0=0$ GeV, $m=0.5$ GeV, and $\mu = 1$ GeV.  The $y$-axes of both plots are dimensionless.  
    }
    \label{f:asympcomparison}
\end{figure}
%_______________________________________________________________

We seek an asymptotic expansion for terms that look like products of cosines and a modified Bessel function of the second kind.  Taking for simplicity only one cosine for the moment, generically we seek an expansion in inverse powers of $\lambda_2\sim\sqrt{\vec p^2L^2}$ of
\begin{multline}
    I(\lambda_1,\lambda_2,\alpha)\equiv \int_0^1 dx\ess\cos(2\pi\ess x\ess \lambda_1) \\ \times K_0\big( 2\pi \sqrt{x(1-x)+\alpha^2}\lambda_2 \big),
\end{multline}
where $\alpha^2\equiv m^2/\vec p^2$ and $\lambda_1\sim p^i L_i$.

The approach we'll take is to employ Laplace's method \cite{Bender:1978}.  To get the integral in the correct form, in which an integrand is multiplied by a decaying exponential, we change integration variables and use the integral representation of the Bessel function.  First, take $y'\equiv\sqrt{x(1-x)+\alpha^2}$.  The inverse of $y'$ is double-valued, so we must separately consider the original integral from $x=0$ to $1/2$ and from $x=1/2$ to $1$:
\begin{align}
    \label{e:cov}
    I_\pm&(\alpha,\lambda_1,\lambda_2)
    \equiv \int_\alpha^{\frac12\sqrt{1+4\alpha^2}}dy' \frac{2y'}{\sqrt{1+4\alpha^2-4y'^2}}\nonumber\\
    & \times\cos\big( 2\pi\frac12(1\pm\sqrt{1+4\alpha^2-4y'^2}) \lambda_1\big)K_0(2\pi\ess\lambda_2y),\\
    I&(\lambda_1,\lambda_2,\alpha) = I_-(\lambda_1,\lambda_2,\alpha) + I_+(\lambda_1,\lambda_2,\alpha);
\end{align}
note the crucial relative sign difference in the arguments of the cosines.  To make the application of Laplace's method easier, we shift the integration variable to $y\equiv y'+\alpha$; at the same time, introduce one of the standard integral representations of the Bessel function \cite{Olver:2010} to yield
\begin{align}
    I_\pm&(\alpha,\lambda_1,\lambda_2) = \int_0^\infinity dt\ess e^{-2\pi\ess\alpha\ess\lambda_2\cosh(t)}\int_0^{\frac12\sqrt{1+4\alpha^2}-\alpha}dy\nonumber\\
    & \times f_\pm(\alpha,\lambda_1;y)e^{-2\pi\ess\alpha\ess\lambda_2\cosh(t)y}, \\[5pt]
    f_\pm&(\alpha,\lambda_1;y) \equiv \nonumber\\
    & \frac{2(y+\alpha)}{\sqrt{1-8\alpha\ess y-4y^2}}\cos\big(\pi(1\pm\sqrt{1-8\alpha\ess y-4y^2})\lambda_1\big).\nonumber
\end{align}
Laplace's method tells us that
\begin{align}
    \label{e:laplace}
    I_\pm&(\alpha,\lambda_1,\lambda_2) \sim \int_0^\infinity dt\ess e^{-2\pi\ess\alpha\ess\lambda_2\cosh(t)}\int_0^{\infinty}dy\nonumber\\
    & \times \big[f_\pm(\alpha,\lambda_1;0) + f'_\pm(\alpha,\lambda_1;0)y + \ldots \big]e^{-2\pi\ess\alpha\ess\lambda_2\cosh(t)y},
\end{align}
where $\sim$ indicates that exponentially suppressed terms $\sim\exp(-\lambda_2)$ have been dropped; note especially the change in the upper bound of the $y$ integration limit.  One has in general that
\begin{align}
    \int_0^\infinity dy \ess y^{n-1} e^{-2\pi\lambda_2\cosh(t)y} = \frac{\Gamma(n)}{(2\pi\ess\lambda_2)^{n}}\sech^{n}(t).
\end{align}
One may then perform the integral over $t$, yielding the Bickley function \cite{Olver:2010}
\begin{align}
    \operatorname{Ki}_{n}(x) \equiv \int_0^\infinity dt\ess e^{-x\ess\cosh(t)}\sech^{n}(t),\quad x>0.
\end{align}
%The Bickley functions often occur in p
Clearly $\operatorname{Ki}_0(x) = K_0(x)$.  One may find an explicit expression for $\operatorname{Ki}_1(x)$ by integrating under the integral sign to find
\begin{align}
    \operatorname{Ki}_1(x) & = \int_x^\infinity dx'K_0(x') \nonumber\\
    & = \frac\pi2 - \frac{\pi\ess x}{2}\big( K_0(x)L_{-1}(x) + K_1(x)L_0(x) \big).
\end{align}
Further explicit representations of the Bickley function in terms of modified Bessel functions of the second kind and the modified Struve functions can be found through the recursion relation \cite{Olver:2010}
\begin{multline}
    \alpha \ess \operatorname{Ki}_{\alpha+1}(x) + x\ess\operatorname{Ki}_\alpha(x) \\ +(1-\alpha)\operatorname{Ki}_{\alpha-1}(x) -x\ess\operatorname{Ki}_{\alpha-2}(x) = 0.
\end{multline}
Like the modified Bessel functions of the second kind, the Bickley functions decrease monotonically with $x$, and further decrease monotonically with order.  One finds for large $x$ that
\begin{align}
    \operatorname{Ki}_\alpha(x) \sim K_\alpha(x)\sim \sqrt{\frac{\pi}{2x}}\exp(-x).
\end{align}

One readily computes that
\begin{align}
    f_-(\alpha,\lambda_1;0) & = 2\alpha \\
    f_+(\alpha,\lambda_1;0) & = 2\alpha\cos(2\pi\ess\lambda_1).
\end{align}
Thus we see how the phase factors in \cref{e:renormfiniteV} lead to the non-trivial oscillations in \cref{f:comparison}.  

For increased numerical accuracy, we may go to the next order.  One finds that
\begin{align}
    f'_-(\alpha,\lambda_1;0) & = (1+4\alpha^2) \\
    f'_+(\alpha,\lambda_1;0) & = (1+4\alpha^2)\cos(2\pi\ess\lambda_1) + 4\pi\alpha^2\lambda_1\sin(2\pi\lambda_1).
\end{align}
If we drop the higher order in $\alpha$ contributions, which is well-justified in the large $p^2$ limit, a significant simplification occurs; we find that
\begin{align}
    \label{e:asymp}
    I(\lambda_1,\lambda_2,\alpha) \sim  \frac{\alpha}{\pi\ess\lambda_2}\big(1+\cos(2\pi\ess\lambda_1)\big)K_1(2\pi\ess\alpha\ess\lambda_2).
\end{align}

If we apply \cref{e:asymp} to \cref{e:renormfiniteV} (see also \cref{e:moreefficient}) we find \emph{after integrating over} $x$ that
\begin{align}
    \label{e:fullasymp}
    \overline V&(p^2,\{L_i\};\mu) \sim \overline V_\infinity(p^2;\mu)-\frac12\frac{1}{(4\pi)^2}\bigg\{ \nonumber\\
    &\sum_{s\in2^{[3]}}2^{|s|+1}\sum_{\substack{m_i=1 \\ i\in s}}^\infinity\bigg( 1 + \prod_{i\in s}\Big( \cos(2\pi\ess \ess m_i L_i p^i) \Big) \bigg) \nonumber\\
    &\times\frac{m}{\pi\ess p^2\sqrt{\sum (m_i^2 L_i^2)}} K_1\bigg(2\pi \ess m\sqrt{\sum (m_i^2 L_i^2)} \bigg)\bigg\},
\end{align}
where
\begin{align}
    \overline V_\infinity(p^2;\mu)
    & \equiv -\frac12\frac{1}{(4\pi)^2} \int_0^1 dx\,\ln\frac{\mu^2}{\Delta^2} \nonumber\\
    & = -\frac12\frac{1}{(4\pi)^2}\bigg[ 2 + \ln\frac{\mu^2}{m^2} \nonumber\\ 
    & \qquad-2\sqrt{1+\frac{4m^2}{p^2}}\tanh^{-1}\Big( \sqrt{\frac{p^2}{p^2+4m^2}} \Big) \bigg]
\end{align}
is the infinite volume contribution to NLO $2\rightarrow2$ scattering.

We compare the asymptotic expansion of \cref{e:fullasymp} to the full result \cref{e:renormfiniteV} in \cref{f:asympcomparison}.  The comparison is much better for the scan in $p$ as $\alpha\propto1/\surd p^2$.  Including more terms in the asymptotic expansion improves the convergence.

%%%%%%%%%%%%%%%%%%%%%%%%%%%%%%%%%%%%%%%%%%%%%%%%%%%%%%%%%%%%%%%%%%%%%%%%%%%%%%%%%%%%%%%%%%%%%
\subsection{Small Argument Analysis}
%%%%%%%%%%%%%%%%%%%%%%%%%%%%%%%%%%%%%%%%%%%%%%%%%%%%%%%%%%%%%%%%%%%%%%%%%%%%%%%%%%%%%%%%%%%%%
We wish to understand the behavior of the finite system size correction as $\vec p\rightarrow\vec 0$ and/or $L_i\rightarrow0$.

For $p^2\rightarrow0$ and $m>0$, we may safely ignore the phases and trivially integrate over $x$.  Then
\begin{multline}
    \overline V(p^2,\{L_i\};\mu) \approx \overline V_\infinity(p^2;\mu) \\ -\frac12\frac{1}{(4\pi)^2} %\bigg\{ \\
    2\sideset{}{'}\sum_{\vec m\in\mathbb Z^3}K_0\Big( 2\pi\ess m\sqrt{\sum m_i^2L_i^2} \Big). 
\end{multline}
One can see in \cref{f:comparison} how in the small $p^2$ limit for $m>0$ that the finite system size correction converges to a finite value.

%_______________________________________________________________
\begin{figure}[!htbp]
    \centering
    \includegraphics[width=\columnwidth]{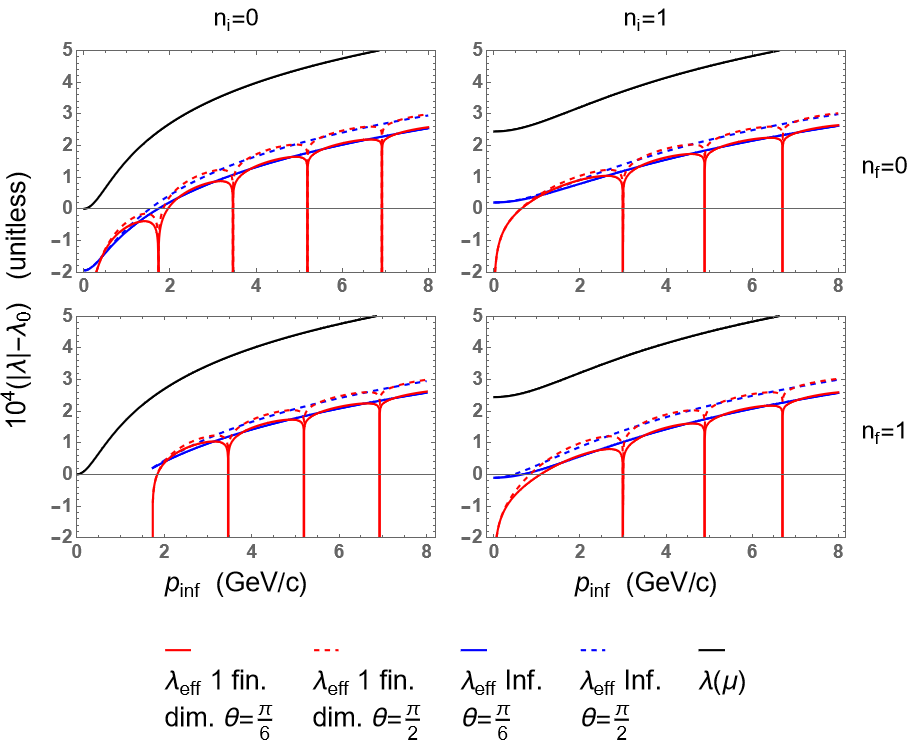} \\[0.1in]
    \includegraphics[width=\columnwidth]{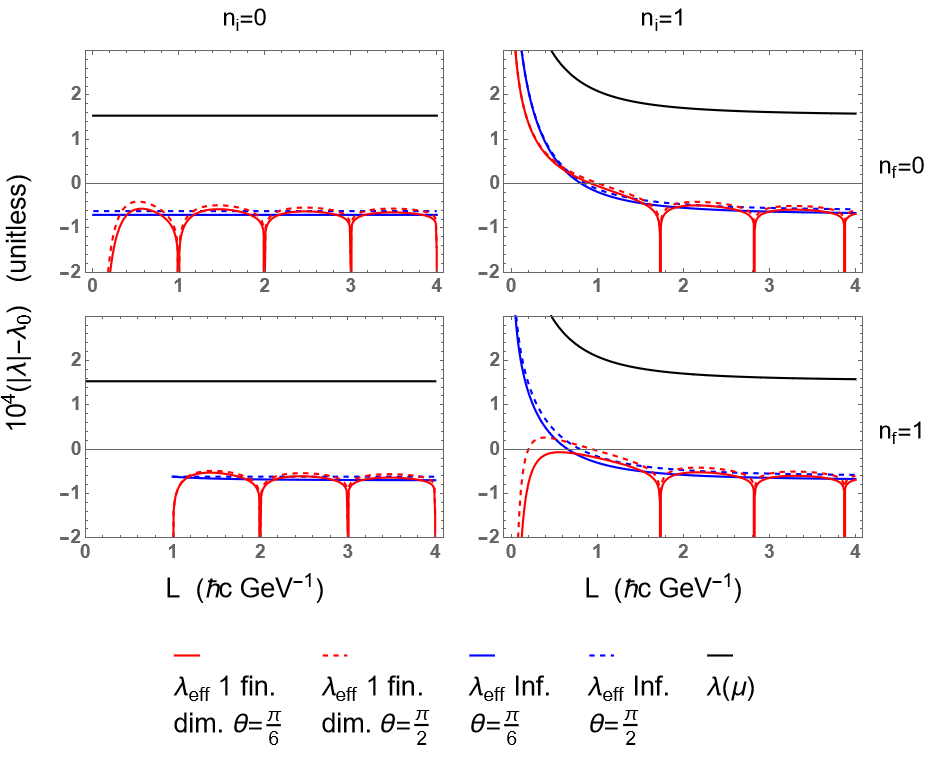}
    \caption{
    (Top) A comparison of the running coupling $\lambda(\mu)$ with $\mu^2=s$ and $\mu_0=2 m$ (black) and the absolute value of the infinite volume effective coupling $\lambda_{\mathrm{eff}}(s,t)$ (blue) with $m=0.5$ GeV and scattering angle $\theta=\pi/6,\pi/2$ (solid,dashed) and the absolute value of the finite volume effective coupling $\lambda_{\mathrm{eff}}(s,t,\{L_i\})$ (red) with $L=1/\surd 3$ GeV$^{-1}$ for $n=1$ compact dimension as a function of $p_\text{inf}$ for renormalized coupling with $\lambda_0=0.1$ at $\mu=\mu_0$.  (Bottom) Same comparison but as a function of $L$ (GeV$^{-1}$) for $p_\text{inf}=1$ GeV.  For both plots, the $y$-axes are dimensionless and the dips actually take the coupling all the way to 0 (note that the y-axis has been shifted and rescaled, to compare the corrections to the effect of the running of the coupling).
    }
    \label{f:1Defflam}
\end{figure}
%_______________________________________________________________

%_______________________________________________________________
\begin{figure}[!t]
    \centering
    \includegraphics[width=\columnwidth]{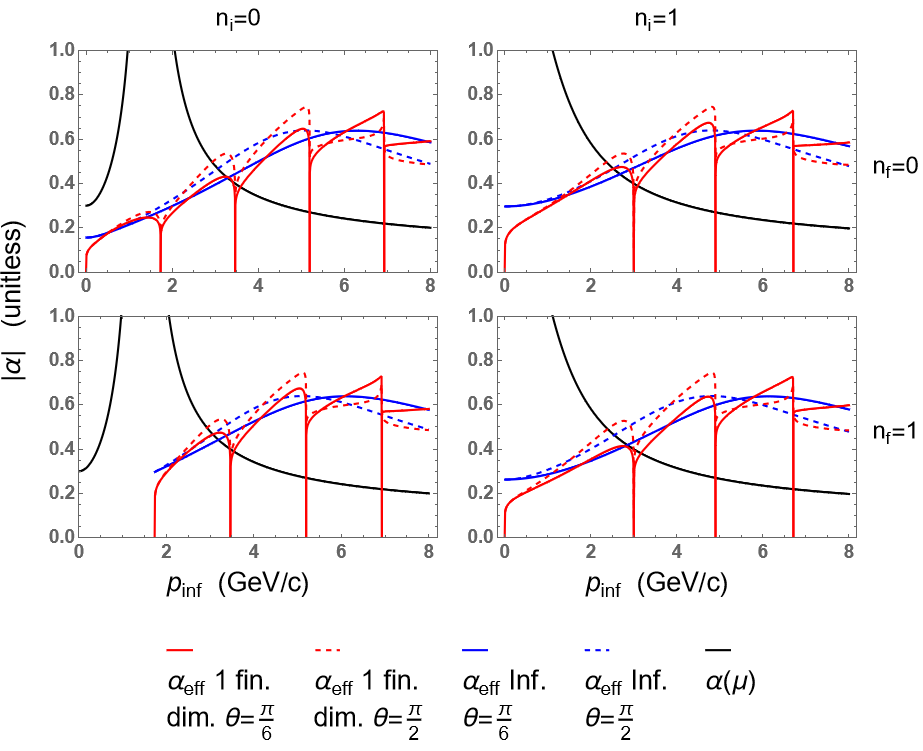} \\[0.1in]
    \includegraphics[width=\columnwidth]{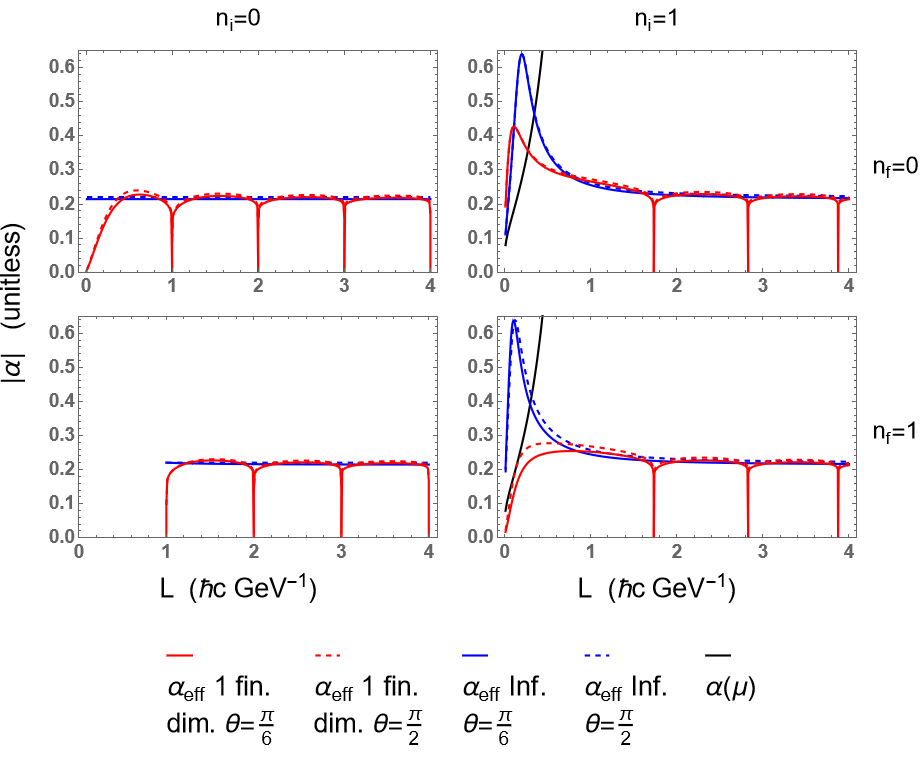}
    \caption{
    (Top) A comparison of the running coupling $\alpha(\mu)$ with $\mu^2=s$ and $\mu_0=2m$ (black) and the absolute value of the infinite volume effective coupling $\alpha_{\mathrm{eff}}(s,t)$ (blue) with $m=0.5$ GeV and scattering angle $\theta=\pi/6,\pi/2$ (solid, dashed) and the absolute value of the finite volume effective coupling $\alpha_{\mathrm{eff}}(s,t,\{L_i\})$ (red) with $L=1/\surd 3$ GeV$^{-1}$ for $n=1$ compact dimension  as a function of $p_\text{inf}$ for renormalized coupling with $\alpha_0=0.3$ at $\mu=\mu_0$.  (Bottom) Same comparison but as a function of $L$ (GeV$^{-1}$) for $p_\text{inf}=1$ GeV.  For both plots, the $y$-axes are dimensionless and the dips again take the coupling exactly to 0.
    }
    \label{f:1Deffalf}
\end{figure}
%_______________________________________________________________

\begin{figure*}[!htb]
    \centering
    \includegraphics[width=0.85\textwidth]{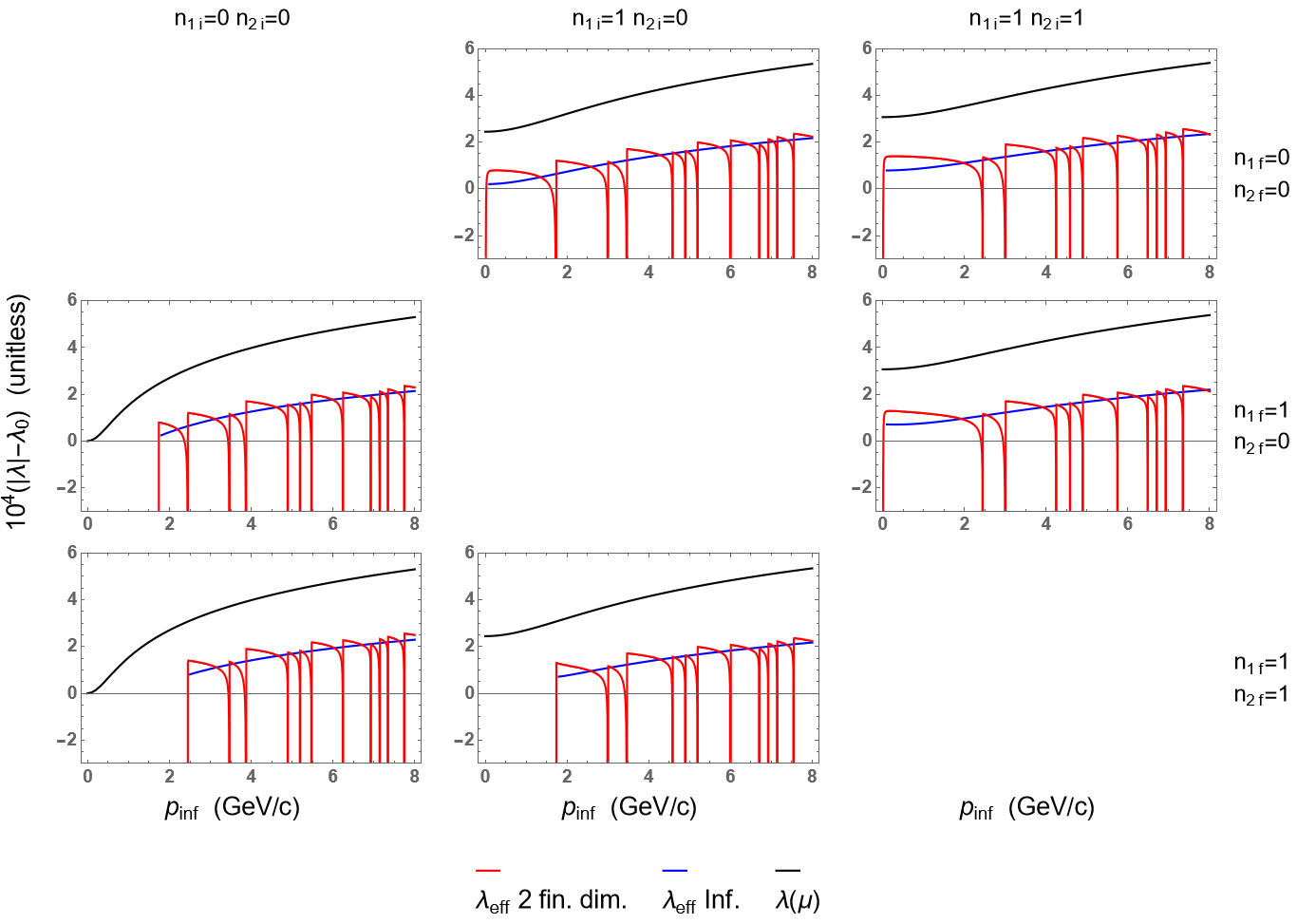}
    \caption{
    A comparison of the running coupling $\lambda(\mu)$ with $\mu^2=s$ and $\mu_0=2m$ (black) and the absolute value of the infinite volume effective coupling $\lambda_{\mathrm{eff}}(s,t)$ (blue) with $m=0.5$ GeV  and the absolute value of the finite volume effective coupling $\lambda_{\mathrm{eff}}(s,t,\{L_i\})$ (red) with $L_1=L_2=L=1/\surd 3$ GeV$^{-1}$ for $n=2$ compact dimensions as a function of $p_\text{inf}$ for renormalized coupling $\lambda_0=0.1$ at $\mu=\mu_0$.  The $y$-axes are dimensionless and the coupling is exactly 0 at the dips.
    }
    \label{f:2Defflamp}
\end{figure*}

If $m=0$ and we take $p^2\rightarrow 0$, or if $L_i\rightarrow 0$ (and $m\ge0$), then we may still ignore the phases but must consider the integration over $x$ in the finite system size correction.  For small argument $K_0(z) = \ln (2/z) - \gamma_E + \mathcal O(z)$ \cite{Olver:2010}.  If we perform our change of variables as in \cref{e:cov} we may integrate over $x$ to find
\begin{multline}
    \int_0^1dx\,K_0(2\pi\sqrt{x(1-x)+\alpha^2}\lambda_2) \\ = \frac{1}{2}\big(1-\gamma_E - \ln(\pi\ess\lambda_2)\big) + g(\alpha) + \mathcal O(\lambda_2),
\end{multline}
where $g(\alpha) = \mathcal O(\alpha^2)$ is a complicated expression that smoothly goes to 0 as $m\rightarrow0$ and may be safely ignored for the range of $m$ and $p^2$ we consider here.  We thus see that in the small $L$ limit or in the small $p^2$ limit (for $m=0$) the finite system size correction diverges logarithmically.  The divergence is slightly faster than logarithmic in the sense that as the argument of the Bessel function decreases, more and more terms from the sum over $m_i$ contribute; one must sum over $m_i\lesssim1/\pi\sqrt{p^2\sum L_i^2}$.  One can see in \cref{f:comparison} how in the small $L_i$ limit the finite system size correction diverges slightly faster than logarithmically.

We leave a more detailed analytic exploration of the small argument asymptotics for future work.

\section{Finite Size Effective Coupling}

\begin{figure*}[!ht]
    \centering
    \includegraphics[width=0.85\textwidth]{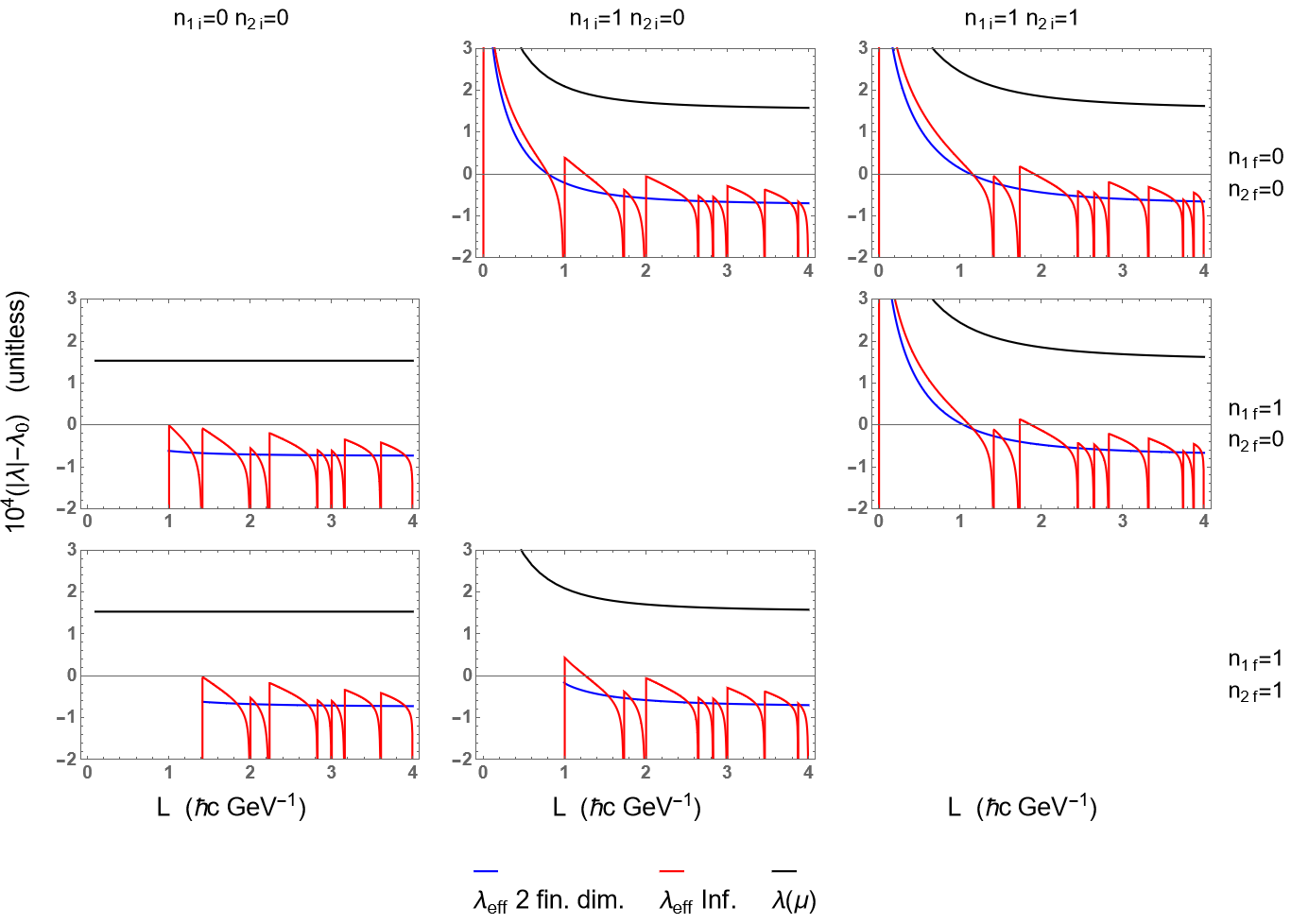}
    \caption{
    Same comparison as in \cref{f:2Defflamp} but as a function of $L$ (GeV$^{-1}$) for $p_\text{inf}=1$ GeV.  The $y$-axes are dimensionless and the coupling is exactly 0 at the dips.
    }
    \label{f:2DefflamL}
\end{figure*}

Having explored the finite system size correction to $\overline V(p^2, \{L_i\};\mu)$, we wish to understand quantitatively the finite system size effects on the effective coupling.  The finite system size effective coupling is the obvious generalization of \cref{e:effectivecoupling}, replacing the infinite volume $\overline V(p^2;\mu)$ with the finite volume $\overline V(p^2, \{L_i\};\mu)$:
\begin{multline}
    -i\lambda_{\mathrm{eff}}(s,t,\{L_i\}) = 
    \\ \frac{-i\lambda(\mu)}{1 - \lambda(\mu)\big( \overline V(s,\{L_i\}) + \overline V(t,\{L_i\}) + \overline V(u,\{L_i\}) \big)}.
\end{multline}

Since $\overline V$ picks up an imaginary part for the $s$ channel for $|\vec p|>0$, both in the infinite volume and for the finite volume correction (as required by unitarity), the effective coupling becomes complex.  A complex effective coupling is a bit unusual but not a fundamental problem \cite{Horowitz:2010yg}: physical quantities always involve taking the modulus squared of the (potentially complex) amplitude.

We show in \cref{f:1Defflam} the absolute value of the running coupling $\lambda(\mu)$, the infinite volume effective coupling $\lambda_{\mathrm{eff}}$, and the finite volume effective coupling $\lambda_{\mathrm{eff}}$ minus the initial value of the coupling $\lambda_0$ and scaled by a factor of $10^4$ as functions of $p_\text{inf}$ and of the size of the finite dimension, $L$, for $n=1$ compact dimensions.  We choose to subtract the initial value and scale the results in order to make the differences among the results more clearly visible.  We use the same notation as in \cref{sec:vbarplots}; in particular, our momenta are given by \cref{e:onedvariables}.  We take the finite system size direction to have a length $1/\surd 3$ GeV$^{-1}$.  For the running coupling, we use $\mu^2=s$ and $\mu_0=2m$, where $m=0.5$ GeV, and the value of the renormalized coupling at the initial scale is $\lambda_0=0.1$.  We consider scattering at angles $\theta=\pi/2,\pi/6$ among the infinite directions.  One sees that the running coupling captures the leading logarithmic behavior of the effective coupling.  One can further see that the finite system size effective coupling converges to the infinite system size effective coupling as either $p_\text{inf}$ and/or $L$ become large (compared to the other).  We can see that other than the geometric bound states (where the effective coupling goes to 0) there are minimal corrections with one finite dimension. One may consider it odd that in the case of the geometric bound states the coupling goes to zero given that the cross section diverges.  We understand this discrepancy as due to the resummation of the bubble diagrams not capturing the bound state physics; at the geometric bound states, one should rather resum a set of ladder diagrams.  Note that the existence and location of these divergences in $\overline V(p^2,\{L_i\};\mu)$, which lead to the finite system size effective coupling going to 0, are \emph{independent} of the strength of the coupling.

Similar to $\alpha_{EM}$ (in QED) or $\alpha_s$ (in QCD), we may further define the relevant expansion parameter for $\phi^4$ theory, $\alpha\equiv\lambda/(4\pi)^2$ \cite{Kleinert:2001ax}.  We show the comparison between the running coupling $\alpha(\mu)$, the infinite volume effective coupling $\alpha_{\mathrm{eff}}(s,t)$, and the finite volume effective coupling $\alpha_{\mathrm{eff}}(s,t,\{L_i\})$ in \cref{f:1Deffalf}.  The much larger initial $\alpha=0.3$ leads to a much larger change in the coupling as a function of the various scales.  In particular, one can see that the rapid approach of the Landau pole at modest $p\sim2$ GeV means that the running coupling no longer captures the leading logarithmic dependence of the effective coupling.  Interestingly, the effective coupling completely avoids the Landau pole.
One can see that the finite system size effective coupling still tracks the infinite volume effective coupling; for large enough $p$ or $L$ the results converge.  

One can see that, e.g.\ in \cref{f:1Deffalf}, the effective coupling decreases as $p_\text{inf}\times L$ becomes small.  The coupling decreasing in this limit is not due to a geometric bound state.  Rather, the coupling going to zero here is a reflection of the magnitude of the NLO contribution to the amplitude increasing due to the small system size compared to the momentum, $p_\text{inf}\times L\ll 1$.  This decrease of the coupling with $p_\text{inf}\times L$ is consistent with a calculation of the effective coupling in a lattice discretization of $\phi^4$ theory in finite volume \cite{Montvay:1994cy}.  Since $1/L\rightarrow\infinity$ corresponds to a large momentum scale and the beta function for $\phi^4$ theory is positive, it's very interesting that the coupling decreases as $p_\text{inf}\times L$ decreases.

\begin{figure*}[!ht]
    \centering
    \includegraphics[width=0.85\textwidth]{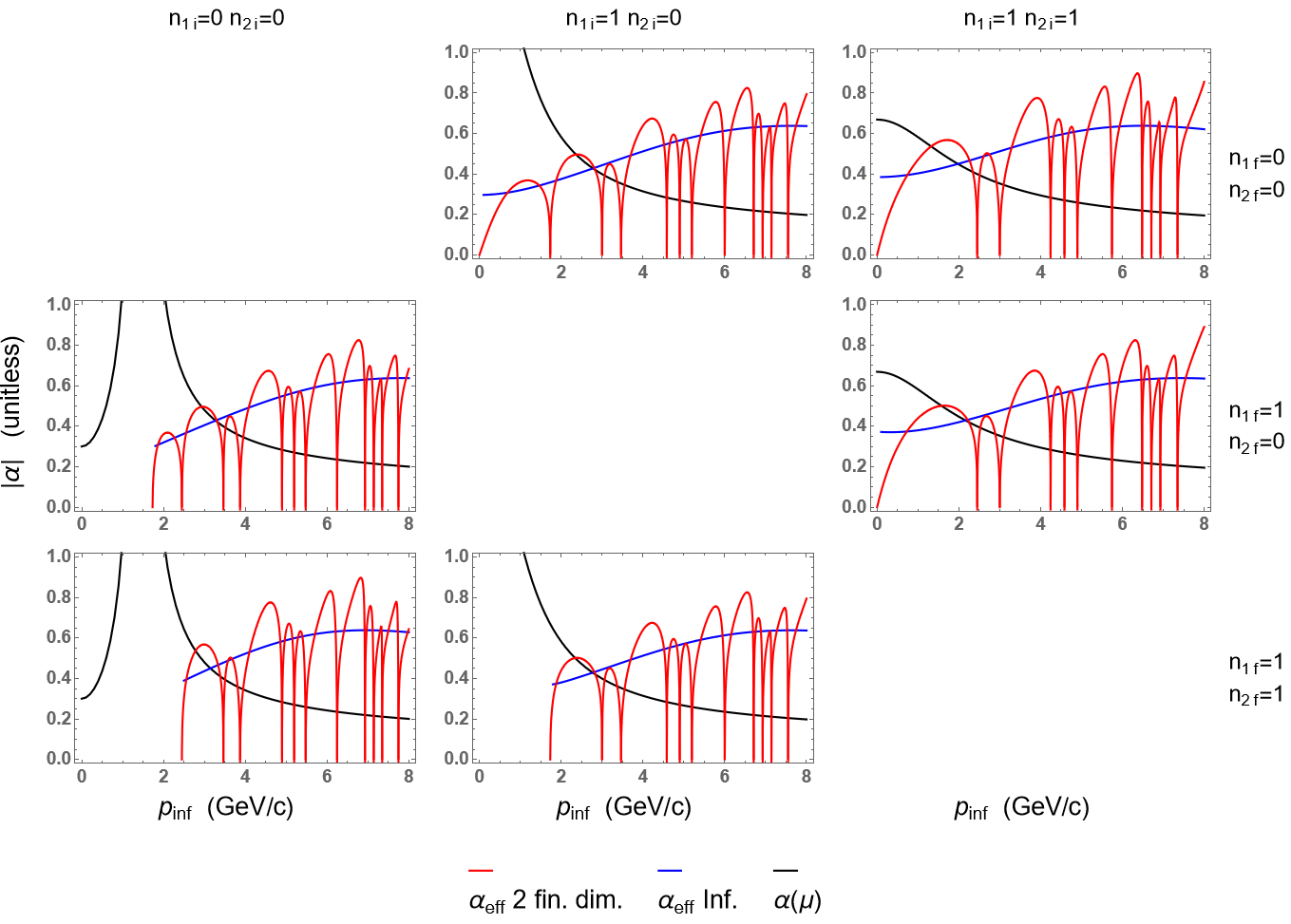}
    \caption{
    A comparison of the running coupling $\alpha(\mu)$ with $\mu^2=s$ and $\mu_0=2m$ (black) and the absolute value of the infinite volume effective coupling $\alpha_{\mathrm{eff}}(s,t)$ (blue) with $m=0.5$ GeV = and the absolute value of the finite volume effective coupling $\lambda_{\mathrm{eff}}(s,t,\{L_i\})$ (red) with $L_1=L_2=L=1/\surd 3$ GeV$^{-1}$ for $n=2$ compact dimensions as a function of $p_\text{inf}$ for renormalized coupling $\alpha_0=0.3$ at $\mu=\mu_0$.  The $y$-axes are dimensionless and the coupling is exactly 0 at the dips.
    }
    \label{f:2Deffalfp}
\end{figure*}

Recall from \cref{sec:1dschannel} that for $p_\text{inf}\times L\rightarrow 0$ the amplitude $\overline V_1(s,L,\mu)$ in fact diverges.  Certainly for $\lambda(\mu)\overline V_1(s,L,\mu)\lesssim 1$ we may trust the geometric sum into the effective coupling.  For $\lambda(\mu)\overline V_1(s,L,\mu)\gtrsim 1$, one may consider the effective coupling as the analytic continuation of the resummation of the geometric series, which agrees with, e.g., the Borel sum of the geometric series.  Taking the analytic continuation seriously, the effective coupling goes to zero in the limit $p_\text{inf}\times L\rightarrow 0$.  It's not clear that one may still physically interpret this effective coupling; it is not immediately obvious that it is sensible for the effective coupling to be zero when the cross section is infinite.  

For $n=2$ compact dimensions, we show in \cref{f:2Defflamp,f:2DefflamL} the absolute value of the running coupling $\lambda(\mu)$, the infinite volume effective coupling $\lambda_{\mathrm{eff}}(s,t)$, and the finite volume effective coupling $\lambda_{\mathrm{eff}}(s,t,\{L_i\})$ minus the initial value of the coupling $\lambda_0$ and scaled by a factor of $10^4$ as functions of $p_\text{inf}$ and of the size of the finite dimension, $L$.  We again choose $\mu^2=s$ and $\mu_0=2m$ for the running coupling with $\lambda_0=0.1$.  The finite sizes of the system are both $L_1 = L_2 = 1/\surd 3$ GeV$^{-1}$.  The running coupling is given by the black curves, the infinite volume effective coupling by the blue curves and the finite volume effective coupling by the red curves.  

\begin{figure*}[!ht]
    \centering
    \includegraphics[width=0.85\textwidth]{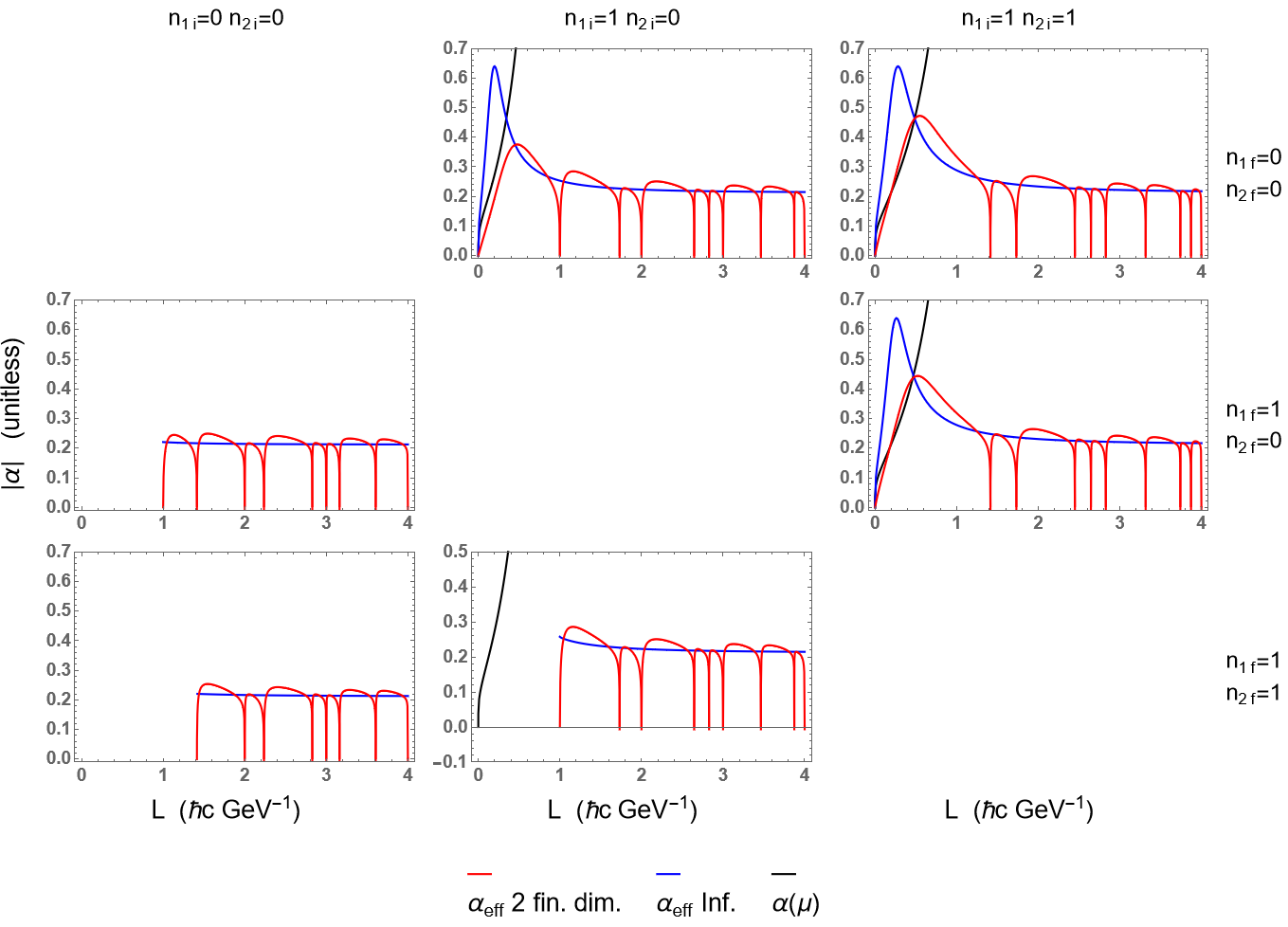}
    \caption{
    Same comparison as in \cref{f:2Deffalfp} but as a function of $L$ (GeV$^{-1}$) for $p_\text{inf}=1$ GeV.  The $y$-axes are dimensionless and the coupling is exactly 0 at the dips.
    }
    \label{f:2DeffalfL}
\end{figure*}

In \cref{f:2Deffalfp,f:2DeffalfL} we show the running coupling $\alpha(\mu)$, the infinite volume effective coupling $\alpha_{\mathrm{eff}}(s,t)$, and the finite volume effective coupling $\alpha_{\mathrm{eff}}(s,t,\{L_i\})$ as functions of $p_\text{inf}$ and of the size of the finite dimensions, $L_1=L_2\equiv L$, for $n=2$ compact dimensions.  We again choose $\mu^2=s$ and $\mu_0=2m$ for the running coupling with $\alpha_0=0.3$.  The running coupling is given by the black curves, the infinite volume effective coupling by the blue curves and the finite volume effective coupling by the red curves.  

One can see that the size of the finite system size corrections increases noticeably compared to the $n=1$ case.  Nevertheless, the finite system size effective coupling asymptotically approaches the infinite volume effective coupling for large $p$ and/or large $L$.  The dips are again to zero coupling, where the approach is smooth from both sides; the coupling is only non-differentiable at the points of exactly zero coupling.  The drop to zero coupling is again due to the divergences of the real and imaginary parts of $\overline V$ in the $s$-channel.

%_______________________________________________________________

\section{Conclusions and Outlook}
In this work, we explored the finite system size corrections to NLO scattering in scalar $\phi^4$ theory.  We first recounted the usual derivation of the NLO $2\to2$ scattering amplitude in massive $\phi^4$ theory in infinite volume using dimensional regularization.  We then determined the running coupling from the Callan-Symanzik equation and the effective coupling from a resummation of bubble diagrams.

Next, we utilized denominator regularization \cite{Horowitz:2022rpp} to regulate the divergences in the NLO correction to $2\to2$ scattering in a spacetime with spatial dimensions compactified into a torus, i.e., with periodic boundary conditions.  The use of denominator regularization allowed us to work on a general, potentially completely asymmetric torus. In order to capture the NLO UV divergence of the theory, we derived an analytic continuation of the generalized Epstein zeta function. Subsequently, we determined the relevant renormalized quantities, including the renormalized NLO scattering amplitude, by applying a natural and minimal modification of the \msbar{} renormalization prescription relevant for denominator regularization.

We observed that the amplitude reduced to the infinite volume case when all lengths were taken to infinity. To further validate our derivation, we performed a non-trivial check by confirming that the NLO scattering amplitude respects unitarity in the form of the optical theorem for $0\le n\le3$ finite spatial dimensions. This check was performed in two ways, one of which involved generalizing a number theoretic result from Ramanujan and Hardy \cite{Hardy1978RamanujanTL}.

Satisfied by these self-consistency checks, we then numerically evaluated the NLO $s$- and $t$-channel contributions to the scattering amplitude for 1 and 2 finite spatial dimensions.  Computing the real part of the $s$-channel contribution for 2 finite spatial dimensions was non-trivial and required the derivation of a novel dispersion relation.  We found divergences in the $s$-channel contribution to the scattering amplitude when the momenta satisfied certain quantization conditions, corresponding to resonances associated with ``geometric bound states.''  The limits of the finite system size corrections for $p_\text{inf}\to0$ and/or $L\to0$ such that $p_\text{inf} L\ll1$ are complicated, and are channel and number of compact dimensions specific.  With 1 finite dimension the effective coupling, outside of geometric bound states, receives little correction for moderate to large scales ($p L\gtrsim 1$).  In 2 finite dimensions the behavior becomes even more non-trivial, but---away from the geometric bound states---we observe corrections up to order $40\%$.  

One can see that the effective coupling defined here is identically 0 at the geometric bound states.  That the effective coupling is 0 at these bound states is perhaps non-intuitive since the cross section diverges at these states.  We note, however, that we only include the resummation of the bubble diagrams in the effective coupling shown here; our effective coupling does not include the resummation of ladder diagrams necessary to fully capture the relevant physics of the geometric bound states.  Nevertheless, that the effective coupling decreases around the geometric bound states can be trusted in our work as the coupling times the NLO contribution remains small, $\lambda \overline V \ll1$, until the momenta are extraordinarily close to the resonance, $|p-p_{\mathrm{resonance}}|/p_{\mathrm{resonance}}\sim10^{-10}$.

Ignoring questions about the applicability of LSZ reduction, for 3 finite spatial dimensions the $s$ channel scattering amplitude is either identically 0 when the incoming and/or outgoing momenta are unphysical (i.e.\ not an integer mode dictated by the sizes of the spatial dimensions) or $\infinity$ if the incoming and outgoing momenta are physical.  We further examined analytically the asymptotically small and large argument limits of the $t$-channel contribution when all three spatial directions are finite.  We found good agreement between the numerics and the analytics.

For 1 and 2 finite spatial dimensions, we found that the coupling drops to zero as $p L\rightarrow0$.  This asymptotic behavior is consistent with a calculation of the effective coupling in a lattice discretization of $\phi^4$ theory in finite volume \cite{Montvay:1994cy}.  Since $1/L\rightarrow\infinity$ corresponds to a large momentum scale and the beta function for $\phi^4$ theory is positive, it's very interesting that the coupling decreases as the length of the finite size dimension becomes small compared to the momenta. It therefore seems that the system size can not simply be seen as the introduction of a dimensionful scale, but rather has a significantly less trivial effect on the physics.  In particular, it's unclear how the coupling of a gauge theory will behave.  Naively, one might expect that placing QCD in a small enough spatially confined system would yield a weak coupling; the small length scale of the confined system translates into a large momentum scale, and the beta function for QCD is negative.  But our results for $\phi^4$ theory show the exact opposite trend.  It will therefore be very interesting to explicitly perform the calculation for QCD in a small system to see whether the coupling grows or shrinks.

Interesting additional future work includes applying denominator regularization to additional systems, theories, and processes.  Denominator regularization has already been applied at one loop to QED \cite{Horowitz:2022uak} and to Higgs production \cite{Bansal:2022juh}.  It will be especially interesting to compute the trace anomaly for massive $\phi^4$ theory in a spatially confined system.  In addition to thermal QCD in a finite system, the finite system size corrections might have relevance for energy loss calculations in QCD in small collision systems \cite{Adil:2006ei,Kolbe:2015rvk,Faraday:2023mmx}.  

\section*{Acknowledgments}
WAH wishes to thank the South African National Research Foundation and the SA-CERN Collaboration for support and New Mexico State University for its hospitality, where part of the work was completed. JFDP wishes to thank the SA-CERN Collaboration for financial support. The authors wish to thank Matt Sievert, Alexander Rothkopf, Bowen Xiao, Jonathan Shock, and Andrea Shindler for valuable discussions.

\appendix
%%%%%%%%%%%%%%%%%%%%%%%%%%%%%%%%%%%%%%%%%%%%%%%%%%%%%%%%%%%%%%%%%%%%%%%%%%%%%%%%%%%%%%%%%
\section{Analytic Continuation of the Generalized Epstein Zeta Function}
%%%%%%%%%%%%%%%%%%%%%%%%%%%%%%%%%%%%%%%%%%%%%%%%%%%%%%%%%%%%%%%%%%%%%%%%%%%%%%%%%%%%%%%%%
\label{s:appsum}
We would like to analytically continue the generalized Epstein zeta function.  We follow the starting steps of \cite{Elizalde:1997jv}.  One begins from the Poisson summation formula (see, e.g., \cite{Stein:1971}).  Suppose that for some $p\in\mathbb N$
\begin{align}
    \tilde F(\vec k) & = \int d^p x \ess e^{2\pi\ess i \ess \vec k \cdot \vec x} f(\vec x) \\
    f(\vec x) & = \int d^p k \ess e^{-2\pi \ess i \ess \vec x\cdot \vec k}\tilde F(\vec k)
\end{align}
with $f$ and $\tilde F$ continuous and for some $\delta>0$
\begin{align}
    |f(\vec x)| \le A(1+|\vec x|)^{-p-\delta}; \quad
    |\tilde F(\vec k)| \le A(1+|\vec k|)^{-p-\delta}. \nonumber
\end{align}
Then
\begin{align}
    \label{e:psf1}
    \sum_{\vec n\in\mathbb Z^p}f(\vec x + \vec n) = \sum_{\vec m\in\mathbb Z^p} \tilde F(\vec m)e^{2\pi\ess i \ess \vec m \cdot \vec x}.
\end{align}
In particular, one has the Poisson summation formula
\begin{align}
    \label{e:psf}
    \sum_{\vec n\in\mathbb Z^p}f(\vec n) = \sum_{\vec m\in\mathbb Z^p} \tilde F(\vec m).
\end{align}
Note that the four series in \cref{e:psf1,e:psf} converge absolutely.

We now wish to apply the Poisson summation formula \cref{e:psf} to the generalized Eptein zeta function
\begin{align}
    \zeta(\{a_i\},\{b_i\},c;s) & \equiv \sum_{\vec n \in \mathbb Z^p} \big[ a_i^2 n_i^2 + b_i n_i + c \big]^{-s},
\end{align}
where repeated indices are summed over, e.g.\ $a_i^2 n_i^2 \equiv \sum_{i=1}^p a_i^2 n_i^2$, and we assume that $a_i,\,b_i,\,c\in\mathbb R$.  In order to apply the Poisson summation formula, we need to evaluate the Fourier transform of the summand of the generalized Epstein zeta function.  In order to make contact with \cref{e:infsum} consider the case in which we subtract a small imaginary part from $c$ such that we avoid the possibility of integrating through any poles.  Then we must consider for $\varepsilon>0$
\begin{multline}
    \int d^px\ess e^{2\pi\ess i \vec k\cdot \vec x}\big(a_i^2x_i^2+b_ix_i+c-i\varepsilon\big)^{-s} = \\
    e^{-2\pi \ess i \sum\limits_{i=1}^p\frac{k_i\ess b_i}{2a_i^2}}\frac{1}{\prod\limits_{i=1}^pa_i}\int d^p x' e^{2\pi \ess i \ess \vec k \cdot \vec x'}\big( \vec x'^2 + c' -i\varepsilon\big)^{-s}, \nonumber
\end{multline}
where $x'_i \equiv x_i + \frac{b_i}{2a_i^2}$ and $c' \equiv c - \sum_{i=1}^p \frac{b_i^2}{4a_i^2}$.
The remaining integral may be split into radial and angular parts,
\begin{align}
    \int & d^p x' e^{2\pi \ess i \ess \vec k \cdot \vec x'}\big( \vec x'^2 + c' -i\varepsilon\big)^{-s} \nonumber\\
    & = \int_0^\infty \rho^{p-1}d\rho(\rho^2+c'-i\varepsilon)^{-s}\int d\Omega_{p-1}e^{2\pi\ess i \ess k\ess \rho \cos\theta},
\end{align}
where $\rho\equiv|\vec x'|$, $k\equiv|\vec k|$, and
    $\Omega_p = 2\pi^{\frac{p+1}{2}}/\Gamma\big(\frac{p+1}{2}\big)$
is the solid angle of a $p$-dimensional sphere; $\Omega_2 = 4\pi$.  The angular integration evaluates for $k\ess\rho>0$ (which is always satisfied in our case) and $p>1$ to
\begin{align}
    \int & d\Omega_{p-2}\int_0^\pi\sin^{p-2}\theta d\theta e^{2\pi\ess i \ess k \ess \rho \cos\theta} \nonumber\\
    & = \Omega_{p-2}\sqrt\pi\Gamma\big(\frac{p-1}{2}\big) \frac{{}_0F_1\big(;\frac{p}{2};-(\pi\ess k \ess \rho)^2\big)}{\Gamma(p/2)} \nonumber\\
    & = \frac{2\pi^{p/2}}{\Gamma(p/2)}{}_0F_1\big(;\frac{p}{2};-(\pi\ess k \ess \rho)^2\big),
\end{align}
where ${}_0F_1(;a;z)$ is a usual generalized hypergeometric function.  One may check that for $p=2$ the above correctly reproduces $2\pi J_0(2\pi\ess k \ess \rho)$, where $J_\nu(z)$ is the usual Bessel function of the first kind.  One may then complete the evaluation through the use of
\begin{multline}
    \label{e:epsteinfourier}
    \int_0^\infty \rho^{p-1}d\rho(\rho^2+c'-i\varepsilon)^{-s}\frac{2\pi^{p/2}}{\Gamma(p/2)}{}_0F_1\big(;\frac{p}{2};-(\pi\ess k \ess \rho)^2\big) \\
    =\frac{2\pi^2}{\Gamma(s)}\Big( \frac{c'-i\varepsilon}{\vec k^2} \Big)^{\frac p4-\frac s2}K_{s-\frac p2}(2\pi |\vec k|\sqrt{c'-i\varepsilon})
\end{multline}
for $\Re s > p/2$, $d>1$, $c'\in\mathbb R$, $\varepsilon>0$, and $|\vec k|>0$, where $K_\nu(z)$ is the usual modified Bessel function of the second kind.

\cref{e:epsteinfourier} doesn't have an obvious $\vec k = \vec 0$ limit.  We must therefore separately evaluate the $\vec k = \vec 0$ Fourier mode.  One finds that for $\Re s > d/2 > 0$, $c'\in\mathbb R$, and $\varepsilon > 0$
\begin{align}
    \int & d^p x' (\vec x'^2 + c' - i\varepsilon)^{-s} \nonumber\\
    & = \Omega_{p-1}\int_0^\infty \rho^{p-1}d\rho(\rho^2+c'-i\varepsilon)^{-s} \nonumber\\
    & = \Omega_{p-1}\frac{\Gamma\big(\frac p2\big)\Gamma\big(s-\frac p 2\big)}{2\Gamma s}(c'-i\varepsilon)^{\frac p2-s} \nonumber\\
    & = \pi^{p/2}\frac{\Gamma\big(s-\frac p2\big)}{\Gamma s}(c'-i\varepsilon)^{\frac p2-s}.
\end{align}

Putting the pieces together we arrive at our master formula for the analytic continuation of the generalized Epstein zeta function:
\begin{widetext}
    \begin{multline}
        \label{e:epsteinanalytic}
        \sum_{\vec n\in\mathbb Z^p} (a_i^2 n_i^2 + b_i n_i + c-i\varepsilon)^{-s} = \frac{1}{a_1\cdots a_p}\frac{1}{\Gamma(s)} \bigg[ 
        \pi^{p/2}\Gamma\big( s - \frac p2\big) \Big( c - \sum \frac{b_i^2}{4a_i^2} - i\varepsilon \Big)^{\frac p2 - s} \\
        +2\pi^s\sideset{}{'}\sum_{\vec m\in\mathbb Z^p} e^{-2\pi\ess i \sum \frac{m_i b_i}{2a_i^2}}\bigg( \frac{c - \sum\frac{b_i^2}{4a_i^2} - i\varepsilon}{\sum \frac{m_i^2}{a_i^2}} \bigg)^{\frac p4-\frac s2} 
        K_{s-\frac p2}\bigg(2\pi \sqrt{(c-\sum\frac{b_i^2}{4a_i^2} - i\varepsilon)\big(\sum \frac{m_i^2}{a_i^2}\big)} \bigg)
        \bigg],
    \end{multline}
\end{widetext}
where $\sideset{}{'}\sum\limits_{\vec m\in\mathbb Z^p}$ indicates a sum over all integers $\vec m\in\mathbb Z^p$ \emph{except} for $\vec m = \vec 0$ and where the suppressed limits on the sums run from $i=1\ldots p$.  Notice that the contribution from $\vec m = \vec 0$ isolates the pole as we analytically continue $s\rightarrow p/2$.

One may numerically evaluate the $\sideset{}{'}\sum\limits_{\vec m\in\mathbb Z^p}$ in \cref{e:epsteinanalytic} more efficiently by combining the phases into cosines.  The speedup comes from evaluating a pure real expression and from drastically reducing the total number of summed terms.  The result is a sum over all the subsets of the set of numbers $\{1,\ldots,p\}$, known as the power set, $2^{[p]}$:
\begin{widetext}
    \begin{multline}
        \label{e:moreefficient}
        \sideset{}{'}\sum_{\vec m\in\mathbb Z^p} e^{-2\pi\ess i \sum \frac{m_i b_i}{2a_i^2}}\bigg( \frac{c - \sum\frac{b_i^2}{4a_i^2} - i\varepsilon}{\sum \frac{m_i^2}{a_i^2}} \bigg)^{\frac p4-\frac s2} 
        K_{s-\frac p2}\bigg(2\pi \sqrt{(c-\sum\frac{b_i^2}{4a_i^2} - i\varepsilon)\big(\sum \frac{m_i^2}{a_i^2}\big)} \bigg) \\
        = \sum_{s\in2^{[p]}}2^{|s|+1}\sum_{\substack{m_i=1 \\ i\in s}}^\infinity\prod_{i\in s}\Big( \cos(2\pi\ess x \ess m_i L_i p^i) \Big)\bigg( \frac{c - \sum\frac{b_i^2}{4a_i^2} - i\varepsilon}{\sum \frac{m_i^2}{a_i^2}} \bigg)^{\frac p4-\frac s2} K_{s-\frac p2}\bigg(2\pi \sqrt{(c-\sum\frac{b_i^2}{4a_i^2} - i\varepsilon)\big(\sum \frac{m_i^2}{a_i^2}\big)} \bigg)
        ,
    \end{multline}
\end{widetext}
where $|s|$ is the length of the current set of indices being summed over and the sums with the suppressed limits are over $i\in s$.

%%%%%%%%%%%%%%%%%%%%%%%%%%%%%%%%%%%%%%%%%%%%%%%%%%%%%%%%%%%%%%%%%%%%%%%%%%%%%%%%%%%%%%%%%
\section{Alternative Check of Unitarity}
%%%%%%%%%%%%%%%%%%%%%%%%%%%%%%%%%%%%%%%%%%%%%%%%%%%%%%%%%%%%%%%%%%%%%%%%%%%%%%%%%%%%%%%%%
\label{s:altunitarity}
For this unitarity analysis we will need to derive a generalization of the formula
\begin{multline}
    \label{eq:Hardy}\sum_{0\leq n<x}\frac{r_2(n)}{\sqrt{x-n}} =2\pi\sqrt{x}+\sum_{n=1}^\infty\frac{r_2(n)}{\sqrt{n}}\sin(2\pi\sqrt{n x}),\\
    x>0,x\not\in\mathbb{Z}
\end{multline}
originally proposed by Ramanujan and expanded on by Hardy in Eq.\ (2) on page 82 of \cite{Hardy1978RamanujanTL}.

We will generalize \cref{eq:Hardy} by considering the rectangular lattice sum
\begin{align}
    \sum_{\vec{k}\in\Lambda}\sinc(2\pi\|\vec{k}\|), \nonumber
\end{align}
where $\Lambda$ is a lattice of dimension $\text{dim}(\Lambda)=n$ and determinant $|\Lambda|$, the volume obtained as the product of the lattice spacings.

We will be employing the Poisson summation formula for rectangular lattice sums \cite{cohn2013formal}, given by
\begin{equation}\label{eq:PSF}
    \sum_{\vec{k}\in\Lambda}f(\vec{k})=\frac{1}{\left|\Lambda\right|}\sum_{\ell\in\Lambda^*}F(\vec{\ell}),
\end{equation}
where $\Lambda^*$ is the dual lattice, the lattice with the same dimension with spacings inverted of that of the original (so $|\Lambda|=|\Lambda^*|^{-1}$), and $F$ is the Fourier transform of $f$, given by
\begin{equation}
    F(\vec{\ell})\equiv\int d^n k\,e^{-2\pi i\vec{k}\cdot\vec{\ell}}f(\vec{k}).
\end{equation}
As in \cite{JohnsonMcDaniel2012ADC} we can exploit the radial nature of the function to use the formula
\begin{equation}
    F(\ell)=2\pi\ell^{\frac{2-n}{2}}\int_0^\infty dr f(r)J_{\frac{n-2}{2}}(2\pi\ell r)r^{\frac{n}{2}}. \nonumber
\end{equation}
We therefore need to evaluate
\begin{align}
     F(\ell)&=2\pi\ell^{\frac{2-n}{2}}\int_0^\infty dr \sinc(2\pi r)J_{\frac{n-2}{2}}(2\pi\ell r)r^{\frac{n}{2}}\label{eq:domcon}\\
     &=\frac{\pi^{\frac{1-n}{2}}}{2\Gamma\left(\frac{3-n}{2}\right)}\sqrt{1-\ell^2}^{1-n}\theta(1-\ell^2),\label{eq:ft}
\end{align}
which is valid for $0<n<3$ and $\ell\geq 0$. Inserting \cref{eq:ft} into \cref{eq:PSF} we find the generalization of \cref{eq:Hardy},
\begin{equation}\label{eq:SincSumGen}
    \sum_{\vec{k}\in\Lambda}\sinc(2\pi\|\vec{k}\|)=\frac{\pi^{\frac{1-n}{2}}\left|\Lambda^*\right|}{2\Gamma\left(\frac{3-n}{2}\right)}\sideset{}{^*}\sum_{\substack{\vec{\ell}\in\Lambda^* \\ \ell^2\leq 1}}\sqrt{1-\ell^2}^{1-n},
\end{equation}
where $\sideset{}{^*}\sum$ denotes that terms with $\ell^2=1$ have a weight of $\theta(0)$, whose value depends on convention. If one considers the $n=1$ dimensional lattice with unit spacing case, one can show that the only self-consistent choice is $\theta(0)=\frac12$, which agrees with the convention used in \cref{eq:Hardy} and is corroborated by explicitly computing \cref{eq:domcon} with $n=1,\ell=1$ and equating it with \cref{eq:ft}. \cref{eq:SincSumGen} with $\theta(0)=\frac12$ is then our generalziation of Hardy and Ramanujan's \cref{eq:Hardy}. One can carefully check that \cref{eq:SincSumGen} is also valid for $n=0$ (trivially gives $1=1$), but one should be careful with \cref{eq:domcon}, since the dominated convergence theorem fails at $n=0,\ell=0$, so we cannot interchange a limit to $n=0,\ell=0$ with the integral (indeed, doing this gives an incorrect answer). We also consider the $n=3$ case below. \cref{eq:SincSumGen} of course also holds for more general lattices (than just the rectangular lattices we are interested in), where the determinant is then not, in general, simply given by the product of lattice spacings.

We can now also consider some special cases on square lattices with spacings given by some positive $R$. These come up in our numerical explorations, which considers the case where all finite dimensions have equal length scale.

Considering a 1D lattice with spacing $R$, \cref{eq:SincSumGen} gives
\begin{align}
    \sum_{k=-\infty}^\infty\sinc(2\pi R k)&=\frac{R^{-1}}{2}\sideset{}{^*}\sum_{-R\leq\ell\leq R}1\nonumber\\
    &=\begin{cases}
    \frac{\lfloor R\rfloor}{R}+\frac{1}{2R} & r_1(R^2)=0\\
    1 & R\in\mathbb{Z}
    \end{cases}\label{eq:1Dsinc}
\end{align}

Considering a 2D lattice with both spacings $R$; \cref{eq:SincSumGen} gives
\begin{align}
    \sum_{\vec{k}\in\mathbb{Z}^2}\sinc(2\pi R \|\vec{k}\|)&=\frac{R^{-2} }{2\pi}\sideset{}{^*}\sum_{\substack{\vec{\ell}\in\mathbb{Z}^2 \\ \ell^2\leq R^2}}\frac{1}{\sqrt{1-\frac{\ell^2}{R^2}}}.\\
    \intertext{2D radial sums can be simplified into a 1D sum using the sum of squares function $r_2$, giving}
    \sum_{k=0}^\infty r_2(k)\sinc\left(2\pi R\sqrt{k}\right)&=\frac{1}{2\pi R}\sideset{}{^*}\sum_{0\leq l\leq R^2}\frac{r_2(l)}{\sqrt{R^2-l}}
    \end{align}
In order to connect with \cref{eq:Hardy} we consider $R^2\not\in\mathbb{Z}$ and we write out $\sinc(x)=\sin(x)/x$
    \begin{align}
     2\pi R+\sum_{k=1}^\infty \frac{r_2(k)}{\sqrt{k}}\sin\left(2\pi R\sqrt{k}\right)&=\sum_{0\leq l< R^2}\frac{r_2(l)}{\sqrt{R^2-l}}.\label{eq:Rhardy}
\end{align}
Note that we have dropped the star from the $\sideset{}{^*}\sum_{0\leq l< R^2}$ and dropped the equality in the range of the sum, since we have $R^2\not\in\mathbb{Z}$. We can then see that we get \cref{eq:Hardy} by setting $x=R^2$ in \cref{eq:Rhardy}. This shows that \cref{eq:SincSumGen} is a direct generalization to \cref{eq:Hardy}.

To consider \cref{eq:SincSumGen} with $n=3$ (on a potentially assymetric lattice), we need to think of \cref{eq:SincSumGen} as an analytic continuation. The left hand side of \cref{eq:SincSumGen} evaluated numerically with Cesaro summation seems to yield $0$ for $R\not\in\mathbb{Z}$ and diverge to positive infinity for $R\in\mathbb{Z}$. To consider the right hand side, we need to return to the Fourier transform integral in \cref{eq:ft}, which does not converge for $n=3$. Let us then consider, for some small $\epsilon>0$
\begin{align}
    F(\ell)\Big|_{n=3}&\equiv2\pi\ell^{-\frac{1}{2}}\int_0^\infty dr \sinc(2\pi r)J_{\frac{1}{2}}(2\pi\ell r)r^{\frac{3}{2}}e^{-\epsilon r}\label{eq:3dfour}\\
    &=\frac{8\pi\epsilon}{16\pi^4(1-\ell^2)^2+8\pi^2(1+\ell^2)\epsilon^2+\epsilon^4}\nonumber\\
    &=\begin{cases}
        \frac{1}{2\pi\epsilon}+\mathcal{O}(\epsilon)&\ell^2=1\\
        \frac{\epsilon}{2\pi^3(1-\ell^2)^2}+\mathcal{O}(\epsilon^2)&\ell^2\neq1
    \end{cases}
\end{align}
In the $\epsilon\to0$ limit, we see this continuation agrees with numerics that for a $3$ dimensional lattice $\Lambda$
\begin{equation}
    \sum_{\vec{k}\in\Lambda}\sinc(2\pi\|\vec{k}\|)=\begin{cases}
        \infty&\vec{\ell}\in\Lambda^*\text{ with }\ell^2=1\\
        0&\text{otherwise}.
    \end{cases}\label{s:threecompact}
\end{equation}
One might then wonder if the divergences on the two sides of \cref{eq:SincSumGen} that happens for an $n=3$ dimensional lattice $\Lambda$ with some $\vec{\ell}\in\Lambda^*\text{ with }\ell^2=1$ are equal. To think about two infinities being equal, we need to have some limit of a ratio of diverging quatities limit to 1. There are two ways we can consider these infinities to be equal. The first is to take the expression in \cref{eq:3dfour} seriously, by considering
\begin{align}
    \lim_{\epsilon\to 0^+}\frac{\sum_{\vec{k}\in\Lambda}\sinc(2\pi\|\vec{k}\|)e^{-\epsilon\|\vec{k}\|}}{\sum_{\vec{\ell}\in\Lambda^*}\frac{8\pi\epsilon}{16\pi^4(1-\ell^2)^2+8\pi^2(1+\ell^2)\epsilon^2+\epsilon^4}}=1
\end{align}
by the Poisson summation formula. Similarly we know (if we consider \cref{eq:SincSumGen} to be an analytic continuation for non-integer $0<n<3$ dimensional lattice sums, neglecting to even attempt having a concrete notion of a non-integer dimensional lattice) that
\begin{align}
    \lim_{n\to3^-}\frac{\sum_{\vec{k}\in\Lambda}\sinc(2\pi\|\vec{k}\|)}{\frac{\pi^{\frac{1-n}{2}}\left|\Lambda^*\right|}{2\Gamma\left(\frac{3-n}{2}\right)}\sideset{}{^*}\sum_{\substack{\vec{\ell}\in\Lambda^* \\ \ell^2\leq 1}}\sqrt{1-\ell^2}^{1-n}}=1
\end{align}
by \cref{eq:SincSumGen}.

We may then push the interpretation of \cref{eq:SincSumGen} as an analytic continuation in $n$ even further. We note that
\begin{equation}
    \lim_{\epsilon\to0^+}\sum_{\vec{k}\in\Lambda}\sinc(2\pi\|\vec{k}\|)e^{-\epsilon\|\vec{k}\|}\nonumber\\
\end{equation}
is the Abel summation of the left hand side of \cref{eq:SincSumGen}. Retracing our previous derivation we then find that in order to employ the Poisson summation formula \cref{eq:PSF} we need to compute
\begin{align}
     F(\ell)&=2\pi\ell^{\frac{2-n}{2}}\int_0^\infty dr \sinc(2\pi r)e^{-\epsilon r}J_{\frac{n-2}{2}}(2\pi\ell r)r^{\frac{n}{2}}\nonumber\\
     &=2\pi\ell^{\frac{2-n}{2}}\int_0^\infty dr \sinc(2\pi r)e^{-\epsilon r}\label{eq:domcomfail}\\
     &\times\left[\sum_{m=0}^\infty\frac{(-1)^m}{2^{2m+\frac{n-2}{2}}\Gamma\left(\frac{n}{2}\right)\Gamma\left(m+\frac{n}{2}\right)}(2\pi\ell r)^{2m+\frac{n-2}{2}}\right] r^{\frac{n}{2}}\nonumber\\
     &=2\pi\ell^{\frac{2-n}{2}}\sum_{m=0}^\infty\frac{(-1)^m}{2^{2m+\frac{n-2}{2}}\Gamma\left(\frac{n}{2}\right)\Gamma\left(m+\frac{n}{2}\right)}\nonumber\\
     &\times \ell^{2m+\frac{n-2}{2}}(2\pi)^\frac{4m+n}{2}(4\pi^2+\epsilon^2)^\frac{1-2m-n}{2}\Gamma(2m+n-1)\nonumber\\&\times\sin\left((2m+n-1)\arctan\left(\frac{2\pi}{\epsilon}\right)\right)
\end{align}
for $1>\ell\geq0$ and $\Re n>0$. It seems plausible that another method, other than employing the power series of the Bessel function, would extend the range of validity even further. We then find by seemingly safely taking the limit and simplifying to find
\begin{align}
    \lim_{\epsilon\to0^+}F(\ell)&=\frac{1}{2\pi^\frac{n-1}{2}}\sum_{m=0}^\infty\frac{(-\ell^2)^m}{m!}\nonumber\\
    &\times\Gamma\left(m+\frac{n-1}{2}\right)\sin\left(\pi\left(m+\frac{n-1}{2}\right)\right)\nonumber\\
    &=\frac{\pi^\frac{1-n}{2}}{2\Gamma\left(\frac{3-n}{2}\right)}\sqrt{1-\ell^2}^{1-n},
\end{align}
where we performed the sum by assuming $0\leq\ell<1$. For $\ell=1$ we have a bit more difficulty, but find that
\begin{align}
    F(1)&=\begin{cases}
        0 & \Re n<1\\
        \frac14 & \Re n=1\\
        \infty & \Re n>1,
    \end{cases}
\end{align}
where the $\infty$ will, for general $n$ with $\Re n>1$, have some prefactor in the sense of characterizing the divergences on both sides of \cref{eq:SincSumGen} to be equal. For $\ell>1$, the dominated convergence theorem fails to hold at \cref{eq:domcomfail}, so one cannot simply interchange the integral and sum, and indeed we find inconsistent results if we do. For $0<n\leq 3$ we know \cref{eq:domcomfail} evaluates to $0$ for $\ell>1$, and it seems reasonable that this is generally true for $\Re n>0$, which would make \cref{eq:SincSumGen} valid for all $\Re n>0$ in the Abel summation sense.

If we consider \cref{e:opticalLHS1}, and apply a uniform rescaling by a factor of $\frac12\sqrt{s}\tilde Q$ of the lattice $\Lambda$, \cref{e:opticalLHS1} is in a form where we can then apply \cref{eq:SincSumGen} to find an expression equivalent to \cref{e:opticalRHS} under its proper rescaling (the dual lattice $\Lambda^*$ needs to be uniformly rescaled by a factor of $\left(\frac12\sqrt{s}\tilde Q\right)^{-1}$ in order to be the dual lattice of the rescaled $\Lambda$).

%%%%%%%%%%%%%%%%%%%%%%%%%%%%%%%%%%%%%%%%%%%%%%%%%%%%%%%%%%%%%%%%%%%%%%%%%%%%%%%%%%%%%%%%%
\section{s channel dispersion relations}
\label{s:dispersion}
%%%%%%%%%%%%%%%%%%%%%%%%%%%%%%%%%%%%%%%%%%%%%%%%%%%%%%%%%%%%%%%%%%%%%%%%%%%%%%%%%%%%%%%%%

We are interested in evaluating the $s$ channel contribution to the NLO amplitude
\begin{align}
    \overline{V}_n&(s,\{L_i\},\mu)=-\frac{1}{32\pi^2}\int_0^1 dx\Bigg\{\ln\frac{\mu^2}{m^2-x(1-x)s-i\varepsilon}\nonumber\\
    &+2\sideset{}{'}\sum_{\vec{k}\in\mathbb{Z}^n} K_0\left(2\pi \sum L_i k_i \sqrt{m^2-x(1-x)s-i\varepsilon}\right)\Bigg\},\label{eq:Vsapp}
\end{align}
which in its current form is numerically ill-behaved. For simplicity, let us define
\begin{align}
    f(\sigma,\varepsilon)&=\int_0^1 dx \ln\left(\frac{1}{1-x(1-x)\sigma-i\varepsilon}\right)\label{eq:ffunc}
\end{align}
\begin{multline}
    g_n(\Lambda,\sigma,\varepsilon)=f(\sigma,\varepsilon)+2\sideset{}{'}\sum_{\vec{k}\in\Lambda}\int_0^1 dx \\\times K_0\left(2\pi\|\vec{k}\|\sqrt{1-x(1-x)\sigma-i\varepsilon}\right)\label{eq:gn}
\end{multline}
where $\Lambda$ is an $n$-dimensional lattice. We note that using the dominated convergence theorem \cite{inpcdomcon} one can show that, as long as there are no $\vec{\ell}\in\Lambda^*$ with $\ell^2=\frac\sigma4-1$, one can safely take both the $\sigma\to\infty$ or lattice spacings of $\Lambda$ to $\infty$ limits using that $K_0\sim e^{-x}/\sqrt{x}$ to show that $g_n(\Lambda,\sigma,\varepsilon)\sim f(\sigma,\varepsilon)$ in these limits.
\cref{eq:ffunc,eq:gn} then allow us to write \cref{eq:Vsapp} as
\begin{multline}
    \overline{V}_n(s,\{L_i\},\mu)=-\frac{1}{32\pi^2}\Bigg[\ln\left(\frac{\mu^2}{m^2}\right)\\+g_n\left(\Lambda(\{L_i m\}),\frac{s}{m^2},\varepsilon\right)\Bigg],
\end{multline}
where $\Lambda(\{L_i m\})$ is the $n$-dimensional rectangular lattice with lattice spacings given by the $L_i m$. In the special case of all finite lengths being equal,
\begin{multline}\label{eq:vbarg}
    \overline{V}_n(s,L,\mu)=-\frac{1}{32\pi^2}\Bigg[\ln\left(\frac{\mu^2}{m^2}\right)\\+g_n\left(\Lambda(L m),\frac{s}{m^2},\varepsilon\right)\Bigg],
\end{multline}
where $\Lambda(L m)$ is then the $n$-dimensional square lattice with all lattice spacings equal to $L m$.

\subsection*{Imaginary part}
We will try to exploit the complex structure of these functions to find a numerically well behaved equivalent form for \cref{eq:Vsapp}. We have chosen to self-consistently take the $\arg$, $\log$, $K_0$, and the square root branch cuts along the negative real axis. This straightforwardly gives that both $f$ and $g$ have a branch cut along the positive real axis from $\sigma=4$ to $\sigma=\infty$. The real part of both $f$ and $g$ is continuous across the branch cut, and the imaginary part changes sign. We further need to make the assumption that there are no poles for $g$ in the complex $\sigma$ plane off of the positive real axis. This assumption seems reasonable and is supported by numerics. We do however note that there are physical situations where the $S$-matrix has poles violating this assumption \cite{Moroz_2019}. Something we can easily calculate and evaluate is the imaginary part of $f$ as follows:
\begin{align}
    \Im f(\sigma,\varepsilon)&=\int_0^1 dx \Im\ln\left(\frac{1}{1-x(1-x)\sigma-i\varepsilon}\right)\nonumber\\
    &=\int_0^1 dx \pi\theta\left(x(1-x)\sigma-1\right)+\mathcal{O}(\varepsilon)\nonumber\\
    \Im f(\sigma,0^+)&=\pi\sqrt{1-\frac4\sigma},\label{eq:fim}
\end{align}
where $\varepsilon$ helped us avoid the branch cut and choose the sign of the imaginary part. We will use $0^+$ to denote that our $\varepsilon$ is some infinitesimal positive real number, but since $\varepsilon$ only selects a branch cut we will neglect to write it. This does mean we need to be continually mindful of branch cuts. We can then continue onward to calculate
\begin{align}
    I&\equiv\Im \int_0^1 dx K_0\left(2\pi\|\vec{k}\|\sqrt{1-x(1-x)\sigma-i\varepsilon}\right)\nonumber\\
    &=\int_0^1 dx \Im K_0\left(2\pi\|\vec{k}\|\sqrt{1-x(1-x)\sigma-i\varepsilon}\right)\nonumber
    \intertext{We again use the $\varepsilon$ to avoid the branch cut and indicate the sign of the imaginary part, giving us}
    I&=\frac{\pi}{2}\int_0^1 dx J_0\left(2\pi\|\vec{k}\|\sqrt{x(1-x)\sigma-1}\right)\nonumber\\&\qquad\qquad\times\theta\left(x(1-x)\sigma-1\right)+\mathcal{O}(\varepsilon).\nonumber
    \intertext{We can then discard the $\mathcal{O}(\varepsilon)$ terms. We can also use the symmetry of the integral about $x=\frac12$ and the step function to find}
    I&=\pi\int_\frac12 ^{\frac12+\frac12\sqrt{1-\frac4\sigma}}dx J_0\left(2\pi\|\vec{k}\|\sqrt{x(1-x)\sigma-1}\right).\label{eq:joh}
    \intertext{We can further simplify \cref{eq:joh} by using the substitution $y\equiv\frac{x-\frac12}{\frac12\sqrt{1-\frac4\sigma}}$ to find}
    I&=\frac\pi2\sqrt{1-\frac4\sigma}\int_0^1 dy J_0\left(2\pi\|\vec{k}\|\sqrt{\frac\sigma4-1}\sqrt{1-y^2} \right).\nonumber
    \intertext{We then use $\int_0^1 dy\,J_0(a\sqrt{1-y^2})=\sinc(a)$ to find}
    I&=\frac\pi2\sqrt{1-\frac4\sigma}\sinc\left(2\pi\|\vec{k}\|\sqrt{\frac\sigma4-1}\right).\label{eq:gim}
\end{align}

We can combine \cref{eq:fim,eq:gim} to find
\begin{align}
    \Im g_n&(\Lambda,\sigma,0^+)&=\pi\sqrt{1-\frac4\sigma}\sum_{\vec{k}\in\Lambda}\sinc\left(2\pi\sqrt{\frac\sigma4-1}\|\vec{k}\|\right)\label{eq:imgsinc}
\end{align}
which now allows us to use a uniformly rescaled \cref{eq:SincSumGen} in order to write \cref{eq:imgsinc} as
\begin{align}
    &\Im g_n(\Lambda,\sigma,0^+) \nonumber\\
    &=\pi\sqrt{1-\frac4\sigma}\frac{\pi^\frac{1-n}{2}|\Lambda^*|}{2\Gamma\left(\frac{3-n}{2}\right)\sqrt{\frac\sigma4-1}^n}
        \sideset{}{^*}\sum_{\substack{\vec{\ell}\in\Lambda^* \\ \ell^2\leq \frac\sigma4-1}}\sqrt{1-\frac{\ell^2}{\frac\sigma4-1}}^{1-n}\nonumber\\
    &=\frac{\pi^\frac{3-n}{2}|\Lambda^*|}{\sqrt{\sigma}\Gamma\left(\frac{3-n}{2}\right)}\sideset{}{^*}\sum_{\vec{\ell}\in\Lambda^*}
        \left(\frac\sigma4-1-\ell^2\right)^\frac{1-n}{2}\theta\left(\frac\sigma4-1-\ell^2\right).
\end{align}

\subsection*{Complex Analysis}
For asymptotically large $|\sigma|$, we find that $\left|\frac{g_n(\Lambda,\sigma,\varepsilon)}{\sigma^2}\right|$ falls of faster than $|\sigma|^{-1}$ for $n<3$ dimensional lattices. We can use this as follows. Consider the Cauchy integral formula
\begin{align}
    \label{eq:cauchy}
    \frac{d}{d\sigma}\left(g_n(\Lambda,\sigma,0^+)\right)&=\frac{1}{2\pi i}\int_\gamma dz\frac{g_n(\Lambda,z,0^+)}{(z-\sigma)^2},
\end{align}
where $\gamma$ is some small counter-clockwise contour around the pole at $z=\sigma$.

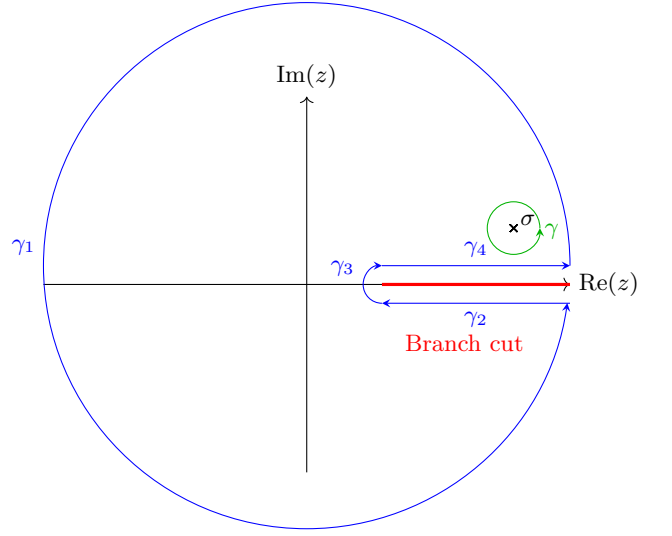
\begin{figure}[t]
    \centering
\begin{tikzpicture}[scale=0.5]
      % Axis
      \draw[->] (-7,0) -- (7,0) node[right] {$\mathrm{Re}(z)$};
      \draw[->] (0,-5) -- (0,5) node[above] {$\mathrm{Im}(z)$};
   
      % Branch cut
      \draw[very thick, red] (2,0) -- (7,0);
   
      % Contour path
      \draw[blue,->,>=stealth] (2,0.5) -- (7,0.5) node[midway, above] {$\gamma_4$};
      \draw[blue,->,>=stealth] (7,0.5) arc (0:351.8288:7) node[midway, left] {$\gamma_1$};
      \draw[blue,->,>=stealth] (7,-0.5) -- (2,-0.5) node[midway, below] {$\gamma_2$};
      \draw[blue,->,>=stealth] (2,-0.5) arc (270:90:0.5) node[midway, above left] {$\gamma_3$};

      % Labels
      % \node[above left, text=red] at (6,-1.75) {Branch cut};
      \node[above left, text=red] at (6,-2) {Branch cut};

      % Pole
      \draw plot[mark=x, mark options={color=black, line width=0.3pt, scale=2}] (5.5,1.5);
      \node[above right, text=black] at (5.45,1.35) {$\sigma$};

      %gamma contour
      \draw[darkgreen,->,>=stealth] (6.2,1.5) arc (0:360:0.7);
      \node[above right, text=darkgreen] at (6.1,1) {$\gamma$};
    \end{tikzpicture}
    \caption{The analytic structure of \cref{eq:cauchy}.}
    \label{fig:cont}
\end{figure}

Looking at the complex structure as discussed above and shown in \cref{fig:cont}, we can see that we can deform our contour around the branch cut just under the real axis (offset due to $\varepsilon=0^+$). Then since $\frac{g_n(\Lambda,\sigma,0)}{\sigma^2}$ falls off faster than $\sigma^{-1}$, the integral along $\gamma_1$ does not contribute. It's also possible to show that the integral along $\gamma_3$ does not contribute. This leaves us with $\gamma_2,\gamma_4$.

$g_n(\Lambda,\sigma,0)$ has a real part that is continuous across its branch cut, and the imaginary part simply changes sign, so since $\gamma_2,\gamma_4$ run in opposite directions along opposite sides of the branch cut, we find
\begin{align}
    &\frac{d}{d\sigma}\left(g_n(\Lambda,\sigma,0^+)\right)\Bigg|_{\sigma=\sigma_0+i\eta}\nonumber \\
    &= \frac{1}{\pi}\int_4^\infty dz  \frac{\Im g_n(\Lambda,z,0^+)}{(z-\sigma_0-i\eta)^2}\\
    &=\frac{\pi^{\frac{1-n}{2}}|\Lambda^*|}{\Gamma\left(\frac{3-n}{2}\right)}\sum_{\ell\in\Lambda^*}\int_4^\infty dz\frac{\theta\left(\frac{z}{4}-1-\ell^2\right)}{\sqrt{z}\left(\frac{z}{4}-1-\ell^2\right)^{\frac{n-1}{2}}}\nonumber\\
    &\pushright{\times\frac{1}{(z-\sigma_0-i\eta)^2}.\quad}
    \end{align}
    We have pushed the pole we introduced through the Cauchy integral formula into the upper-half plane, such that it is inside the contour we are considering, and not exactly on the branch cut itself, since our contour needs to run between the two.
     
    Integrating both sides w.r.t.\ $\sigma_0$ (note that this is a contour integral, so some care needs to be taken) then gives
    \begin{align}
    &g_n(\Lambda,\sigma+i\eta,0^+)-g_n(\Lambda,\sigma_0+i\eta,0^+)\nonumber\\
    &=\frac{\pi^{\frac{1-n}{2}}|\Lambda^*|}{\Gamma\left(\frac{3-n}{2}\right)}\sum_{\ell\in\Lambda^*}\int_4^\infty dz\frac{\theta\left(\frac{z}{4}-1-\ell^2\right)}{\sqrt{z}\left(\frac{z}{4}-1-\ell^2\right)^{\frac{n-1}{2}}}\nonumber\\
    \label{eq:mastern}
    &\pushright{\times\left[\frac{1}{z-\sigma-i\eta}-\frac{1}{z-\sigma_0-i\eta}\right].\quad}
\end{align}

\subsection*{Case: \texorpdfstring{$n=0$}{n=0}}
In principle, the $n=0$ case is trivial as there is no sum to perform.  However, we will examine the $n=0$ case for three reasons.  First, we would like to confirm that \cref{eq:mastern} reproduces the original \cref{eq:gn}, which in this case trivially reduces to \cref{eq:ffunc}.  Second, we will find that even in this simple case we will utilize a method that will be extremely useful in the higher $n$ cases.  Third, we will use an asymptotic form of \cref{eq:ffunc}, which will prove invaluable for the higher $n$ cases.

Simply setting $n=0$, there is no sum to be performed in \cref{eq:gn}, and we obtain
\begin{multline}
    g_0(\{\vec{0}\},\sigma+i\eta,0^+)-g_0(\{\vec{0}\},\sigma_0+i\eta,0^+)\\
    =\int_4^\infty dz\sqrt{1-\frac{4}{z}}\left[\frac{1}{z-\sigma-i\eta}-\frac{1}{z-\sigma_0-i\eta}\right]
\end{multline}
Integrating and then safely taking $\eta\to0$
\begin{widetext}
    \begin{multline}
        g_0(\{\vec{0}\},\sigma,0^+)-g_0(\{\vec{0}\},\sigma_0,0^+)
        = 2\sqrt{1 - \frac{4}{\sigma}}\Bigg[\arctanh\left(\sqrt{1-\frac{4}{\sigma}}\right)+\arctanh\left(\frac{2-\sigma}{\sqrt{\sigma (\sigma - 4)}}\right)\Bigg)\\
        -2\sqrt{1 - \frac{4}{\sigma_0}}\Bigg(\arctanh\left(\sqrt{1-\frac{4}{\sigma_0}}\right)+\arctanh\left(\frac{2-\sigma_0}{\sqrt{\sigma_0 (\sigma_0 - 4)}}\right)\Bigg].\label{eq:arctanhs}
    \end{multline}
    We can rearrange \cref{eq:arctanhs} to obtain
        \begin{multline}
        g_0(\{\vec{0}\},\sigma,0^+) 
        =\Bigg[g_0(\{\vec{0}\},\sigma_0,0^+) 
        -2\sqrt{1 - \frac{4}{\sigma_0}}\left(\arctanh\left(\sqrt{1-\frac{4}{\sigma_0}}\right)+\arctanh\left(\frac{2-\sigma_0}{\sqrt{\sigma_0 (\sigma_0 - 4)}}\right)\right)\Bigg]\\
        +2\sqrt{1 - \frac{4}{\sigma}}\left(\arctanh\left(\sqrt{1-\frac{4}{\sigma}}\right)+\arctanh\left(\frac{2-\sigma}{\sqrt{\sigma (\sigma - 4)}}\right)\right).
    \end{multline}

    Since the left hand side is explicitly independent of $\sigma_0$, we know the first term on the right-hand side must be $\sigma_0$ independent and is simply a constant that needs to be determined.  (We will exploit this method of isolating the $\sigma_0$ independent constant in the higher $n$ cases.)  We therefore define
    \begin{align}
        \label{eq:a0}
        a_0 \equiv g_0(\{\vec{0}\},\sigma_0,0^+)-2\sqrt{1 - \frac{4}{\sigma_0}}\left(\arctanh\left(\sqrt{1-\frac{4}{\sigma_0}}\right)+\arctanh\left(\frac{2-\sigma_0}{\sqrt{\sigma_0 (\sigma_0 - 4)}}\right)\right)
    \end{align}
    in order to obtain
    \begin{align}
        g_0(\{\vec{0}\},\sigma,0^+) = a_0 + 2\sqrt{1 - \frac{4}{\sigma}}\left(\arctanh\left(\sqrt{1-\frac{4}{\sigma}}\right)+\arctanh\left(\frac{2-\sigma}{\sqrt{\sigma (\sigma - 4)}}\right)\right).
    \end{align}
\end{widetext}

For the higher $n$ cases, we'll need to take $\sigma_0\rightarrow\infinity$ and evaluate an equation similar to \cref{eq:a0} numerically.  For the $n=0$ case we may analytically evaluate $a_0$.  We first note that $g_0(\{\vec{0}\},\sigma,0^+) = f(\sigma,0^+)$, and thus for $\sigma>4$
\begin{align}
    \label{eq:g00}
    g_0(\{\vec{0}\},\sigma,0^+) = 2-2\sqrt{1-\frac{4}{\sigma}}\arccoth\left(\sqrt{1-\frac{4}{\sigma}}\right)
\end{align}
from an explicit analytic evaluation of \cref{eq:ffunc}.  One may then evaluate $a_0$ in \cref{eq:a0} by taking $\sigma\rightarrow\sigma_0$ in \cref{eq:g00} and then simplifying to find that
\begin{align}
    a_0 = 2.
\end{align}
Therefore
\begin{multline}
    \label{eq:g01}
    g_0(\{\vec{0}\},\sigma,0^+) = 2 + 2\sqrt{1 - \frac{4}{\sigma}}\Bigg[ \\
    \arctanh\left(\sqrt{1-\frac{4}{\sigma}}\right)+\arctanh\left(\frac{2-\sigma}{\sqrt{\sigma (\sigma - 4)}}\right)\Bigg],
\end{multline}
and we have successfully found an explicit expression for $g_0$ using our alternative method of derivation.  We note that for $\sigma>4$ one can show analytically that \cref{eq:g01} is equivalent to \cref{eq:g00}, and thus \cref{eq:g01} is equivalent to \cref{eq:gn} for $n=0$.

For the higher $n$ cases we'll need the following asymptotic form for \cref{eq:ffunc}, which readily comes from the large $\sigma$ expansion of \cref{eq:g00}:
\begin{equation}
    f(\sigma,0^+)=2+i\pi-\ln(\sigma)+\mathcal{O}(\sigma^{-1}).
\end{equation}

\subsection*{Case: \texorpdfstring{$n=1$}{n=1}}
We may now consider the first non-trivial case, $n=1$.  In this case we may compare a brute force evaluation of \cref{eq:gn} with a numerical evaluation of an equivalent form that we now derive with the methods we have developed so far.

Considering $\Lambda$ to be a 1D lattice
\begin{align}
    &g_1(\Lambda,\sigma+i\eta,0^+)-g_1(\Lambda,\sigma_0+i\eta,0^+)\nonumber\\
    &=|\Lambda^*|\sum_{\ell\in\Lambda^*} \int_4^\infty dz\frac{\theta\left(\frac{z}{4}-1-\ell^2\right)}{\sqrt{z}}\nonumber\\
    &\pushright{\times\left[\frac{1}{z-\sigma-i\eta}-\frac{1}{z-\sigma_0-i\eta}\right]\qquad\qquad}\\
    &=|\Lambda^*|\sum_{\ell\in\Lambda^*}\int_{4 \left(1+\ell^2\right)}^\infty \frac{dz}{\sqrt{z}}\left[\frac{1}{z-\sigma-i\eta}-\frac{1}{z-\sigma_0-i\eta}\right]\\
    &=2|\Lambda^*|\sum_{\ell\in\Lambda^*}\Bigg(\frac{\arctanh\left(\frac{1}{2} \sqrt{\frac{\sigma +i \eta }{1+l^2}}\right)}{\sqrt{\sigma +i \eta }}\nonumber\\ &\pushright{-\frac{\arctanh\left(\frac{1}{2}  \sqrt{\frac{\sigma_0 +i \eta }{1+l^2}}\right)}{\sqrt{\sigma_0 +i \eta }}\Bigg)\qquad\qquad}\label{eq:arctanhsum}
\end{align}

We now have to be careful how we split up the sum in \cref{eq:arctanhsum}, to ensure we don't lose convergence. If we look at the asymptotic behaviour of each term, we see that both asymptotically decay like $1/|l|$, so we can add and subtract any function that has the same asymptotic form in order to separate the $\sigma$ and $\sigma_0$ dependent parts respectively. A convenient and natural choice then is simply $1/\sqrt{1+l^2}$. We also note that the $\eta$ is there to ensure the correct branch cut choice. Since $\sigma>4$, we may therefore safely take $\eta=0$ in the denominator of \cref{eq:arctanhsum}. We can then split our sum into two convergent sums:
\begin{widetext}
    \begin{multline}
        g_1(\Lambda,\sigma+i\eta,0^+)-g_1(\Lambda,\sigma_0+i\eta,0^+)
        =|\Lambda^*|\sum_{\ell\in\Lambda^*}\left(\frac{2}{\sqrt{\sigma}}\arctanh\left(\frac{1}{2}\sqrt{\frac{\sigma}{1+l^2}} +i \eta \right)-\frac{1}{\sqrt{1+\ell^2}}\right)\\
        -|\Lambda^*|\sum_{\ell\in\Lambda^*}\left(\frac{2}{\sqrt{\sigma_0}}\arctanh\left(\frac{1}{2}\sqrt{\frac{\sigma_0}{1+l^2}} +i \eta \right)-\frac{1}{\sqrt{1+\ell^2}}\right)\label{eq:g2bigboy}
    \end{multline}
    We can then rearrange \cref{eq:g2bigboy} as
    \begin{multline}
        g_1(\Lambda,\sigma,0^+)
        =\Bigg[g_1(\Lambda,\sigma_0,0^+)
        -|\Lambda^*|\sum_{\ell\in\Lambda^*}\left(\frac{2}{\sqrt{\sigma_0}}\arctanh\left(\frac{1}{2}\sqrt{\frac{\sigma_0}{1+l^2}} +i \eta \right)-\frac{1}{\sqrt{1+\ell^2}}\right)\Bigg] \\
        +|\Lambda^*|\sum_{\ell\in\Lambda^*}\left(\frac{2}{\sqrt{\sigma}}\arctanh\left(\frac{1}{2}\sqrt{\frac{\sigma}{1+l^2}} +i \eta \right)-\frac{1}{\sqrt{1+\ell^2}}\right).
    \end{multline}
    Since the left hand side is again independent of $\sigma_0$, we know the first term must be a function of only the lattice.
    Noticing that for asymptotically large $\sigma_0$ we have $g_1(\Lambda,\sigma_0,0^+)\sim f(\sigma_0,0^+)\sim 2+i\pi-\ln(\sigma)$, where the first asymptotic can be seen from \cref{eq:gn} and the second can be computed easily from \cref{eq:ffunc}. We can now numerically and asymptotically evaluate
    \begin{equation}
    \label{eq:a1}
    a_1(\Lambda)\equiv2+i\pi+\lim_{\sigma_0\to\infty}\left[-\ln(\sigma_0)-|\Lambda^*|\sum_{\ell\in\Lambda^*}\left(\frac{2}{\sqrt{\sigma_0}}\arctanh\left(\frac{1}{2}\sqrt{\frac{\sigma_0}{1+l^2}} +i \eta \right)-\frac{1}{\sqrt{1+\ell^2}}\right)\right]
\end{equation}
in order to obtain
\begin{equation}\label{eq:g1new}
    g_1(\Lambda,\sigma+i\eta,0^+)=a_1(\Lambda)+|\Lambda^*|\sum_{\ell\in\Lambda^*}\left(\frac{2}{\sqrt{\sigma}}\arctanh\left(\frac{1}{2}\sqrt{\frac{\sigma}{1+l^2}} +i \eta \right)-\frac{1}{\sqrt{1+\ell^2}}\right).
\end{equation}
\end{widetext}

\begin{figure}[!t]
    \centering
    \includegraphics[width=0.5\textwidth]{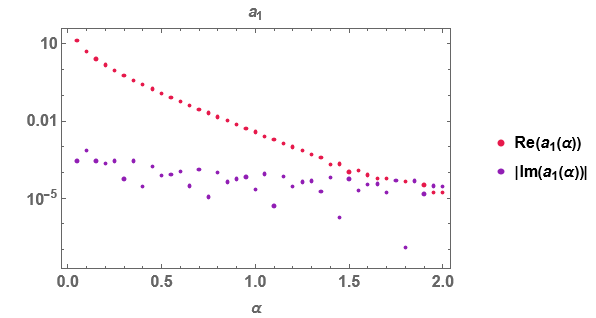}
    \caption{The real and imaginary parts of $a_1$ as a function of the lattice spacing $\alpha$ on a log scale. The absolute value of the imaginary part was taken, since it oscillates around $0$.}
    \label{fig:a1}
\end{figure}

\begin{figure*}
    \centering
    \includegraphics[width=0.95\textwidth]{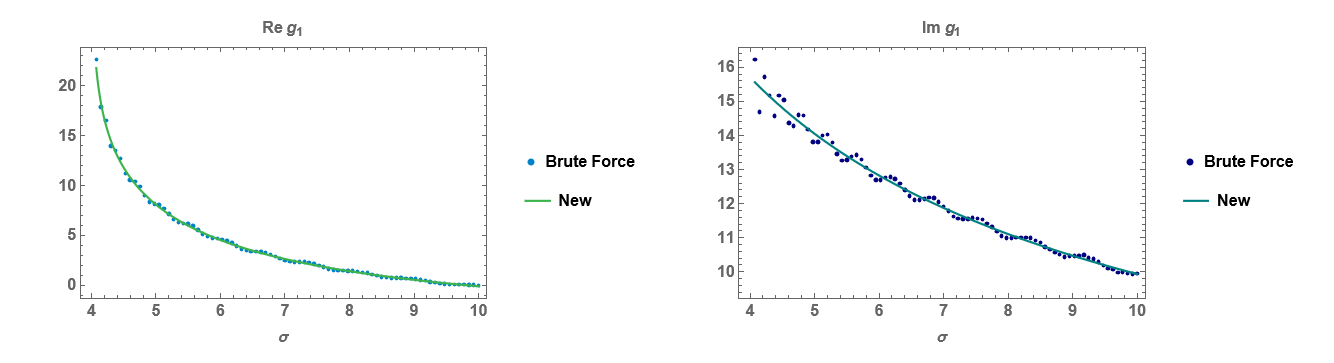}
    \includegraphics[width=0.95\textwidth]{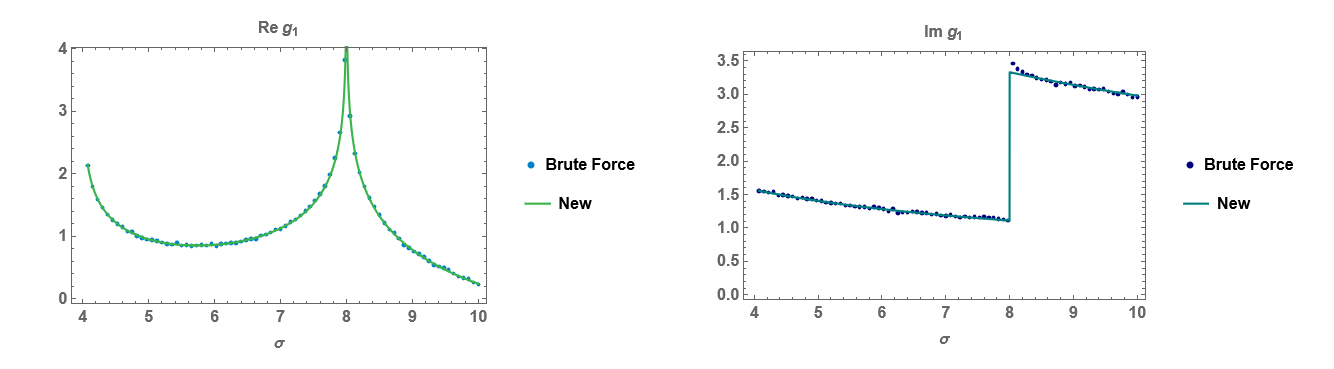}
    \includegraphics[width=0.95\textwidth]{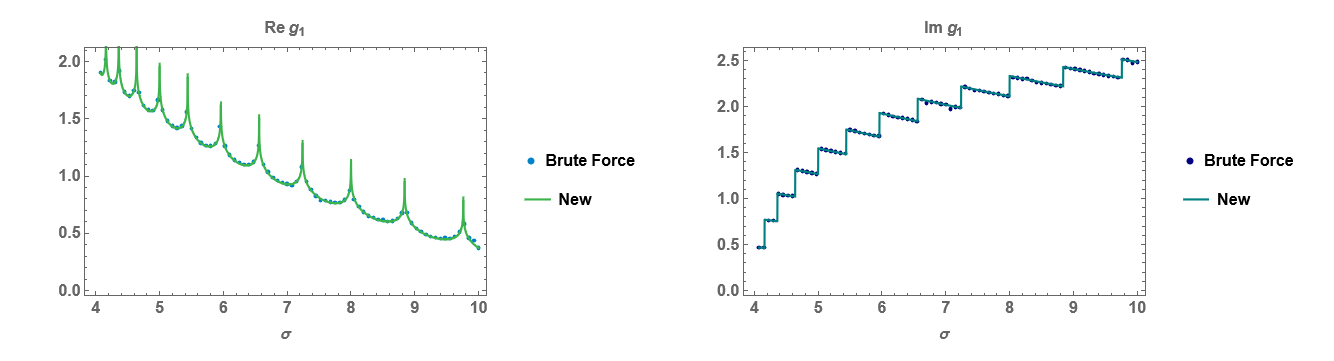}
    \caption{Comparison of real and imaginary parts of $g_1$ calculated with \cref{eq:g1new} (``New") versus \cref{eq:gn} with $n=1$ (``Brute Force"), with lattice spacing $=0.1,1,10$ from top to bottom.}
    \label{fig:g1}
\end{figure*}

\begin{figure*}
    \centering
    \includegraphics[width=0.95\columnwidth]{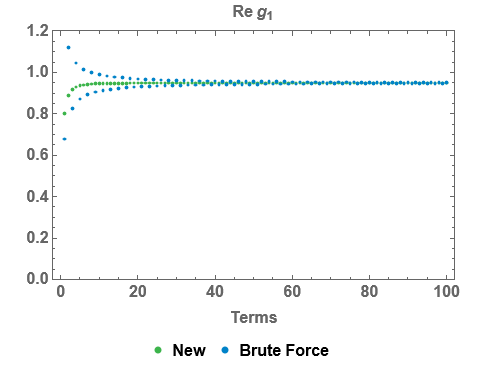}
    \includegraphics[width=0.95\columnwidth]{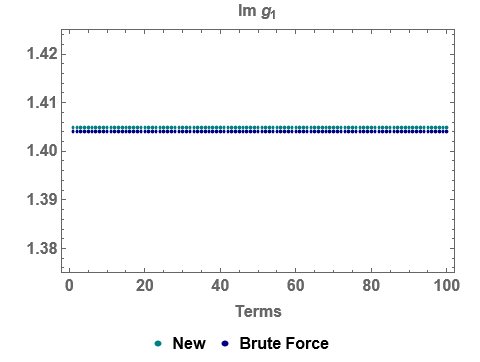}
    \caption{Convergence of partial sums of the real (left) and imaginary (right) parts of $g_1$ from a ``Brute Force'' evaluation of \cref{eq:gn} (blue) compared to the ``New'' method of \cref{eq:g1new} (green) on a lattice with lattice spacing unity, and at $\sigma=5$.}
    \label{fig:g1conv}
\end{figure*}

We plot $a_1(\Lambda)$ as a function of the lattice spacing $\alpha$ of the $n=1$ dimensional lattice $\Lambda$ in \cref{fig:a1}. We can then compare $g_1$ with an $n=1$ dimensional lattice at different lattice spacings calculated with \cref{eq:g1new} versus \cref{eq:gn} in \cref{fig:g1}. We note that there is a subtlety in the ordering of the limits in \cref{eq:a1} that affects numerics. There is no general way to move the $\log(\sigma_0)$ inside the sum in \cref{eq:a1}, and so we must take the limit of the upper bound of the infinite sum before we take the $\sigma_0\to\infty$ limit.  For some $\sigma_0$ we need to sum all vectors $\vec{\ell}$ that don't give vanishing $\sqrt{\sigma_0/(1+l^2)}$, but since we have a logarithmic divergence in $\sigma_0$, we need to have $\sigma_0\sim10^5$ to get precision on the order of $10^{-1}$ for $a_1$. Since the real part of $a_1$ is significantly larger than the imaginary part, the real part is much less sensitive to this limiting issue than the imaginary part, as can be seen in \cref{fig:a1}. This lack of precision in the imaginary part of $a_1$ ends up not mattering, since the imaginary part coming from the rest of $g_1$ in \cref{eq:g1new} is orders of magnitude larger than that coming from $a_1$. We can confirm that the lack of numerical precision of $a_1$ is irrelevant by examining the comparison of $g_1$ computed by brute force evaluation of \cref{eq:gn} and by evaluating the newly derived \cref{eq:g1new} in \cref{fig:g1}. We can furthermore see how much faster \cref{eq:g1new} converges when compared to \cref{eq:gn} with an $n=1$ dimensional lattice in \cref{fig:g1conv}.

\subsection*{Case: \texorpdfstring{$n=2$}{n=2}}
Considering $\Lambda$ to be a 2D lattice
\begin{align}
    &g_2(\Lambda,\sigma+i\eta,0^+)-g_2(\Lambda,\sigma_0+i\eta,0^+)\nonumber\\
    &=\frac{|\Lambda^*|}{\pi}\sum_{\ell\in\Lambda^*} \int_4^\infty dz\frac{\theta\left(\frac{z}{4}-1-\ell^2\right)}{\sqrt{z}\left(\frac{z}{4}-1-\ell^2\right)^{\frac{1}{2}}}\\
    &\pushright{\times\left[\frac{1}{z-\sigma-i\eta}-\frac{1}{z-\sigma_0-i\eta}\right]\qquad\qquad}\nonumber\\
    &=\frac{4|\Lambda^*|}{\pi}\sum_{\ell\in\Lambda^*} \left(\frac{\arcsin\left(\frac{1}{2}   \sqrt{\frac{\sigma +i \eta }{1+\ell^2}}\right)}{\sqrt{\sigma +i \eta } \sqrt{4 \ell^2- (\sigma+i \eta-4)}}\right.\nonumber\\
    &\left.\qquad- \frac{\arcsin\left(\frac{1}{2}\sqrt{\frac{\sigma_0 +i \eta }{1+\ell^2}}\right)}{\sqrt{\sigma_0 +i \eta } \sqrt{4 \ell^2- (\sigma_0+i \eta-4)}}\right).
\end{align}
We can again carefully split up the terms, where an appropriate and natural term to add and subtract is $1/1+\ell^2$,
\begin{widetext}
    \begin{multline}
        g_2(\Lambda,\sigma+i\eta,0^+)-g_2(\Lambda,\sigma_0+i\eta,0^+)
        =\frac{|\Lambda^*|}{\pi}\sum_{\ell\in\Lambda^*}\left(\frac{2\arcsin\left(\frac{1}{2}   \sqrt{\frac{\sigma}{1+\ell^2}}+i \eta \right)}{\sqrt{\sigma} \sqrt{\ell^2+1-\frac\sigma4-i \eta }}-\frac{1}{1+\ell^2}\right)\\
        -\frac{|\Lambda^*|}{\pi}\sum_{\ell\in\Lambda^*}\left(\frac{2\arcsin\left(\frac{1}{2}   \sqrt{\frac{\sigma_0}{1+\ell^2}}+i \eta \right)}{\sqrt{\sigma_0} \sqrt{\ell^2+1-\frac{\sigma_0}{4}-i \eta }}-\frac{1}{1+\ell^2}\right)\label{eq:arcsins}
    \end{multline}
    We can rearrange \cref{eq:arcsins} as
    \begin{multline}
        g_2(\Lambda,\sigma,0^+)=\Bigg[g_2(\Lambda,\sigma_0,0^+)
        -\frac{|\Lambda^*|}{\pi}\sum_{\ell\in\Lambda^*}\left(\frac{2\arcsin\left(\frac{1}{2}   \sqrt{\frac{\sigma_0}{1+\ell^2}}+i \eta \right)}{\sqrt{\sigma_0} \sqrt{\ell^2+1-\frac{\sigma_0}{4}-i \eta }}-\frac{1}{1+\ell^2}\right)\Bigg]\\
        +\frac{|\Lambda^*|}{\pi}\sum_{\ell\in\Lambda^*}\left(\frac{2\arcsin\left(\frac{1}{2}   \sqrt{\frac{\sigma}{1+\ell^2}}+i \eta \right)}{\sqrt{\sigma} \sqrt{\ell^2+1-\frac\sigma4-i \eta }}-\frac{1}{1+\ell^2}\right).
    \end{multline}
    Since the left hand side is independent of $\sigma_0$, we know the first term must be a function of only the lattice.
    With the same reasoning as for $n=1$, we can now introduce
    \begin{equation}
    \label{eq:a2}
    a_2(\Lambda)\equiv2+i\pi+\lim_{\sigma_0\to\infty}\left[-\ln(\sigma_0)-\frac{|\Lambda^*|}{\pi}\sum_{\ell\in\Lambda^*}\left(\frac{2\arcsin\left(\frac{1}{2}   \sqrt{\frac{\sigma_0}{1+\ell^2}}+i \eta \right)}{\sqrt{\sigma_0} \sqrt{\ell^2+1-\frac{\sigma_0}{4}-i \eta }}-\frac{1}{1+\ell^2}\right)\right]
\end{equation}
in order to obtain
\begin{equation}\label{eq:g2new}
    g_2(\Lambda,\sigma+i\eta,0^+)=a_2(\Lambda)+\frac{|\Lambda^*|}{\pi}\sum_{\ell\in\Lambda^*}\left(\frac{2\arcsin\left(\frac{1}{2}   \sqrt{\frac{\sigma}{1+\ell^2}}+i \eta \right)}{\sqrt{\sigma} \sqrt{\ell^2+1-\frac\sigma4-i \eta }}-\frac{1}{1+\ell^2}\right).
\end{equation}
\end{widetext}

\begin{figure}[!t]
    \centering
    \includegraphics[width=\columnwidth]{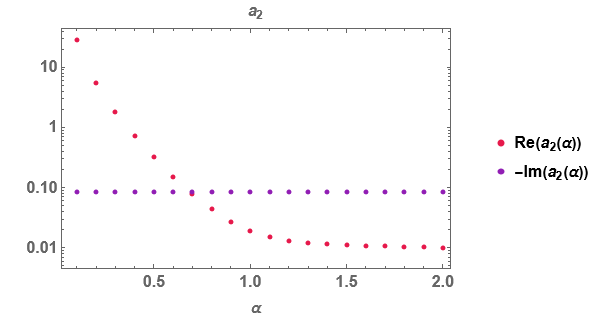}
    \caption{The real and (minus the) imaginary parts of $a_2$ for a square lattice as a function of the lattice spacing $\alpha$.}
    \label{fig:a2}
\end{figure}

\begin{figure}[!t]
    \centering
    \includegraphics[width=\columnwidth]{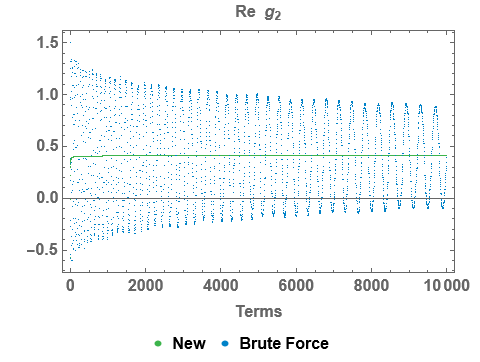}\\
    \includegraphics[width=\columnwidth]{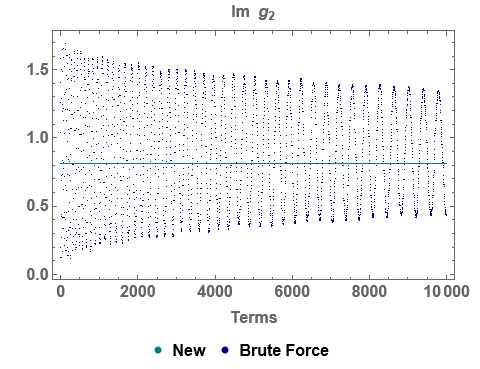}
    \caption{Convergence of partial sums of the real (top) and imaginary (bottom) parts of $g_2$ from a ``Brute Force'' evaluation of \cref{eq:gn} (blue) compared to the ``New'' method of \cref{eq:g2new} (green) on a lattice with lattice spacing unity, and at $\sigma=5$.
    }
    \label{fig:g2conv}
\end{figure}

We plot $a_2(\Lambda)$ as a function of the lattice spacing $\alpha$ for a square $n=2$ dimensional lattice in \cref{fig:a2}. In contrast to $a_1$, we find that \cref{eq:a2} is more numerically stable. The order $\mathcal{O}(\ell)$ in the denominator of \cref{eq:a2}, means that terms vanish for large $\ell$ no matter the value of $\sigma_0$.

One can see how much faster \cref{eq:g2new} converges compared to \cref{eq:gn} in an $n=2$ dimensional lattice in \cref{fig:g2conv}.

\begin{figure}
    \centering
    \includegraphics[width=\columnwidth]{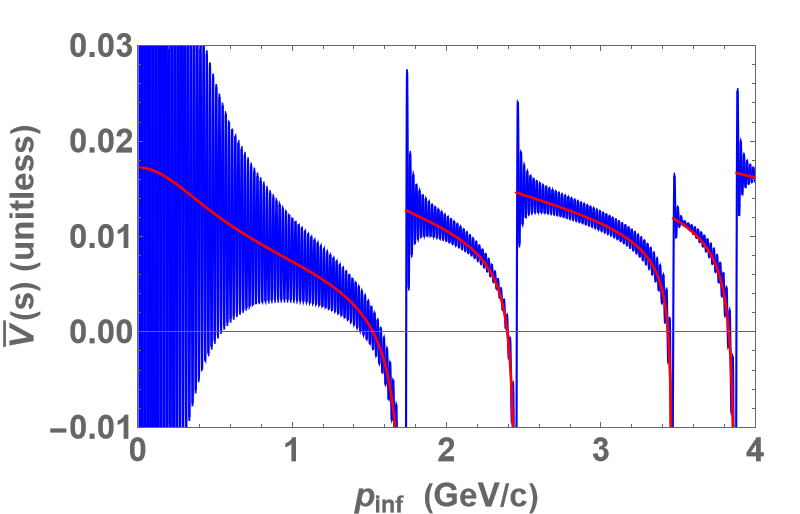}\\
    \includegraphics[width=0.5\columnwidth]{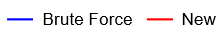}\\
    \includegraphics[width=\columnwidth]{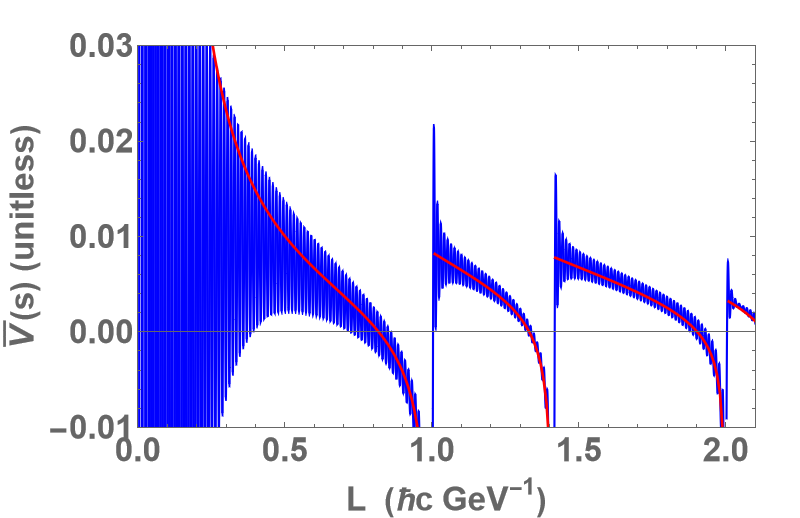}\\
    \includegraphics[width=0.5\columnwidth]{vscomplegend.png}
    \caption{(Top) $\Re \overline{V}_2(s,L,\mu)$ with $s=4(m^2+p^2)$, $m=0.5$ GeV, $\mu=1$ GeV, $L=1/\surd3$ GeV$^{-1}$ and scanning in $p$ naively summed to $10^4$ terms in blue (note that what might appear to be solid filled in, is in fact high-resolution oscillations, due to bad numerical behaviour) compared to the $1000$ terms of the newly derived result in red. (Bottom) Same as above, but with $p=1$ GeV and scanning in $L$.}
    \label{fig:vscomp}
\end{figure}

We can now compare the summation of $\sim10^4$ terms of \cref{eq:Vsapp} directly for the real part of the $s$-channel at $p=1$ GeV and scanning in $L$ (the length of both finite directions), as well as for constant $L_1=L_2=1/\surd3$ GeV$^{-1}$ scanning in $p$, to the summation of only $10^3$ terms of \cref{eq:g2new}. In \cref{fig:vscomp} we can see this comparison, making it clear why this detour to derive a more numerically friendly form was necessary. We understand the apparent disagreement of the two methods at small $L$ in \cref{fig:vscomp} as due to the exceptionally slow numerical convergence of the ``Brute Force'' method of evaluation of \cref{eq:Vsapp}, which does not accurately capture the divergence of $\overline V_2(s,L,\mu)$.  Intuitively, as $L$ decreases, the number of terms required to accurately determine $\overline V_2(s,L,\mu)$ using the ``Brute Force'' method diverges.  The advantage of the ``New'' method is that it sums over the dual lattice, so as $L$ decreases, the method becomes more and more accurate for any fixed number of terms included.  (In principle, then, the ``New'' method becomes inaccurate at large enough $L$; we however have not encountered yet a value of $L$ for which the method fails numerically.)

\clearpage

\end{document}